\documentclass[twocolumn,twocolappendix,numberedappendix]{openjournal}

\usepackage[utf8]{inputenc}
\usepackage[T1]{fontenc}
\usepackage{textcomp}
\usepackage{newunicodechar}
\newunicodechar{−}{-}

\usepackage{graphicx}
\usepackage{amsmath}
\usepackage{booktabs}
\usepackage{makecell}
\usepackage{bm}
\usepackage{xcolor}
\usepackage{orcidlink}
\usepackage{lastpage}
\usepackage{hyperref}
\hypersetup{colorlinks=true,linkcolor=blue,citecolor=blue,
            filecolor=blue,urlcolor=blue}
\usepackage{xurl}

\makeatletter
\def\frontmatter@above@affilgroup{\vspace*{-0.25in}}
\makeatother

\newcommand{\lrest}{\lambda_{\mathrm{rest}}}
\newcommand{\lobs}{\lambda_{\mathrm{obs}}}
\newcommand{\bopt}{\beta_{\mathrm{opt}}}
\newcommand{\buv}{\beta_{\mathrm{UV}}}

\begin{document}

\title{Between Little Red Dots and Star-Forming Galaxies: A Sequence in the Optical Continuum Slope}
\shorttitle{Between LRDs and Star-Forming Galaxies}
\shortauthors{K. Boonmee et al.}
\submitted{To be submitted to the Open Journal of Astrophysics}

\author{Kristiphong Boonmee\,\orcidlink{0009-0001-2570-2127}$^{1*}$,
        Akio K. Inoue\,\orcidlink{0000-0002-7779-8677}$^{1,2,3}$,
        Masafusa Onoue\,\orcidlink{0000-0003-2984-6803}$^{4,5}$, and
        Xingyao Cai\,\orcidlink{0000-0003-3095-7682}$^{2}$}
\email[$^*$E-mail: ]{kris@akane.waseda.jp}
\affiliation{$^{1}$Department of Applied Physics, School of Advanced Science and Engineering, Faculty of Science and Engineering, Waseda University, 3-4-1 Okubo, Shinjuku, Tokyo 169-8555, Japan}
\affiliation{$^{2}$Department of Pure and Applied Physics, Graduate School of Advanced Science and Engineering, Faculty of Science and Engineering, Waseda University, 3-4-1 Okubo, Shinjuku, Tokyo 169-8555, Japan}
\affiliation{$^{3}$Waseda Research Institute of Science and Engineering, Faculty of Science and Engineering, Waseda University, 3-4-1 Okubo, Shinjuku, Tokyo 169-8555, Japan}
\affiliation{$^{4}$Waseda Institute for Advanced Study (WIAS), Waseda University, 1-21-1 Nishi-Waseda, Shinjuku, Tokyo 169-0051, Japan}
\affiliation{$^{5}$Kavli Institute for the Physics and Mathematics of the Universe (WPI), UTIAS, The University of Tokyo, Kashiwa, Chiba 277-8583, Japan}

\begin{abstract}
The “Little Red Dots” (LRDs) uncovered by JWST are typically analyzed as a distinct class of objects; however, whether they are separated from ordinary star-forming galaxies by a genuine boundary has not been tested at the population level. Using a sample of $\sim3,500$ galaxies at $3 \lesssim z \lesssim 8$ with NIRSpec/PRISM spectra from the DAWN JWST Archive, we modeled their rest-frame UV-to-optical continuum emission with power laws and located the sources in the $\buv$--$\bopt$ plane. Alongside a dense locus of star-forming galaxies and a diagonal sequence which can be explained by dust reddening and/or quiescence, we identify a vertical sequence extending from the locus of star-forming galaxies into the LRD region. This sequence is smoothly and continuously populated, with no clear boundary between the two populations. Its middle section, $-1\lesssim\bopt\lesssim0$, hosts an intermediate population of $\sim50$ sources that falls outside the LRD selection boundary and has been largely overlooked; NIRCam imaging shows their rest-frame UV and optical median stacked images lie between point-like LRDs and extended star-forming galaxies. We modeled the vertical sequence as linear mixtures of stacked star-forming spectra and a black hole star (BH*) template derived from “The Cliff”, one of the most compelling BH* candidates, and we infer black hole-to-stellar mass ratios $\eta = M_{\mathrm{BH}}/M_* \approx 0.006-0.1$ at the LRD end. We propose $-1\lesssim\bopt\lesssim0$ as a practical selection window for possible transition candidates and argue that LRDs may be the compact, BH*-dominated extreme of a continuous population in which the relative contributions of the host and BH* vary smoothly.
\end{abstract}


\maketitle

\section{Introduction}
\label{sec:intro}

\subsection{The Nature of Little Red Dots}
\label{subsec:nature_lrd}

One of the most unexpected observational discoveries since the advent of the James Webb Space Telescope \citep[JWST;][]{JWST} is the identification of a population of compact and optically red sources, now commonly known as “Little Red Dots” \citep[LRDs;][]{Kocevski2023, MattheeLRDPioneer, Labbe2023, Barro2024, Greene2024}. These point-like sources are distinguished by a characteristic “V-shaped” spectral energy distribution (SED) \citep{Setton2025}, consisting of a blue rest-frame UV continuum paired with a red rest-frame optical continuum, as well as the presence of broad Balmer emission lines in most cases \citep[e.g.,][]{Harikane2023, Kocevski2023, Lin2024, HvidingRUBIES, Greene2024}. The inferred number densities \citep[$\sim10^{-(4-5)}\, \mathrm{Mpc}^{-3}\,\mathrm{mag}^{-1}$; e.g.,][]{Greene2024, Kocevski2024, Kokorev2024, Akins2025, Zhang2026b} of LRDs imply that this class of objects is remarkably prevalent in the high-redshift universe, persisting to redshifts of at least $z \sim 10$ \citep[e.g.,][]{Taylor2025b, Tanaka2025}. Although they appear to be less abundant in the low-redshift universe, local analogs are observed \citep[e.g.,][]{Chen2026, Lin2026, Ji2026b}, indicating that LRDs exist over a broad span of cosmic time. 

Initial interpretations posited that these sources might be dust-obscured active galactic nuclei (AGN) \citep[e.g.,][]{Kocevski2023, MattheeLRDPioneer, Harikane2023, Noboriguchi2023}, compact massive galaxies \citep[e.g.,][]{Labbe2025, Baggen2024, Perez-Gonzalez2024}, or composite systems in which both an AGN and its host galaxy contribute significantly to the observed emission \citep[e.g.,][]{Wang2024, Ma2025, Ronayne2025}. However, reproducing the full range of observed LRD properties within a single framework remains challenging and typically requires invoking non-standard scenarios that extend beyond conventional models of galaxies or AGNs.

A widely discussed framework for interpreting the spectral properties of LRDs is the “black hole star” or “black hole envelope” scenario \citep[e.g.,][]{Naidu2025, deGraaff2025a, Inayoshi2024, Kido2025, Liu2025, Begelman2025, Rusakov2026, BlackThunder2}. In this configuration, an accreting supermassive black hole is embedded within a dense, dust-poor, neutral gas envelope with a covering factor close to unity \citep[e.g.,][]{Torralba2026b, deGraaff2025a}. This envelope behaves as a pseudo-photosphere with an effective temperature near the Hayashi limit \citep[e.g.,][]{Inayoshi2024, Inayoshi2026, Kido2025}, thus producing an approximately blackbody optical continuum characterized by pronounced Balmer breaks, narrow Balmer absorption features, and emission features such as Fe~{\sc ii} and O~{\sc i} \citep[e.g.,][]{Labbe2024b, DEugenio2025, Tripodi2025, Torralba2026, BlackThunder2}. 

This model successfully accounted for the non-detections in the radio and X-ray regimes \citep[e.g.,][]{Ananna2024, Yue2024, Perger2025}, emission that would ordinarily be anticipated from relativistic jets and a hot corona in a standard AGN, as well as the apparent absence of reprocessed dust emission at infrared to sub-millimeter wavelengths \citep[e.g.,][]{Akins2025, Williams2024, Casey2024, Setton2025b}. Some of the earliest observational cases that motivated the BH* scenario include the discovery of sources such as “The Cliff” and “MoM-BH*-1” \citep{deGraaff2025a, Naidu2025}.

Although this model is capable of reproducing many of the observed spectral features, the physical origin of the broadened Balmer lines and the nature of the UV continuum are still subject to active debate \citep[e.g.,][]{Rusakov2026, MadauMaio2026, Scholtz2026, Brazzini2026, Asada2026}.

\subsection{Linking LRDs to the Star-Forming Population}
\label{subsec:linking_lrds}

Building on the composite picture introduced by \citet{Naidu2025} and subsequently adopted in studies including \citet{deGraaff2025b}, \citet{Barro2026} and \citet{Sun2026}, we represent the spectra of our LRDs as a linear combination of two components: (i) a stellar component associated with the host galaxy, which dominates the UV continuum, and (ii) a blackbody-like component, which dominates the optical continuum. 

As discussed in detail in \S\ref{subsec:linearmixeq}, our analysis is inspired by the framework of \citet{Sun2026}, who assumed that a blackbody-like continuum can be isolated by subtracting the host component that dominates the [O~{\sc iii}] emission line. In \citet{Sun2026}, empirical host-galaxy subtraction applied to a sample of 98 LRDs yields residual spectra that resemble a BH*, with inferred BH* fractions ranging from approximately $40\%$ to $99\%$ in the rest-frame $5500\,$\AA, although the distribution is strongly skewed: $\sim 90\%$ of the sample exhibit BH* fractions of $\gtrsim 60\%$, with only a few sources reaching values as low as $\sim 40\%$.

In this composite picture, the V-shaped continuum is not a unique spectrum but rather a sum of two different spectral shapes in the rest-frame UV-to-optical regime. This suggests that V-shaped selections could have been biased towards higher BH* fractions at $\gtrsim 60\%$, implying that more moderate BH*s may be hidden in brighter hosts and are being missed or classified as other objects such as a blue broad-line AGN. Recently, the work of \citet{Ando2026} showed that the LRD UV spectral slopes are systematically redder and the UV sizes are systematically more compact than star-forming galaxies at similar $M_{UV}$ and redshift. Furthermore, the Balmer break strength correlates with the UV spectral slope and anti-correlates with the UV size, suggesting that the UV-to-optical continuum shape is dependent on the central source.  

If the apparent steep decline in the number of sources with BH* fractions $\lesssim 60\%$ is primarily a selection artifact, one would expect an intermediate population that continuously connects LRDs with the general population of star-forming galaxies. An initial step toward uncovering such systems is presented in \citet{Weibel2026}, who developed a photometric selection strategy based on a composite BH* + host model and identified 241 candidate BH*-dominated sources over an area of $\sim 1,000\,\mathrm{arcmin^2}$ using JWST imaging data. This approach recovers composite BH* + host systems located outside the conventional V-shaped color-selection window adopted in previous LRD studies, thereby supporting the notion that traditional selection techniques are sensitive to only a subset of the underlying population. However, \citet{Weibel2026} restricted their analysis to systems in which the BH* component contributes $>80\%$ of the rest-frame optical SED, which preferentially targets the systems dominated by the central engine. The complementary population of host-dominated systems, including those with intermediate BH* fractions, has yet to be specifically targeted and studied.

Moreover, the inferred duty cycles ($\sim1\%$) and lifetimes ($\sim10-20\, \mathrm{Myr}$) imply that the BH* phase may be transient \citep{Sun2026}. LRDs may therefore represent a short-lived phase that emerges from, or fades back into, the general galaxy population \citep[e.g.,][]{Merida2026, Perez2026b}. By plotting our selected sample of high signal-to-noise galaxies in the optical slope against the UV slope ($\buv$--$\bopt$) plane without any prior selection of LRDs, a well-defined, approximately vertical sequence becomes apparent between the locus of star-forming galaxies and the region occupied by LRDs.

This consideration naturally raises the questions of whether the observed sequence can be reproduced through a linear combination of the spectrum of a BH* template and those of star-forming host galaxies, and whether the associated UV and optical morphologies evolve in a coherent manner along this sequence. Given that spatially extended UV emission has been reported for some LRDs \citep[e.g.,][]{Rinialdi2025, Chen2025, Zhang2025, Killi2024, Ji2026}, it is important to investigate how spatial extent varies along the sequence to assess the presence of systematic trends.

This paper is structured as follows. \S\ref{sec:dataselect} describes the data set along with the criteria used to select the spectroscopic sources. \S\ref{sec:method_and_results} presents the continuum fitting procedure; the construction of dust attenuation tracks and quiescent galaxy tracks using stellar population models; the derivation of linear mixing tracks; and the morphological analysis. \S\ref{discussion} provides a discussion and interpretation of the results obtained in \S\ref{sec:method_and_results}. Finally, \S\ref{conclusions} summarizes the key conclusions of this work. Throughout this paper, we adopt a flat $\Lambda\mathrm{CDM}$ cosmology according to \citet{Planck2018}, and AB magnitudes \citep{OkeGunn}.

\section{Data and Sample}
\label{sec:dataselect}

\subsection{JWST/NIRSpec Data}
\label{subsec:data}

In this study, we utilized all publicly available, highest-quality (\texttt{grade} = 3) Near Infrared Spectrograph \citep[NIRSpec;][]{NIRSpec} PRISM spectra in the version 4.4 data release from the DAWN JWST Archive\footnote{\url{https://dawn-cph.github.io/dja/index.html}} \citep[DJA;][]{BrammerValentino2025, RUBIES, Heintz2024}. The JWST spectroscopic datasets were reduced with the JWST Calibration Pipeline and then with \texttt{msaexp} \citep{Brammer2023}. For further information on MSA data reduction techniques, we refer our reader to \citet{RUBIES}. We only use PRISM spectra provided by DJA due to its broad wavelength coverage of $0.6 < \lobs < 5.3 \,\micron$, which extends to $5.5 \,\micron$ for v4 reductions \citep{Pollock2026}. As a result, we restrict our analysis to PRISM data in order to maximize the number of sources that can be modeled within our spectral fitting regions.

\subsection{JWST/NIRCam Data}
\label{subsec:NIRCam_data}

For the imaging data, we extracted the cutouts using the DJA Web interface in the NIRSpec table\footnote{\url{https://grizli-cutout.herokuapp.com/thumb}}. We note that not all targets have imaging coverage: some NIRSpec sources have only spectroscopic observations without associated imaging, and for others the corresponding Near-Infrared Camera \citep[NIRCam;][]{Rieke2023} data are not available through DJA. The available imaging products are provided as multi-extension FITS files processed with version 1.9.13 of the \texttt{grizli} pipeline \citep{Brammer2023b}, and all images are resampled onto a grid with a pixel scale of 0.05\arcsec{} per pixel. For further details on the NIRCam data reduction process, we refer our reader to \citet{Valentino2023NIRCam}. We extract and analyze only available images for the sources relevant to the vertical trend discussed in \S\ref{morphologyhowto}. The procedures for generating the cutouts and the subsequent data processing steps will be detailed further in that section.

\subsection{Sample Selection}
\label{subsec:sample}

We defined the fitting intervals for the UV and optical continua as $0.15 < \lrest < 0.36\, \micron$ and $0.40 < \lrest < 0.60 \, \micron$, respectively. Each spectrum was transformed to the rest-frame using the best redshift estimate provided by DJA. These wavelength ranges were chosen to minimize the increase in noise at shorter wavelengths and to exclude the Balmer break, thus improving the robustness of the continuum fitting. The choice of fitting interval ranges varies in the LRD literature. In Appendix~\S\ref{apdx:fittingregion}, we examine two additional continuum-fitting configurations: a broken power law anchored at the Balmer limit over $0.15$--$0.60\,\micron$, and the exact fitting range employed by \citet{deGraaff2025b} ($0.12$--$0.70\,\micron$) to demonstrate how different choices of fitting region affect the resulting distribution of sources in the $\buv$--$\bopt$ plane.

In our study, we adopted these fitting intervals in both the UV and optical regimes because our sample comprises evolved galaxies that exhibit prominent Balmer breaks and sources such as LRDs that exhibit blackbody-like radiation. For evolved galaxies, the continuum slope in the optical regime is more reliably parameterized when measured from $\lrest \sim 0.40 \, \micron$ onward. Furthermore, since the intrinsic spectral curvature of LRDs differs in concavity from the simple power-law parameterization $F_{\lambda} = \alpha \times \lrest^{\beta}$ employed in this study, especially for LRDs exhibiting deep Balmer breaks \citep[e.g.;][]{Naidu2025, deGraaff2025a}, we restricted the fitting region to mitigate systematic errors in our power-law fits, which improves the reliability of the inferred continuum slope measurements.

Furthermore, we excluded objects suffering from NIRSpec detector gaps in the spectra, to ensure that all relevant data points exist within the adopted fitting region. The redshift interval for our sample was carefully chosen to be $3.0 \leq z \leq 8.1$, in order to ensure spectral coverage throughout our defined fitting regions. The selection of our sample was further constrained by the following criteria to ensure that the spectra were well-detected and that the resulting fits were of sufficient quality:

\begin{enumerate}
    \item A minimum of 10 spectral pixels is required within each fitting region after applying the emission-line masking procedure (described in \S\ref{subsec:curvefit}) and removing invalid flux density values (NaN).

    \item The median S/N per pixel in the optical region and the UV region must both exceed the value of 2.
    
    \item The $1\sigma$ standard deviations of both the UV continuum slope ($\buv$) and the optical continuum slope ($\bopt$), derived from the curve-fitting procedure described in \S\ref{subsec:curvefit}, must be less than 1.
    \label{OurCondition}
\end{enumerate}

\begin{figure}
    \centering
    \includegraphics[width=0.95\columnwidth]{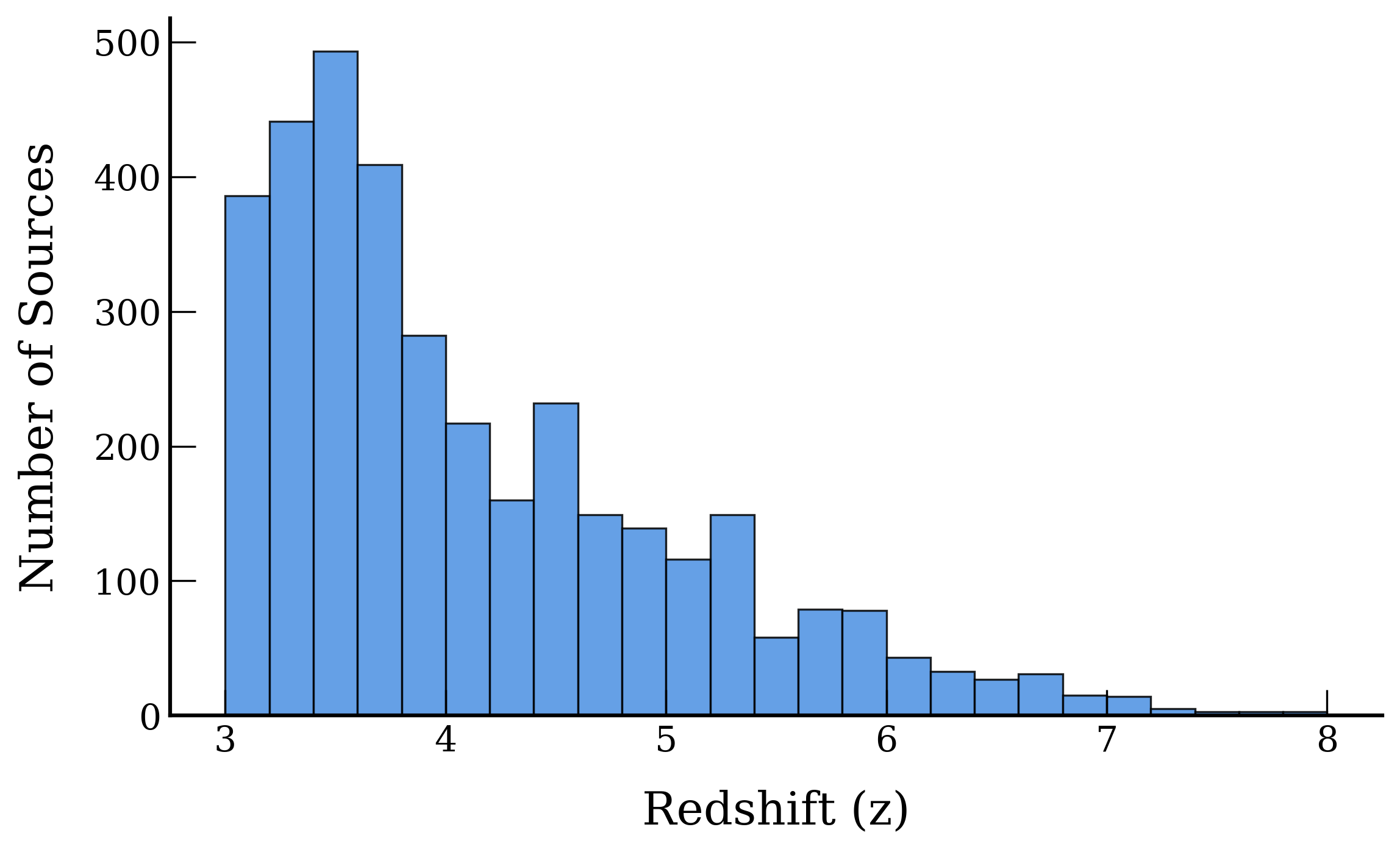} 
    \caption{The redshift distribution of our sample, comprising 3,566 sources, spanning an interval of \(3.0 \leq z \leq 8.1\).}
    \label{fig:plot1}
\end{figure}

The source \texttt{rubies-uds31-v4\_4233\_154183}, commonly known in the literature as “The Cliff” \citep{deGraaff2025a}, was manually added to the sample because it is a crucial source used to determine the linear mixing track (see \S\ref{subsec:linearmixeq}). Although The Cliff exhibits a high median S/N per pixel of 20 throughout the PRISM wavelength coverage and a median S/N per pixel in the optical region of 27, its median S/N per pixel in the UV regime is only 1.69. This relatively low UV S/N is likely due to the intrinsically weak UV continuum observed in The Cliff. After removing duplicate sources within an angular separation of 0.15\arcsec{} and reducing a triply lensed system to a single entry \citep{deGraaff2025b}, the final catalog contains 3,566 unique sources, as shown in Figure~\ref{fig:plot1}. The sample is primarily concentrated at redshifts of approximately $z \sim 3-4$, although the redshift coverage extends to $z \sim 8$.

\section{Methodology and Results}
\label{sec:method_and_results}

\subsection{UV and Optical Spectral Slopes}
\label{subsec:curvefit}

\begin{figure}
    \centering
    \includegraphics[width=0.95\columnwidth]{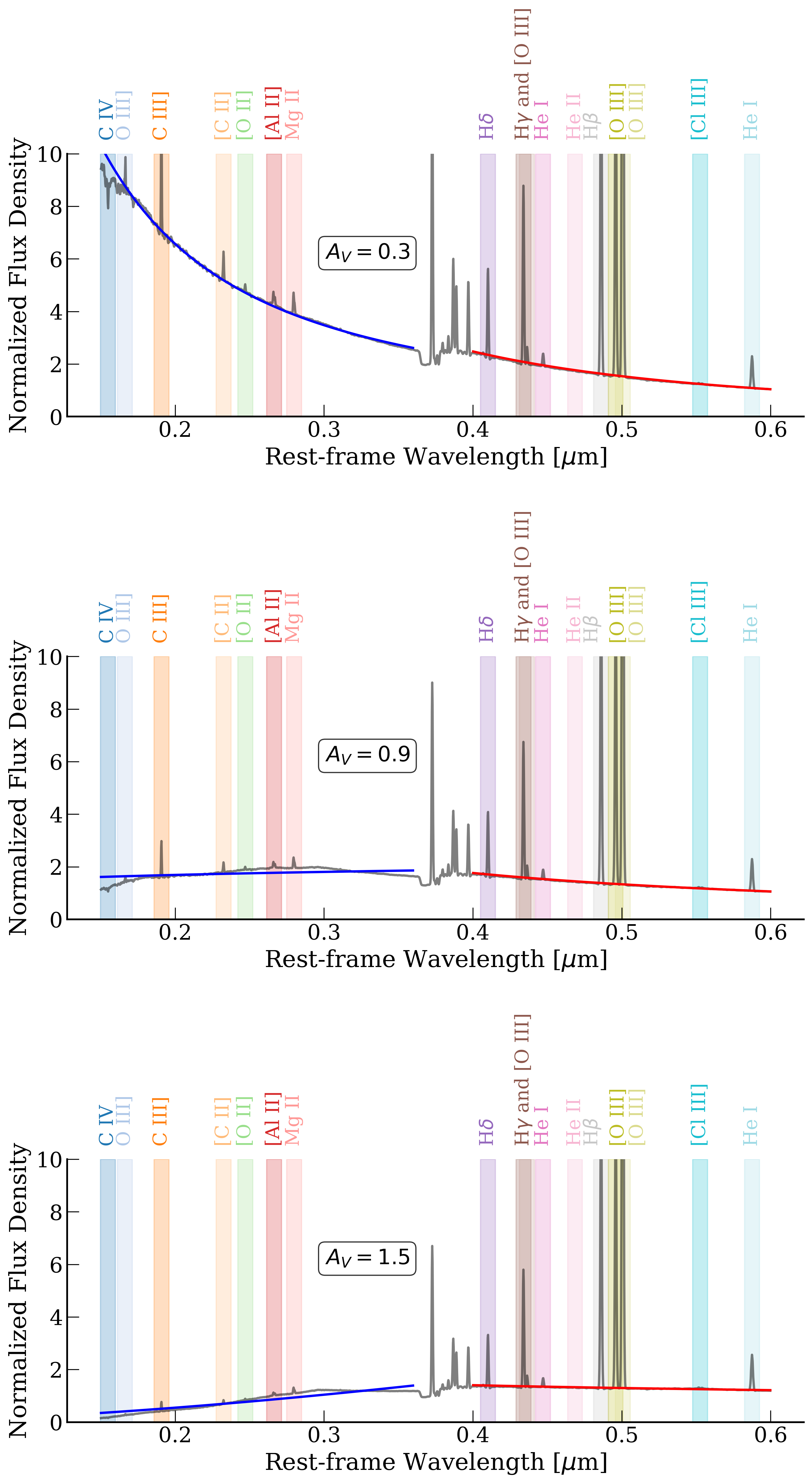} 
    \caption{Rest-frame spectra generated using BAGPIPES with SMC dust attenuation for increasing values of $A_V$ (top to bottom). The continuum is fit with power laws in the UV (blue) and optical (red) regions, excluding emission-line regions. The emission lines are masked using a width of 100\,\AA, with the name of each emission line displayed immediately above its corresponding feature.}
    \label{fig:SMCmasks}
\end{figure}

To measure the spectral slopes in the UV and optical regimes, we assume a power-law function $F_{\lambda} \propto \lrest^{\beta}$, where $\beta$ is the spectral slope, which is obtained by model fitting to the rest-frame spectra converted from the observed-frame based on the redshift.
Prior to model fitting, we normalized the flux by the median flux across the entire spectrum of each source to improve the stability of the fitting procedure.
The emission lines were excluded by applying a uniform masking window with a width of 100\,\AA\ in the rest-frame, which is sufficiently wide even in the low-resolution PRISM data. 
Figure~\ref{fig:SMCmasks} shows the emission-line masks applied to three example model spectra produced by BAGPIPES described in \S\ref{subsec:BAGPIPESDust}. 
After masking the emission lines, we employed the \texttt{curve\_fit} function from the Python package \texttt{scipy} \citep{Virtanen2020} to estimate the optimal rest-frame slope parameters, $\buv$ and $\bopt$, for the UV and optical regimes, respectively, by minimizing the chi-squared $(\chi^2)$ statistic.

\begin{figure}
    \centering
    \includegraphics[width=0.95\columnwidth,height=0.42\textheight,keepaspectratio]{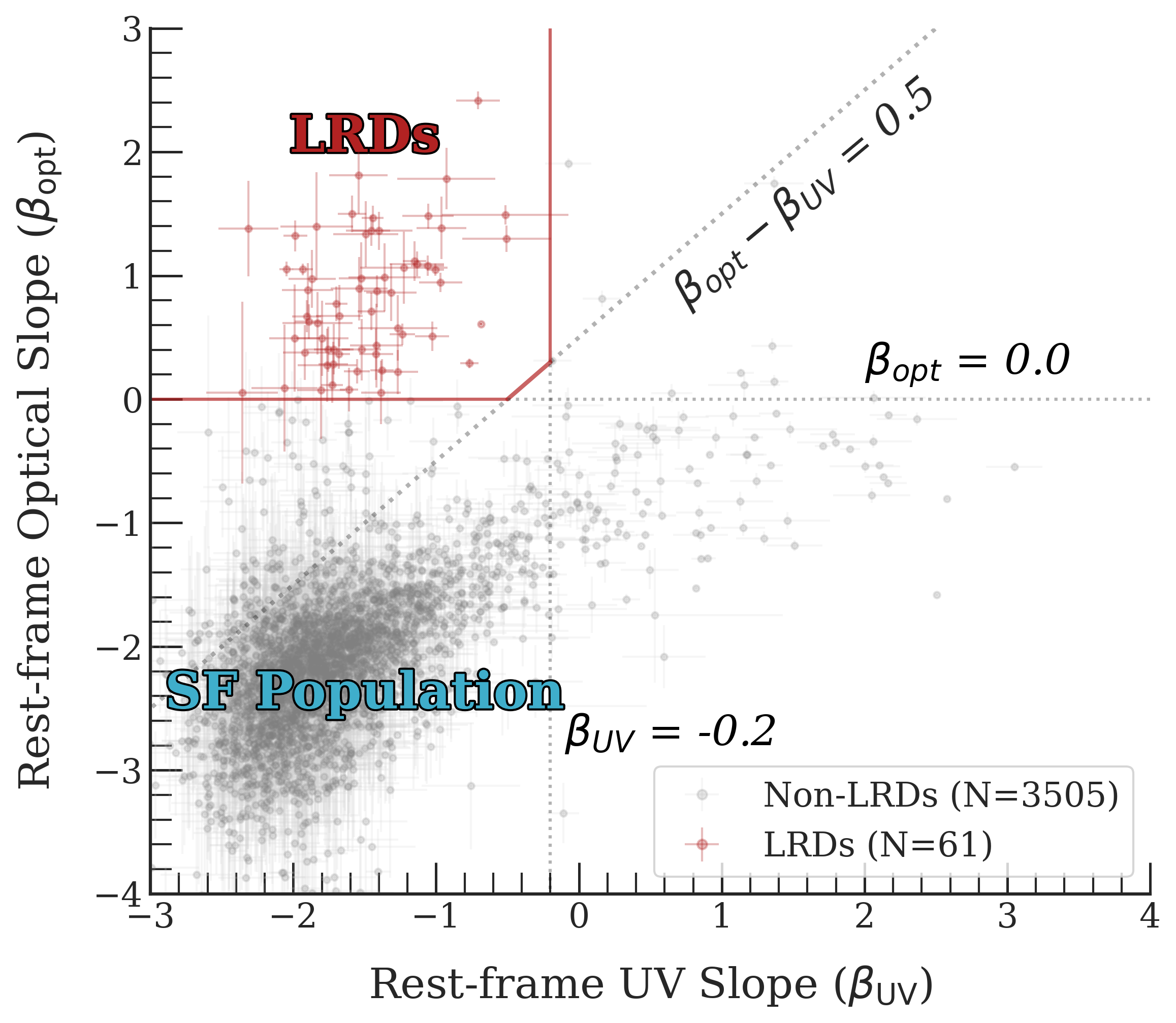} 
    \caption{$\bopt$ versus $\buv$ distribution for all sources in our sample, with the dotted gray lines indicating the V-shaped continuum selection boundaries defined by \citet{HvidingRUBIES} and the solid red lines forming the border of the LRD selection criteria adopted by \citet{deGraaff2025b}. The LRDs are shown as red points, while the gray points denote the Non-LRD sources.} 
    \label{fig:plot2}
\end{figure}

Figure~\ref{fig:plot2} shows the distribution of our measured $\buv$ and $\bopt$ for all sources in the sample. 
We find three trends: a large cluster of data points around $(\buv, \bopt)=(-2,-2)$, a diagonal sequence from the cluster towards redder $\buv$ and $\bopt$, and a vertical sequence from the cluster towards redder $\bopt$. The last sequence ends up in the area where the LRDs are located.

We adhere to the V-shaped continuum selection criteria as outlined by \citet{HvidingRUBIES} and adopted by \citet{deGraaff2025b} for their LRD selection, which are as follows:
\begin{enumerate}
    \item $\bopt > 0.0$
    \item $\buv < -0.2$
    \item $\bopt - \buv > 0.5$.
    \label{deGraaffcondition}
\end{enumerate}
From the parent catalog of 3,566 objects, 61 LRDs were identified. 

When comparing Figure~\ref{fig:plot2} with the corresponding diagram presented by \citet{NL_LRDS}, we observe that the same global trends are already vaguely apparent in their representation. Nevertheless, the application of our signal-to-noise quality cuts and our spectral curve fitting intervals enables these trends to be revealed with greater clarity. An analogous $\buv$--$\bopt$ diagram in \citet{Billand2026} also vaguely exhibits the same global behavior.

\subsection{Gaussian Kernel Density Estimation}
\label{subsec:kde}

\begin{figure}
    \centering
    \includegraphics[width=0.95\columnwidth]{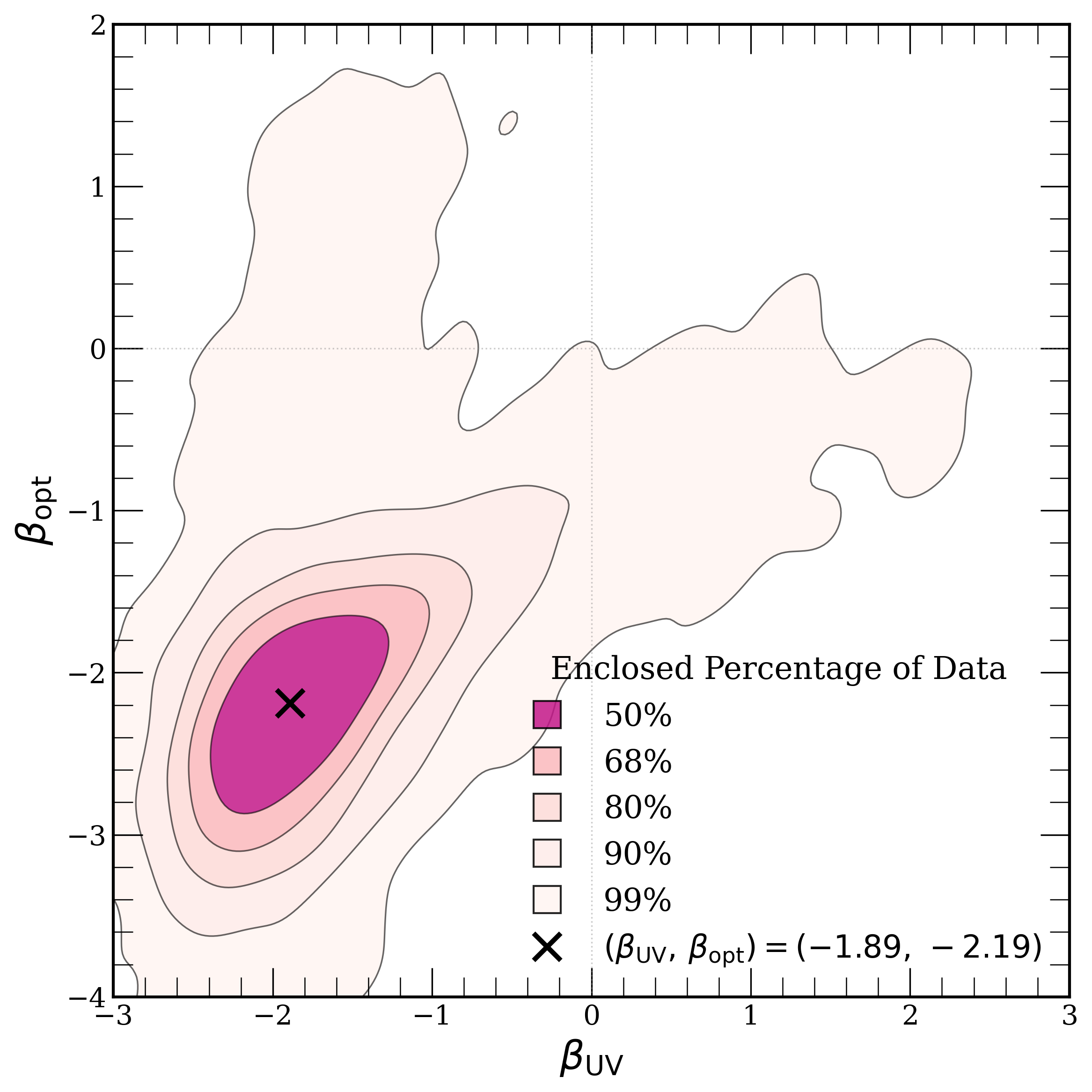} 
    \caption{Gaussian KDE contour, using Silverman's rule, of the distribution of $\buv$ and $\bopt$ for our sample. Contours represent enclosed probability regions of 50\%, 68\%, 80\%, 90\%, and 99\%, with darker colors indicating higher kernel densities.}
    \label{fig:contour}
\end{figure}

To better visualize the $\buv$ and $\bopt$ distribution of our sample and to determine a typical set of these values that characterize the “average” star-forming galaxy, we employed a non-parametric statistical method, Gaussian Kernel Density Estimation (Gaussian KDE), to construct an empirical probability density function (PDF) \citep{ChaconDoung2018}. 
For the implementation, we used the \texttt{gaussian\_kde} function provided in the \texttt{scipy} library \citep{Virtanen2020}. We first construct a $250 \times 250$ grid which defines the set of evaluation points for the kernels. The samples $\buv$ and $\bopt$ are then passed to the \texttt{gaussian\_kde} function to compute the corresponding kernel density estimate at each point in this grid. The result is shown in Figure~\ref{fig:contour}. We adopt the values of $\buv$ and $\bopt$ at the density peak inferred from Gaussian KDE using the Silverman bandwidth selection criterion\footnote{Silverman's rule derives the width of the Gaussian smoothing kernel from the covariance and the sample size, which scales in accordance with $h \propto n^{-1/6}$, where $h$ is the bandwidth factor and $n$ is the sample size for a 2D Gaussian KDE \citep{Silverman1986}}, as representative of the “average” star-forming galaxy, with $\buv = -1.89$ and $\bopt = -2.19$, respectively.

Visually, the three trends become more prominent: (i) an approximately 2D Gaussian component centered on the representative values of $\buv$ and $\bopt$ (-1.89, -2.19), (ii) a trend that is approximately diagonal, and (iii) a trend that is predominantly vertical.
Trend (i) consists of 80--90\% of the sample, while trends (ii) and (iii) each account for around 5\%.
Interestingly, in vertical trend (iii), the boundary between LRDs and star-forming galaxies is indistinct. The two populations are connected smoothly, suggesting a gradual transition between them.

We interpret the diagonal trend, using the galaxy SED modeling software: BAGPIPES, in terms of increasing dust attenuation and increasing stellar population age, as discussed in \S\ref{subsec:BAGPIPESDust} and \S\ref{subsec:BAGPIPESPassive}. In contrast, we interpret the vertical trend, which extends towards LRDs, using the linear mixing model described in \S\ref{subsec:linearmixeq}.

\subsection{BAGPIPES for Dust Track}
\label{subsec:BAGPIPESDust}

We used the Bayesian Analysis of Galaxies for Physical Inference and Parameter EStimation \citep[BAGPIPES; ][]{Carnall2018} software package to model the SEDs of galaxies and thereby characterize the diagonal trend observed in our $\bopt$ versus $\buv$ diagram. Using BAGPIPES, we generated the SED of a model galaxy defined by the following physical properties:

\begin{enumerate}
    \item Constant star formation history
    \item Stellar Metallicity: $Z=0.2Z_{\odot}$
    \item Ionization parameter: $-3.0 \leq \log(U) \leq -2.0$
    \item Dust attenuation/extinction: $0 \leq A_{V}~[{\rm mag}] \leq 4.5$ using the Calzetti attenuation formula \citep{Calzetti2000} and $0 \leq A_{V}~[{\rm mag}] \leq 2.5$ using the Small Magellanic Cloud (SMC) extinction formula \citep{Pei1992}.
    \label{BAGPIPEScondition}
\end{enumerate}

The ionization parameter $\log(U)$ of the nebular component in high-redshift galaxies spans a broad range, $-3.5 \lesssim \log(U) \lesssim -1.5$ \citep[e.g.,][]{Reddy2023, Cleri2026}. It is also established that the ionization parameter tends to be higher in UV-bluer galaxies than in UV-redder systems \citep[e.g.,][]{UVSlope}.

\begin{figure}
    \centering
    \includegraphics[width=0.95\columnwidth]{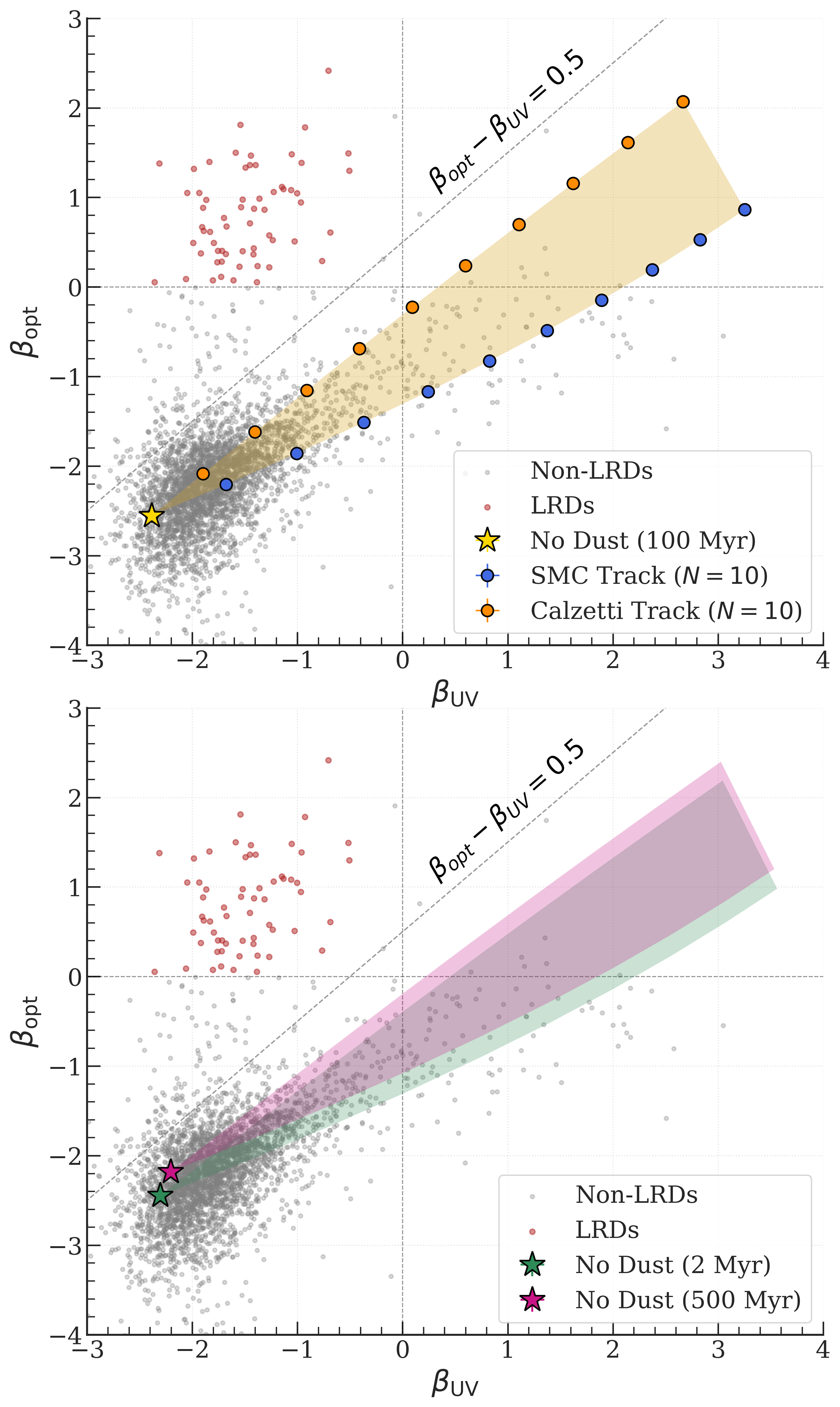} 
    \caption{$\bopt$ versus $\buv$ distribution for our sample and synthetic dust attenuation tracks generated with BAGPIPES. Gray points represent the non-LRDs, while red points indicate the LRDs. The yellow (100 Myr), pink (500 Myr), and green (2 Myr) five-pointed-stars mark the intrinsic dust-free model spectrum used as the starting point for the attenuation tracks. From these sources, dust extinction/attenuation tracks are generated using either an SMC extinction law (blue, top panel) with $A_{V}$ from 0 to 2.5 in steps of 0.25, or the Calzetti attenuation law (orange, top panel) with $A_{V}$ from 0 to 4.5 in steps of 0.45. The shaded region highlights the parameter space spanned by the two laws.}
    \label{fig:dusttrack}
\end{figure}

Therefore, for simplicity, we initially assumed a constant star formation rate over a timescale of 100 Myr and an ionization parameter of $\log(U) = -2.5$. The corresponding result is shown in the top panel of Figure~\ref{fig:dusttrack} depicted as a yellow star. In the bottom panel, we show two models comprising (i) a source with a constant star formation rate over 2 Myr, representing newly formed galaxies, and (ii) a source with a constant star formation rate over 500 Myr, representing more evolved galaxy populations. For the 2 Myr component, we assumed an ionization parameter of $\log(U) = -2.0$; for the 500 Myr component, we adopted the ionization parameter of $\log(U) = -3.0$.

For the metallicity parameter, we initially explored the range $0.05\,Z_{\odot} \leq Z \leq 0.5\,Z_{\odot}$ observed for galaxies in the JADES survey within our redshift range \citep[e.g.,][]{Curti2024}. However, we found that variations within this interval did not induce significant changes in the spectral slope values. Therefore, we adopted a fixed metallicity of $Z = 0.2\,Z_{\odot}$ for all sources. For masking in the slope measurement procedure, we mask the emission lines using a width of 100\,\AA\,corresponding to the peaks shown in Figure~\ref{fig:SMCmasks}.

We used well-known attenuation and extinction laws to characterize the dust tracks that attempt to explain the aforementioned diagonal trend. Galaxies at high redshift exhibit a wide diversity of dust attenuation curves, spanning from Calzetti-like attenuation laws to steeper Small Magellanic Cloud (SMC)–type extinction curves \citep[e.g.,][]{Salim2020, Hopkins2004}. To encompass this range, we constructed two sets of dust tracks: one based on the Calzetti attenuation curve \citep{Calzetti2000} and the other using the SMC extinction curve \citep{Pei1992}.
The region enclosed between these two tracks, as illustrated in Figure~\ref{fig:dusttrack}, is used to represent the plausible range of attenuation laws that could characterize the dust attenuation properties of high-redshift galaxies based on these two attenuation/extinction laws.

\subsection{BAGPIPES for Quiescent Track}
\label{subsec:BAGPIPESPassive}
Similarly, we used BAGPIPES to construct a model track for quiescent galaxies that may also account for the diagonal sequence observed in the $\bopt$ versus $\buv$ diagram. For this track, we adopted a simple exponentially declining ($\tau$) star formation history. We selected this model because it allows us to encompass both nearly instantaneous bursts and approximately constant star formation histories by exploring the limiting cases of the parameter $\tau$. Consequently, we obtain a more comprehensive characterization of the evolutionary tracks associated with quiescent galaxies. Observational studies have shown that quiescent galaxies in the early Universe typically exhibit low dust attenuation \citep[e.g.,][]{CarnallDust, deGraaffDust} and span a wide range of metallicities \citep[$-0.96\lesssim\log(Z/Z_\odot)\lesssim0.35$; ][]{CarnallMetal}. These empirical trends are reflected in the following set of assumptions adopted in the modeling of our quiescent galaxy track:

\begin{enumerate}
    \item Tau-model star formation history ($0.01 \leq \tau \leq 0.4\,$Gyr)
    \item Stellar Metallicity: $Z$ = $Z_{\odot}$
    \item Low dust attenuation: $A_{V}=0.1$
    \item $\log(U) = -3.0$.
    \label{BAGPIPEStau}
\end{enumerate}

\begin{figure}
    \centering
    \includegraphics[width=0.95\columnwidth]{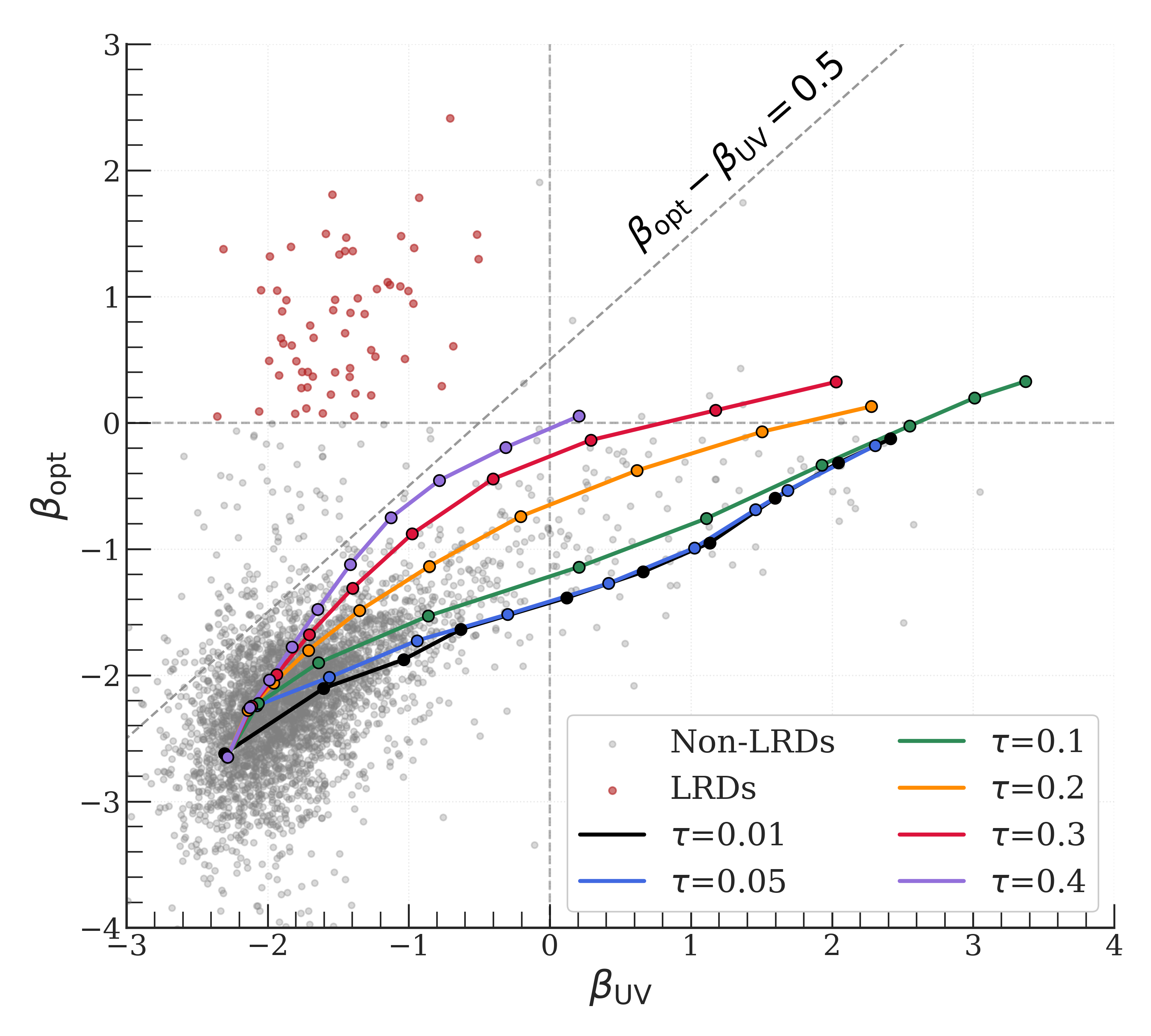} 
    \caption{$\bopt$ versus $\buv$ distribution for the sample population. The gray points represent non-LRDs, while the red points denote LRDs. The sequences illustrate the transition from blue, star-forming systems to redder, more evolved populations as the stellar population ages using the exponentially declining ($\tau$) star formation history model. Each colored track corresponds to a distinct value of $\tau$.}
    \label{fig:passtrack}
\end{figure}

Similar to previous studies of quiescent galaxies \citep[e.g.,][]{CarnallMetal, Ito2026}, we fixed the ionization parameter to $\log(U) = -3.0$ for all model tracks \citep{Byler2017}. We explored a range of star formation ages by varying the time elapsed since the onset of star formation as follows: from 0.01 to 1.01 Gyr for $\tau = 0.01$ Gyr and $0.05$ Gyr, from 0.01 to 1.51 Gyr for $\tau = 0.1$ Gyr and $0.2$ Gyr, and from 0.01 to 2.01 Gyr for $\tau = 0.3$ Gyr and $0.4$ Gyr. The maximum age of 2 Gyr is set by the age of the Universe at $z=3$. We normalized the spectrum by its median flux density and applied the emission-line masking procedure described in \S\ref{subsec:curvefit} before determining the UV and optical continuum slopes. The resulting evolutionary tracks are presented in Figure~\ref{fig:passtrack}.

\subsection{Linear Mixing Equation}
\label{subsec:linearmixeq}

We seek to interpret the approximately vertical trend evident in Figure~\ref{fig:contour} by adopting a model in which the SEDs of LRDs are represented as a linear combination of emission from the host galaxy and from dense gas surrounding the central AGN (BH*), such that
\begin{equation}
    \mathrm{LRD} = \mathrm{Host\ Galaxy} + \mathrm{BH}^*.
\end{equation}

To implement this, we require a stacked source that can be treated as a representative purely star-forming galaxy and another that can be regarded as a representative purely black hole-star-dominated system with a flat UV continuum. These two sources define our linear mixing sequence that attempts to characterize the change in shape of the SED between LRDs and star-forming galaxies. As shown in the work by \citet{Sun2026}, among all currently detected sources in the deep JWST fields, The Cliff is identified as one of the two objects that most closely approximate a pure black hole star (BH*), with the smallest associated uncertainty, $f_{\rm BH*}=99\pm1$\% \citep{Sun2026}. The second object, commonly referred to as “MoM-BH*-1”, is a high-redshift ($z=7.76$) LRD that was excluded from our sample due to our S/N threshold. Our selection of The Cliff as the primary target is motivated by its broad optical spectral coverage with high S/N due to its comparatively low redshift. We assume that The Cliff can be used to construct a BH* template under the condition that its contribution to the UV regime blueward of the Balmer limit is negligible, similar to the methodology employed by \citet{Barro2026}. We impose a flat UV continuum by setting the UV flux density to be constant and equal to the flux density measured at the Balmer limit. 

\begin{figure}
    \centering
    \includegraphics[width=0.95\columnwidth]{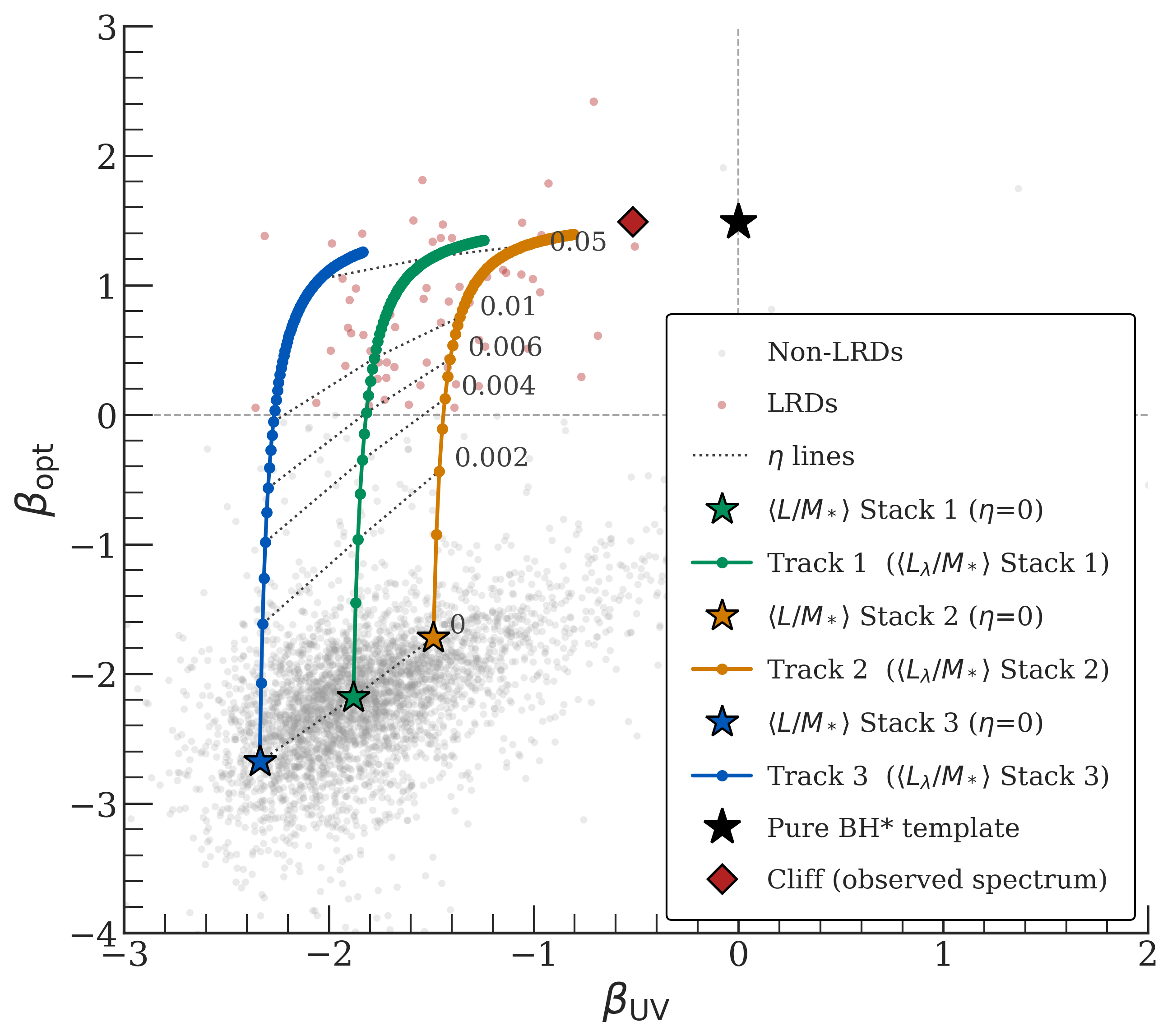} 
    \caption{Distribution of $\buv$ and $\bopt$ spectral slopes. Gray points represent the non-LRDs, while red points denote LRDs. The five-pointed-stars indicate the colors of the pure SF stacks (colored) and the pure BH* template (black). The Cliff's spectrum is separated from the LRD population in this figure and plotted independently as a red diamond point. The solid colored lines represent the mixing tracks as $\eta$ increases from $0$ to $0.1$ in increments of $0.001$. Dotted lines connect equal $\eta$ values along the three tracks, with each line labeled by its value.}
    \label{fig:3mixtracks}
\end{figure}

We construct three stacked spectra for the host galaxy term in order to create three linear mixing tracks that broadly encompass the vertical sequence leading up towards the LRD region. To construct our first stacked spectrum (green star in Figure~\ref{fig:3mixtracks}), we selected galaxies that lie near the highest density point centered at $\buv = -1.89$ and $\bopt = -2.19$, as defined in \S\ref{subsec:kde}. The selection was implemented by applying top-hat windows of $\buv \pm 0.05$ and $\bopt \pm 0.05$. This $\pm 0.05$ interval is an arbitrary selection criterion used solely to identify galaxies exhibiting comparable UV and optical slope values to be included in the stack. The other two stacked spectra are composed of galaxies with high S/N that satisfy $\buv \pm 0.05$ and $\bopt \pm 0.05$ about the central coordinates $(\buv = -1.49, \bopt = -1.79)$ and $(\buv = -2.29, \bopt = -2.59)$, respectively.
These two cases represent redder and bluer host galaxies than the average case as illustrated in Figure~\ref{fig:3mixtracks}.

For spectral stacking (or averaging), we apply proper normalization to avoid the dominance of a few bright sources. Here, we adopt a normalization by the stellar mass as follows: 
\begin{align}
\frac{L_{\lambda,\ \mathrm{comp}}}{M_*} &= \left\langle \frac{L_{\lambda,\ \mathrm{SF\,stack}}}{M_*} \right\rangle_{\mathrm{median}} + \eta \times \frac{L_{\lambda, \mathrm{Cliff}}}{M_{\mathrm{BH, Cliff}}}, 
\label{eq:linearmixing}
\end{align}
where the first term in the right-hand side denotes a median stack of the luminosity density of star-forming galaxies and the second term denotes the BH* term represented by the luminosity density spectrum of The Cliff (but with the modified UV part as explained above). The parameter $\eta=M_{\rm BH}/M_*$, is the ratio of the black hole mass, $M_{\rm BH}$, to the stellar mass, $M_*$, which controls the mixing of the two terms. Namely, $\eta=0$ is the case without (or negligible) BH*, while a larger $\eta$ is a case more dominated by BH*. 
The $M_*$ normalization is almost equivalent to the normalization by the optical luminosity which traces $M_*$ relatively well. However, if we adopted the optical luminosity normalization, the quantitative interpretation of the mixing of the two terms might be obscured. The $M_*$ normalization, on the other hand, offers a clear quantitative interpretation of the mixing through the parameter $\eta$.

We note that equation~\ref{eq:linearmixing} implicitly assumes constant mass-to-light ratios for both components. Specifically, the host term adopts the luminosity per unit stellar mass inferred from the median-stacked star-forming galaxies, whereas the BH* term adopts the luminosity per unit black hole mass inferred from The Cliff. Therefore, the mixing parameter $\eta$ should only be interpreted under the assumption that these average mass-to-light ratios are representative of the population.

The black hole mass of The Cliff, $\log_{10}(M_\mathrm{BH}/M_\odot) = 7.18$, was taken from \citet{deGraaff2025a}, who inferred it from the broad H$\alpha$ emission line using the local single-epoch AGN scaling relation from \citet{GreeneHo2005}.
However, this choice may be a significant overestimation of the black hole mass if the H$\alpha$ broadness is caused by electron or resonant scattering \citep{Rusakov2026}.
We also obtained the individual stellar masses of galaxies to be stacked by fitting the SED with \texttt{BAGPIPES} \citep{Carnall2018} adopting the model configuration and priors according to Table~\ref{tab:bagpriors}.

\begin{table*}
\caption{Model configuration and priors adopted for the \texttt{BAGPIPES} SED fitting of the stacked star-forming galaxies.}
\label{tab:bagpriors}
\centering
\begin{tabular}{llllll}
\toprule
Component & Parameter & Symbol / Unit & Range & Prior & Hyper-parameters \\
\midrule
General & Redshift & $z$ & --- & Fixed & $z = z_{\mathrm{spec}}$ (DJA) \\
\midrule
SFH & Star-formation history & $\mathrm{SFR}(t) \propto t\,e^{-t/\tau}$ & --- & Delayed-$\tau$ model & \\
 & Age since onset of SF & $t_{\mathrm{age}}$ / Gyr & $(0.01,\ t_{\mathrm{univ}}(z))$ & Uniform & \\
 & $e$-folding timescale & $\tau$ / Gyr & $(0.05,\ 10)$ & Uniform & \\
 & Total stellar mass formed & $\log_{10}(M_{\mathrm{formed}}/M_\odot)$ & $(5,\ 12)$ & Uniform & \\
 & Stellar metallicity & $Z$ / $Z_\odot$ & $(0.05,\ 2.5)$ & Uniform & \\
\midrule
Dust & $V$-band attenuation & $A_V$ / mag & $(0,\ 3)$ & Uniform & Calzetti law \\
\midrule
Nebular & Ionization parameter & $\log(U)$ & $(-4,\ -1)$ & Uniform & \\
 & Birth-cloud lifetime & $t_{\mathrm{bc}}$ / Gyr & --- & Fixed & $t_{\mathrm{bc}} = 0.01$ (default) \\
\midrule
Noise & White noise scaling & $a$ & $(1,\ 10)$ & Log-uniform & \\
\bottomrule
\end{tabular}
\end{table*}

For the stacked spectra of star-forming galaxies, we first shifted the observed wavelengths to the rest-frame and converted the flux density to the luminosity density using the best redshift estimates provided by DJA. We then resampled each spectrum onto a common wavelength grid using the Python package \texttt{spectres} \citep{CarnallSpectres}, before normalizing each spectrum with respect to their stellar masses. The stacking was performed by computing the median luminosity density per unit stellar mass at each wavelength across all galaxies used for each stack. After obtaining the three median stacked spectra for the pure star-forming galaxy population, we constructed three linear mixing sequences between the pure star-forming galaxy median stacks and the pure BH* template using equation~(\ref{eq:linearmixing}).

In Figure~\ref{fig:3mixtracks}, we vary $\eta \in [0,\ 0.1]$ in 100 uniformly spaced increments for each track. As $\eta$ increases from zero, the optical slope ($\bopt$) becomes substantially redder, whereas the UV slope ($\buv$) exhibits only minor reddening. Once $\eta$ increases above the value of $\approx0.01$ for all three tracks, the UV slope begins to increase drastically, while the optical slope exhibits only marginal changes. This rightward curvature in the $\buv$--$\bopt$ plane is consistent with the observed LRD population, in which many sources appear to have elevated UV slopes compared to the other intermediate sources along the vertical sequence.

\subsection{Morphological Analysis}
\label{morphologyhowto}
\subsubsection{Cutout Extraction}
\label{extract_cutout}

\begin{figure}
    \centering
    \includegraphics[width=0.95\columnwidth]{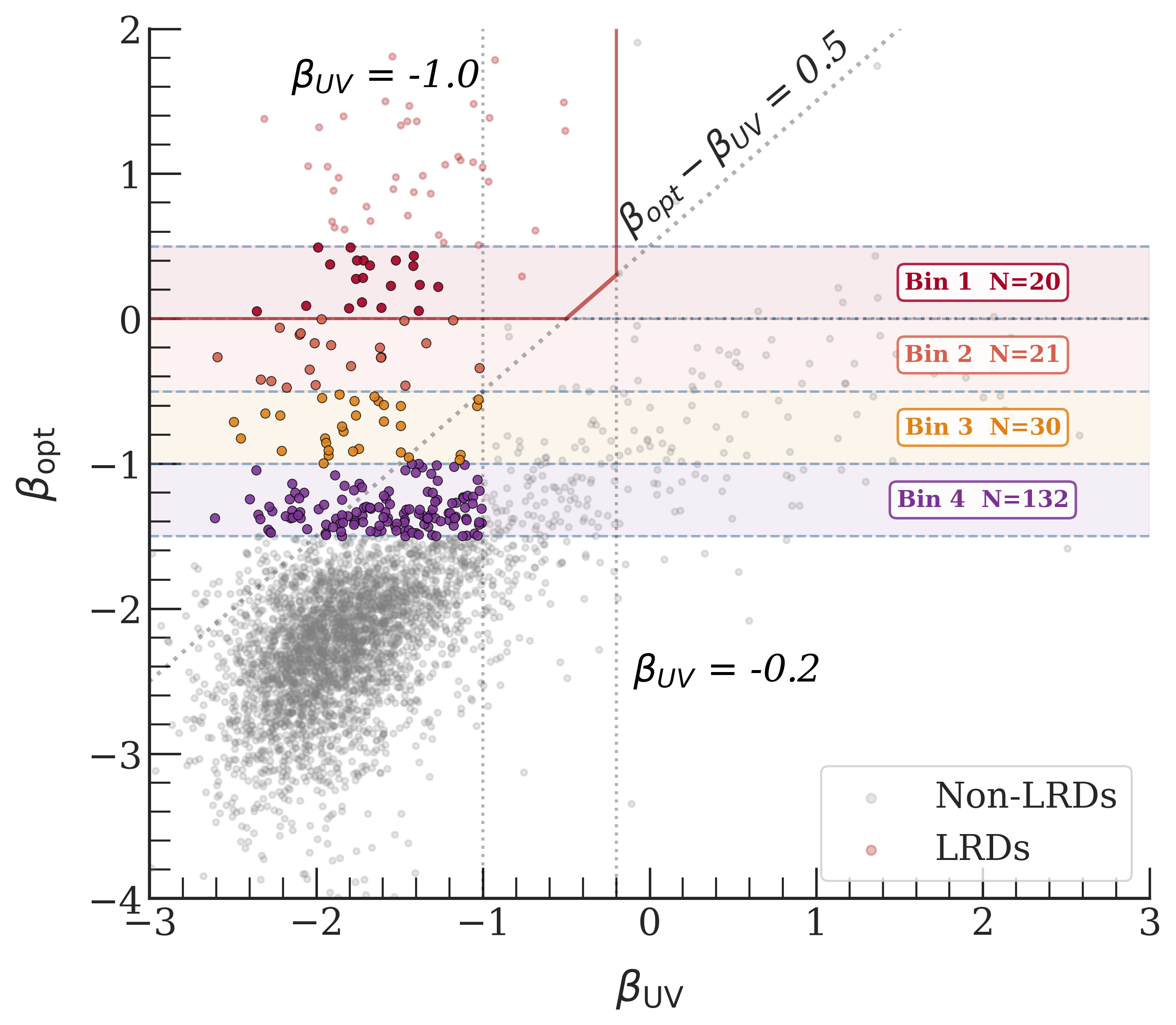} 
    \caption{$\bopt$ versus $\buv$ distribution for our sample divided into 4 equal bins: Bin 1 ranges from $0.0 < \bopt < 0.5$, Bin 2 ranges from $-0.5 < \bopt < 0.0$, Bin 3 ranges from $-1.0 < \bopt < -0.5$, and Bin 4 ranges from $-1.5 < \bopt < -1.0$. An additional criterion of $\buv < -1.0$ was imposed to ensure the selection of sources exclusive to the vertical trend. The number of sources per bin exhibits an increasing trend from Bin 1 through Bin 4.}
    \label{fig:betabin}
\end{figure}

We further examine the vertical trend through a morphological analysis using NIRCam imaging. Before source selection, we partition the sample into four equal bins along the $\bopt$ axis, characterizing the vertical sequence, within the range $-1.5 < \bopt < 0.5$. In addition, we impose a constraint of $\buv < -1.0$ to ensure that we exclusively select sources associated with the vertical trend, as shown in Figure~\ref{fig:betabin}.

\begin{figure}
    \centering
    \includegraphics[width=0.95\columnwidth]{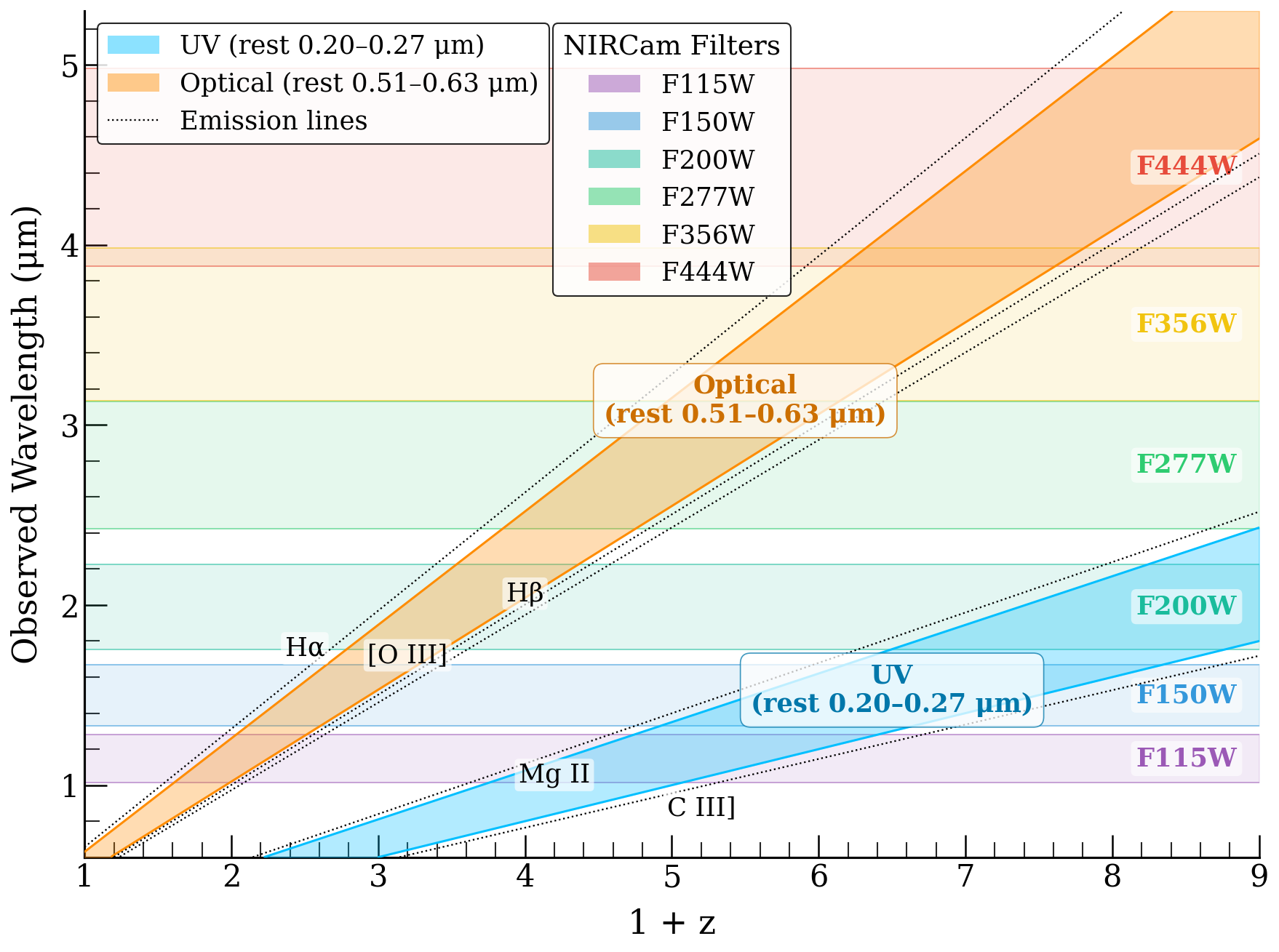} 
    \caption{The observed wavelength as a function of \(1+z\). The shaded orange region illustrates the evolution with redshift of the target optical wavelength range, while the shaded cyan region similarly traces the evolution of the ultraviolet (UV) wavelength range. The black dotted lines denote the locations of strong emission features as they evolve with redshift, which we aim to avoid. The horizontal shaded bands indicate the wavelength ranges covered by the NIRCam broadband filters, corresponding to the FWHM of each filter’s transmission profile.}
    \label{fig:lamobs_vs._1+z}
\end{figure}

To measure the continuum emission sizes of these sources as robustly as possible, we selected two rest frame wavelength intervals, one in the UV of $0.20 < \lambda_{\rm rest} < 0.27\,\micron$ and the other in the optical of $0.51 < \lambda_{\rm rest} < 0.63\,\micron$, which are relatively free of strong emission lines. 
We rely primarily on broad-band filters because medium-band filters are considerably shallower and are not as readily available as broad-band filters in many JWST observational programs. However, using broad-band filters makes it difficult to find sources that are completely free of strong emission lines, as shown in Figure~\ref{fig:lamobs_vs._1+z}. Consequently, we accept the risk of contamination from strong Balmer emission lines ($\mathrm{H}\alpha$ and $\mathrm{H}\beta$) and oxygen emission lines ([O~{\sc iii}]) in the optical regime and from prominent AGN emission features (e.g. C~{\sc iii}] and Mg~{\sc ii}) in the UV regime, in exchange for a significantly larger sample size. 

Therefore, we constructed two subsamples: (i) a Clean sample that is free of significant contamination from strong emission lines, and (ii) a Full sample which comprises all the selected imaging files, including those that allow some degree of emission-line contamination. The Clean sample comprises sources free of emission lines by requiring that all relevant emission features lie outside the full width at half maximum (FWHM) of the filter used. The blue and red wavelength limits of the FWHM are obtained from the JWST documentation\footnote{\url{https://jwst-docs.stsci.edu/jwst-near-infrared-camera/nircam-instrumentation}} for each filter. For the Full sample, we require that the FWHM coverage range of the selected filter encompasses at least half of the target observed-wavelength interval. After defining the selection criteria, we retrieved all available broad-band filter images with a size of $2\arcsec{}\times 2\arcsec{}$ for each source from the DJA.

\subsubsection{Extendedness Ratio}
\label{subsec:_ER}

We quantify the degree of morphological extension of each source using a metric that we define as the extendedness ratio (ER):
\begin{equation}
    \mathrm{ER} = \frac{\sigma_{\mathrm{fit}}}{\sigma_{\mathrm{PSF}}}, 
\label{eq:ER}
\end{equation}
where $\sigma_{\rm fit}$ and $\sigma_{\rm PSF}$ are the standard deviations of the Gaussian functions that approximately describe the radial profile of surface brightness of the source and the point-spread function (PSF), respectively.
The value of $\sigma_{\rm PSF}$ is converted from the PSF FWHM in pixels for each filter adopted from JWST Documentation\footnote{\url{https://jwst-docs.stsci.edu/jwst-near-infrared-camera/nircam-performance}}.
A smaller ER corresponds to a more compact source, whereas a larger ER indicates a more spatially extended source.

To fit the surface brightness profile of a source, the definition of the central position is important.
Following a visual inspection of the cutouts, we found that many sources exhibited an average offset of $\sim0.05\arcsec{}$ between the intensity peak and the initially adopted coordinate provided by DJA's NIRSpec table. For the majority of these sources, we manually selected the central RA and Dec values by adopting the intensity peak position. We identify several extended sources in bins 3 and 4 that exhibit multiple intensity peaks, potentially indicative of ongoing or recent merger activity. For such systems, we chose the center to be the peak that is closest to the RA and Dec coordinates as listed in DJA's NIRSpec catalog.

\subsubsection{Stacked Image}
\label{subsec:stacked}

\begin{figure*}
    \centering
    \includegraphics[width=0.85\textwidth]{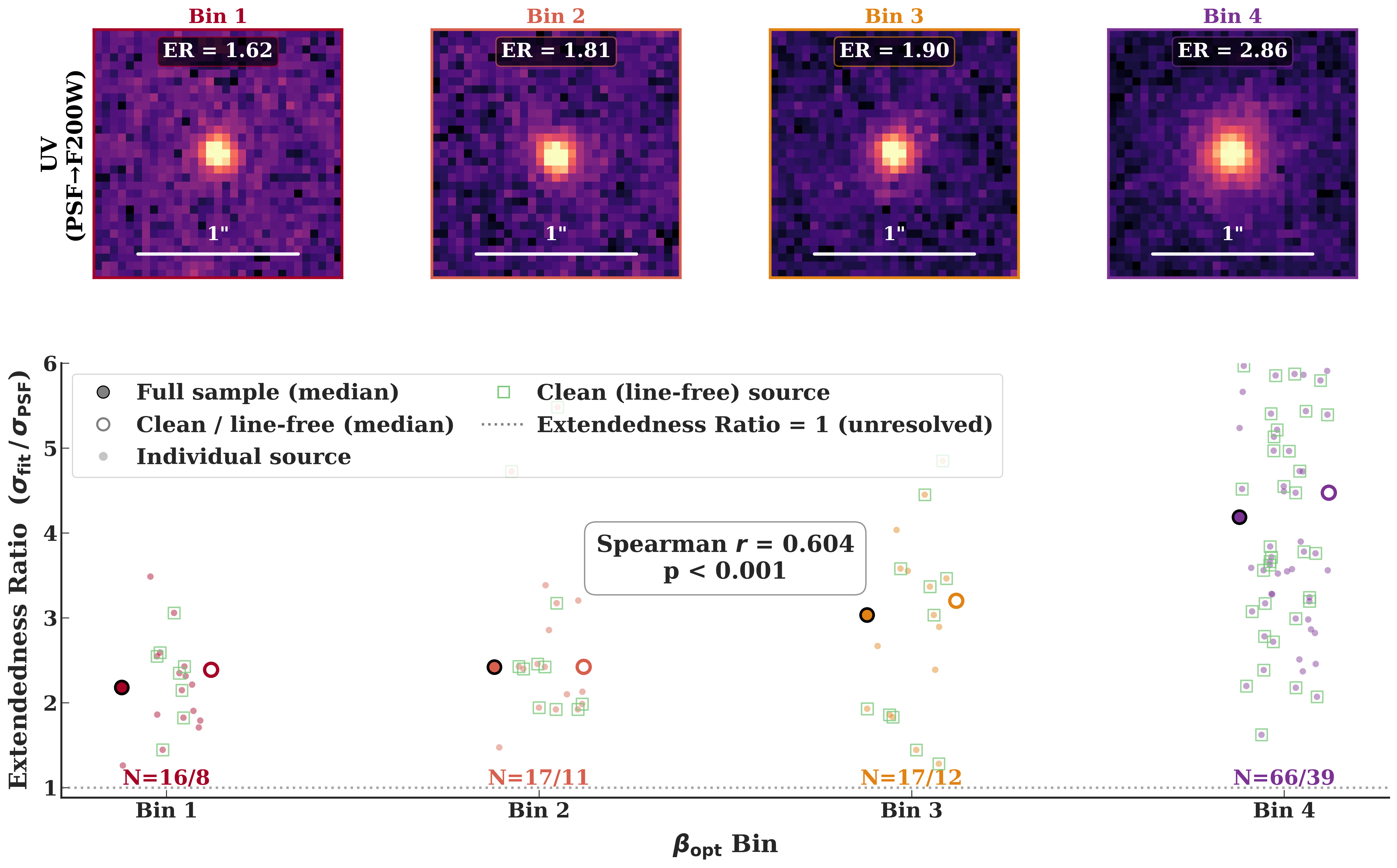}
    \caption{The median stacked image in each bin (top), defined in Figure~\ref{fig:betabin}, together with the corresponding extendedness ratio (ER; bottom). The upper panel displays, for each bin, the median of the stacked images. For each source, we use the broad-band filter that probes effectively identical rest-frame UV range ($0.20 < \lambda_{\rm rest} < 0.27\,\micron$). Prior to stacking, all images are convolved to match the PSF of the F200W filter, thereby homogenizing the PSF across the different filters. The lower panel illustrates the behavior of the ER across bins in the UV regime. Two median circles are displayed: the solid circle represents the “Full” sample, whereas the open circle corresponds to the “Clean” subsample after excluding sources with strong emission-line contamination. The horizontal dotted line denotes the hard lower bound on the ER. The numbers shown at the bottom of the panel indicate the total number of objects in the Full sample (left) and in the Clean subsample (right). The vertical axis has been truncated to enhance the visibility of the median values, therefore certain sources are not represented in the figure. The reason for ER value differences between stacked images and medians of individual measurements is explained in Appendix~\ref{apdx:ERdifference}.}
    \label{fig:ER+Median_Stack_UV}
\end{figure*}

\begin{figure*}
    \centering
    \includegraphics[width=0.85\textwidth]{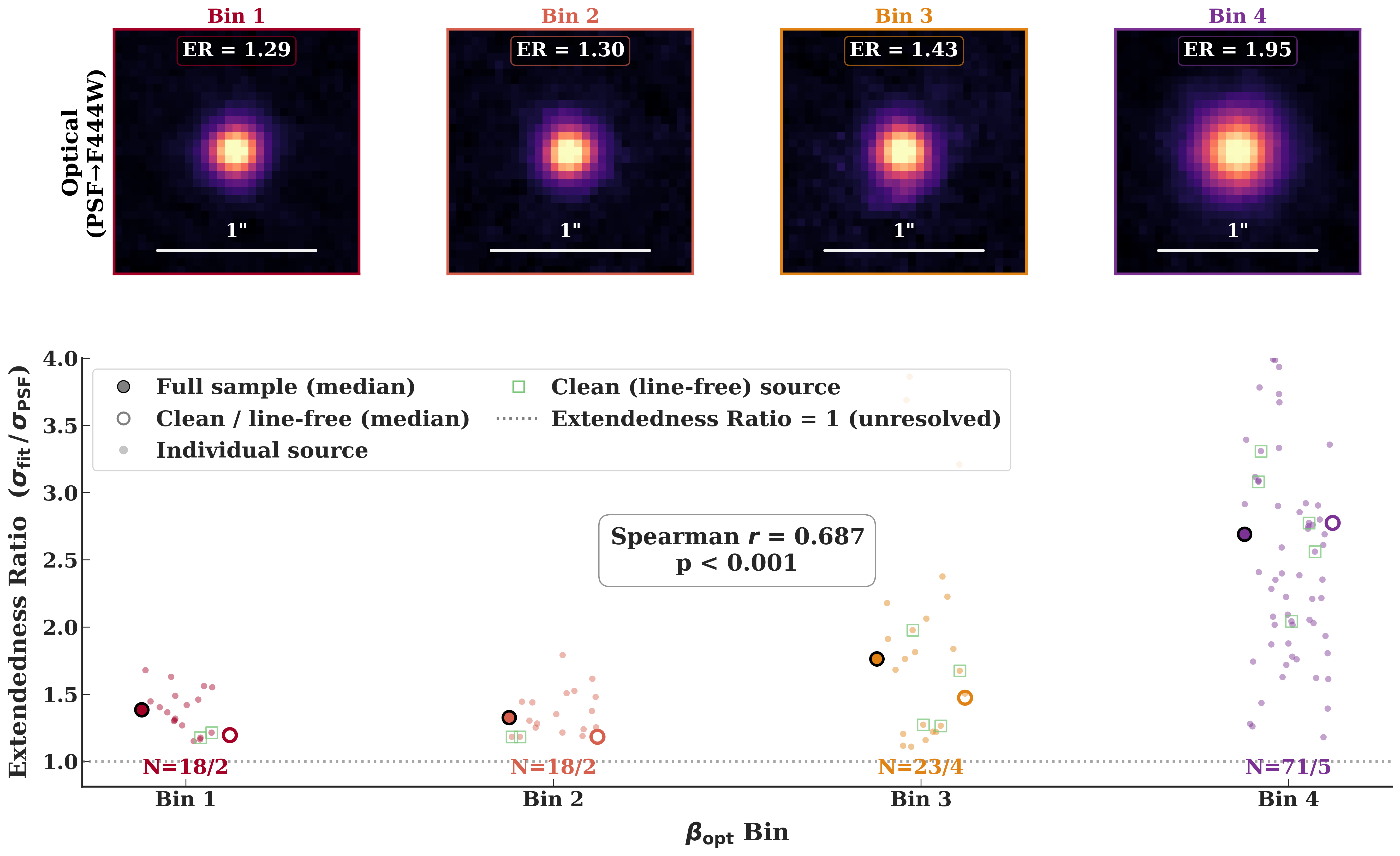} 
    \caption{The same as Figure~\ref{fig:ER+Median_Stack_UV}, but for the optical band. The median stacked images in the top row are created from, for each source, the broad-band filter that samples effectively the same rest-frame optical wavelength ($0.51 < \lambda_{\rm rest} < 0.63\,\micron$), with each image convolved to match the PSF of the F444W filter prior to stacking. See Appendix~\ref{apdx:ERdifference} for the reason of ER value differences between the stacked images and the medians of individual measurements.}
    \label{fig:ER+Median_Stack_Optical}
\end{figure*}

Using the same sample as the previous calculation of the ER, we construct a median-stacked image for each bin, representing the median morphology of the sources as a function of the bin. Since the source images in each bin are composed of a mixture of multiple filters, prior to image stacking, the images are convolved to a common reference PSF, chosen to be that of the longest-wavelength filter in each regime: F200W for the rest-frame UV and F444W for the rest-frame optical.
The Gaussian standard deviation required for this convolution is given by
\begin{equation}
    \sigma_{\mathrm{conv}} = \sqrt{\sigma_{\mathrm{target}}^{2} - \sigma_{\mathrm{source}}^{2}} ,
    \label{eq:convolution}
\end{equation}
where $\sigma_{\text{target}}$ represents the target PSF in each regime (F200W in the rest-frame UV and F444W in the rest-frame optical), and $\sigma_{\text{source}}$ corresponds to the intrinsic PSF associated with the native filter. For each source, we compute the required smoothing kernel, $\sigma_{\text{conv}}$, and convolve its image to the target PSF using the \texttt{gaussian\_filter} function from the \texttt{scipy} \citep{Virtanen2020} library.
Then, we normalize each convolved image by the peak flux in its central region before stacking, generate the median image for each bin, and compute the ER on these stacked images to verify the consistency between the ER trends and the morphology of the median stacks, as illustrated in Figures~\ref{fig:ER+Median_Stack_UV} and \ref{fig:ER+Median_Stack_Optical}. The detection of spatially extended rest-frame UV emission in the LRD bin (Bin 1, ER = 1.62) is consistent with the imaging decomposition analysis by \citet{Zhang2025}, who reported that extended host-galaxy components contribute up to $\sim60\%$ of the rest-frame UV flux in the F115W filter, but only about $12\%$ of the rest-frame optical flux in the F444W filter for a stack of $\sim200$ COSMOS-Web LRDs. In their study, the UV sizes ($\sim390\,\mathrm{pc}$) are significantly larger than the corresponding optical sizes ($\sim210\,\mathrm{pc}$). Furthermore, these findings are broadly consistent with the IFU decomposition of LRDs presented by \citet{Ishikawa2026}, who showed that the blue continuum and narrow emission-line components of their sample originate from spatially extended regions, whereas the red continuum is dominated by a compact, point-like core. 

\section{Discussion}
\label{discussion}

\subsection{Comparison with Other LRD Works}
\label{Other_LRD_Works}
\begin{figure}
    \centering
    \includegraphics[width=0.95\columnwidth]{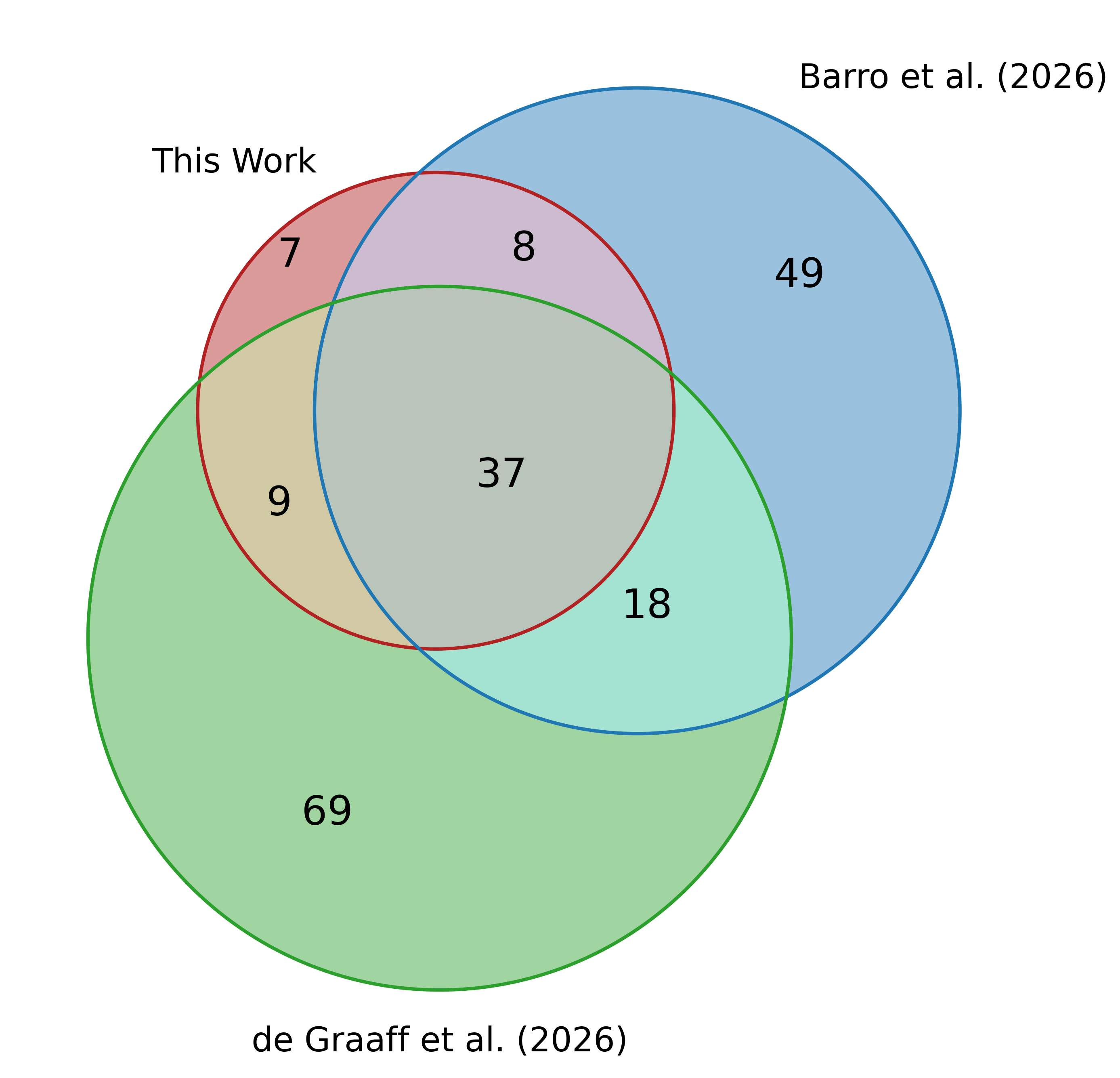} 
    \caption{A Venn diagram illustrating the correspondence between the samples of \citet{deGraaff2025b} and \citet{Barro2026} within the redshift interval of our study ($3.0 \leq z \leq 8.1$). The red region (N = 7) denotes objects classified exclusively as LRDs in this work. The green region (N = 69) marks sources identified as LRDs solely in \citet{deGraaff2025b}, while the blue region (N = 49) shows those identified exclusively as LRDs in \citet{Barro2026}. The remaining colors indicate the LRDs that were identified either in any two of the studies or across all three studies.}
    \label{fig:Venn}
\end{figure}

Given the absence of a strict, universally accepted definition of LRD in the literature---particularly the photometric color cuts, continuum-slope measurements, and compactness criteria---distinct selection criteria can yield different LRD samples even when applied to the same parent data set. To quantitatively assess these differences, we compared our spectroscopically selected LRD sample with two recent LRD compilations, specifically \citet{Barro2026} and \citet{deGraaff2025b}, based on JWST/NIRSpec observations. 

\citet{Barro2026} initiated their analysis from photometric catalogs and applied color cuts in imaging data. Their definition of the V-shaped continuum was based on a color–color selection using the $\mathrm{F115W}$, $\mathrm{F200W}$, and $\mathrm{F444W}$ filters. Specifically, they implemented the following four conditions: $\mathrm{F200W-F444W>1}$, $(\mathrm{F200W-F444W})>(\mathrm{F115W-F200W})+0.25$, $\mathrm{F115W-F200W}>-0.5$, and $\mathrm{F444W}\lesssim27.5$. They adopted a compactness criterion defined by the flux ratio $f_{\mathrm{F444W}}[0.5\arcsec{}]/f_{\mathrm{F444W}}[0.2\arcsec{}] < 1.5$. From their photometric LRD candidates, they selected 157 objects with existing NIRSpec/PRISM observations in DJA and subsequently imposed an additional requirement of an average $\mathrm{S/N} > 2$ in the rest-frame UV regime giving them 118 LRDs (112 within our redshift range). Their aim was to maximize the completeness of LRD detections using photometric data within JWST deep fields.

In contrast, the goal of \citet{deGraaff2025b} was to utilize all available NIRSpec/PRISM spectra in DJA to identify LRDs. They performed broken power-law fits, with the break fixed at the Balmer limit, in order to determine the UV and optical spectral slopes. Their compactness metric was defined as $f_{\mathrm{F444W}}[0.2\arcsec{}]/f_{\mathrm{F444W}}[0.1\arcsec{}] < 1.7$, resulting in a total of 146 unique LRDs (133 within our redshift range) after removing duplicates and reducing one triply lensed source to a single entry.

In our study, the main objective is to isolate sources with a high $\mathrm{S/N}$ that can be reliably modeled under our assumed power-law form (see \S\ref{subsec:sample}). Consequently, we did not optimize for sample completeness. Instead, we prioritized the selection of the highest-quality NIRSpec data to robustly characterize the underlying trends in the $\bopt$ versus $\buv$ plane as discussed in \S\ref{subsec:kde}.

Figure~\ref{fig:Venn} shows the overlap of our work with \citet{deGraaff2025b} and \citet{Barro2026} in the form of a Venn diagram\footnote{We used the most recent version of the catalog released by \citet{deGraaff2025b}, which is based on the DJA v4.5 data release.}. We cross-matched the DJA filenames between all three catalogs and constrained the redshift within the range of $3.0 \leq z \leq 8.1$. As a result of this redshift cut, 197 unique sources are defined as LRDs in these three works\footnote{Within the sample analyzed by \citet{Barro2026}, three sources were determined to be duplicates of sources reported in \citet{deGraaff2025b}.}. The sample of \citet{Sun2026}, is drawn almost entirely from the catalog of \citet{deGraaff2025b} with a small addition of 5 LRDs from the FRESCO surveys \citep{Oesch2023, MattheeLRDPioneer}. It is therefore not treated as an independent catalog here and is omitted from the Venn diagram. Through cross-matching, we found that only 37/197 sources (19\%) were agreed upon as LRDs by all three studies. 69 sources were uniquely identified as LRDs by \citet{deGraaff2025b}, 49 for \citet{Barro2026} and 7 for this work. 

Overall, Figure~\ref{fig:Venn} clearly demonstrates how sensitive LRD identification is to the chosen selection technique. Although we applied the same LRD spectral slope cut to \citet{deGraaff2025b}, we identified 15 LRDs that were not reported in their study. We highlight two sources that were marginally excluded by our selection, but were included in \citet{deGraaff2025b}: \texttt{mom-uds01-v4\_5224\_149501} ($\bopt \approx -0.2$) and \texttt{rubies-uds1-v4\_4233\_40579} ($\buv \approx -0.1$). These two sources were  omitted from our LRD sample because their slope values lie just outside the thresholds defined in our selection criteria (see \S\ref{subsec:curvefit}).

Nonetheless, as the majority of our sources (54/61) overlap with those reported in the other LRD literature, we consider our selection to be consistent with the broadly adopted definition of LRDs. For the remaining 7 sources that are unique to our sample, we quantified their compactness using our ER, as described in \S\ref{subsec:_ER}, and compared their ER values with those of the 54 LRDs identified in previous LRD studies. The results are summarized in Table~\ref{tab:er_data}. It is important to note that of the 54 sources previously reported in \citet{deGraaff2025b} or \citet{Barro2026}, only 47 have available UV imaging and 51 have available optical imaging in the specific filter required for this analysis. Furthermore, among the 7 exclusive sources presented in this work, only 5 have imaging in the desired filter in both the UV and the optical bands.

\begin{table}
\caption{Summary of the groups and wavelength band (UV or optical), including source counts and median extendedness ratio (ER). $N_{\rm src}$ denotes the number of sources with available imaging files required to calculate the ER. The last column reports the median ER of the full sample in Bin 1 for reference, as presented in Figure~\ref{fig:ER+Median_Stack_Optical} for the optical regime and Figure~\ref{fig:ER+Median_Stack_UV} for the UV regime.}
\label{tab:er_data}
\centering
\begin{tabular}{llccc}
\toprule
Group & Band & $N_{\rm src}$ & \makecell{Median \\ ER} & \makecell{Bin 1 \\ ER (ref.)} \\
\midrule
Exclusive in this work & UV      & 5  & 2.55 & 2.18 \\
Exclusive in this work & Optical & 5  & 1.46 & 1.38 \\
\midrule
LRDs already reported  & UV      & 47 & 2.08 & 2.18 \\
LRDs already reported  & Optical & 51 & 1.29 & 1.38 \\
\bottomrule
\end{tabular}
\end{table}

As shown in Table~\ref{tab:er_data}, the median ERs of the LRDs in the UV and optical bands are more compact than the median ER of sources in Bin 1 of Figure~\ref{fig:ER+Median_Stack_UV} (UV) and Figure~\ref{fig:ER+Median_Stack_Optical} (Optical) by around $5\%$ and $7\%$, respectively. In contrast, for sources exclusive to this work (5/7 with available imaging files), their median ERs are larger than the median ER of Bin 1 in the UV and optical regimes by approximately 17\% and 6\%, respectively. The increased degree of spatial extension exhibited by the exclusive objects identified in this work is likely the primary reason they were not included in earlier studies that imposed stricter compactness selection criteria.

It is also important to note that 6 of these 7 sources exclusive to our work are represented in Bin 1 of Figure~\ref{fig:betabin}, which is the bin that contains the lowest $\bopt$ values to be classified as an LRD in terms of the V-shaped continuum criteria adopted by \citet{deGraaff2025b}, while the single remaining source lies just beyond the bin edge ($\bopt = 0.5$) with an optical slope of $\bopt=0.676$---well within the LRD area in the $\buv$--$\bopt$ plane. We therefore conclude that exclusive sources presented here are sources that exhibit V-shaped continua and appear more extended than the typical compactness thresholds adopted in previous LRD works. This suggests that the compactness criterion adopted for LRDs may be overly restrictive and may exclude more spatially extended systems, as these sources are located near the lower $\bopt$ boundary of the LRD population.

\subsection{Locus of Star-Forming Galaxies around the Highest Density Point}
\label{subsec:central_bulge}

\begin{figure}
    \centering
    \includegraphics[width=0.95\columnwidth]{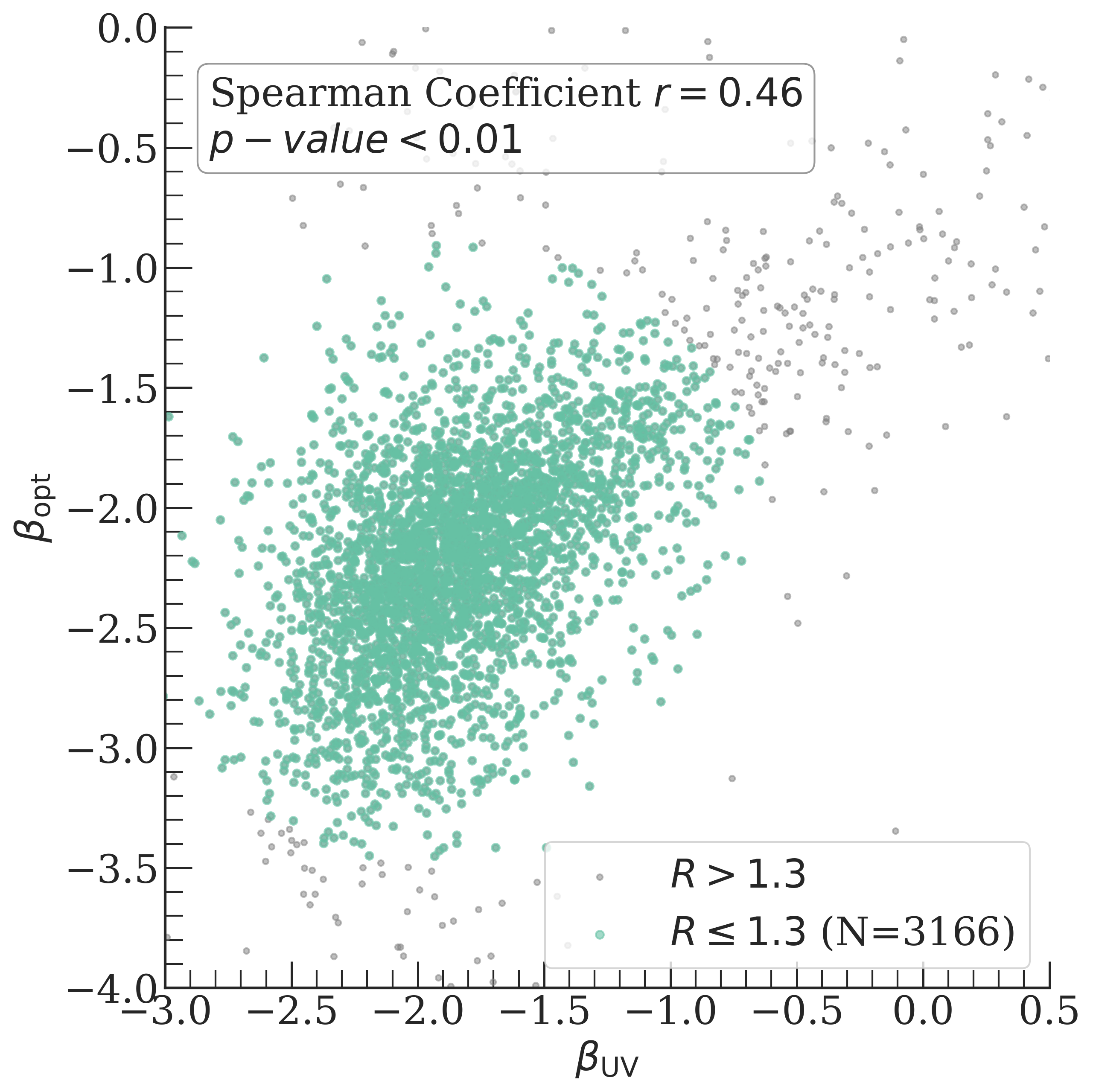} 
    \caption{$\bopt$ versus $\buv$ diagram of our sample. The cyan points indicate points that lie within the radius of 1.3 in the $\buv$--$\bopt$ plane from the highest density point determined in \S\ref{subsec:kde} using the Gaussian KDE function and implementing Silverman's rule. The gray points are the rest of the sources that do not satisfy this criterion.}
    \label{fig:circ_beta_dist}
\end{figure}

From this subsection, we will discuss the features of the $\buv$--$\bopt$ diagram to understand the relationship between the LRD and star-forming galaxy populations by examining the trends one by one. First, we discuss the distribution of galaxies around the region of maximum density. 
To do this, we apply a selection criterion that includes all sources within a radius $r \leq 1.3$ in the $\buv$--$\bopt$ plane of the peak density location, given by $\bopt = -2.19$ and $\buv = -1.89$. This radius was arbitrarily selected to isolate the star-forming locus while excluding intermediate sources extending between the central dense locus of points and the LRD population, as well as the diagonal sequence.

We modeled the distribution of the difference in the slope values $\Delta\beta = \bopt - \buv$, with a Gaussian function, where the $y$-axis represents the number of galaxies (counts) and $\Delta\beta$ is plotted along the $x$-axis. The best-fitting Gaussian exhibits a mean value of $\mu = -0.37$, which is significantly offset from $\Delta\beta = 0$. 
This result provides robust evidence that the average spectrum of star-forming galaxies is characterized by systematically bluer (i.e., more negative) optical slopes ($\bopt$) compared to their UV slopes ($\buv$). This behavior is naturally explained by standard dust attenuation (reddening) laws, which predict a stronger suppression of the flux density in the UV regime than in the optical regime.

From Figure~\ref{fig:circ_beta_dist}, it is evident that $\buv$ and $\bopt$ exhibit a statistically significant correlation and therefore cannot be regarded as independent variables. This is quantitatively supported by a Spearman rank correlation coefficient of 0.46, which indicates a moderate but statistically significant trend between the two quantities. As examined in the following section, the trend observed in the central dense locus of points extends to the diagonal pattern shown in Figure~\ref{fig:contour}, where we interpret this behavior in terms of dust-reddening models and the aging of the galaxy’s stellar population.

\subsection{The Diagonal Trend: Dust Reddening and Passiveness}
\label{subsec:dustandpass}

As shown in Figures~\ref{fig:dusttrack} and \ref{fig:passtrack}, the diagonal trend is well explained by dust reddening or the quiescence of the star formation activity, while we do not reject a combination of the two.
Furthermore, the model grids predicted by these two effects cannot enter the area $\bopt - \buv \geq 0.5$, where the LRDs are distributed.

In the limit of very small $\tau$ in Figure~\ref{fig:passtrack}, the model mimics an instantaneous burst-like star-formation history.
As illustrated by the evolutionary tracks for $\tau=0.01$ (black line) and $\tau=0.05$ (blue line), the tracks converge as $\tau$ decreases, indicating that these values are close to the starburst limit and reproduce the bluest $\bopt$ part of the diagonal trend well. 
For the red track ($\tau = 0.3$) and the purple track ($\tau = 0.4$), the endpoints are fixed at an age of 2.01 Gyr. Given the redshift range of our sample, the corresponding cosmic age spans approximately 600 Myr to 2.1 Gyr \citep{Hogg2000}, which implies that these two tracks already probe the upper age limit relevant for our analysis. 
In the limit $\tau \rightarrow \infty$, the model approaches a constant star formation history, for which ($\buv$, $\bopt$) converges to the positions indicated by the five-pointed-stars in Figure~\ref{fig:dusttrack}.
Consequently, within the age of the universe, the model cannot reproduce the vertical trend in the $\bopt$ versus $\buv$ plane towards LRDs, even when adopting our least steep attenuation law (Calzetti) because its attenuation vector has a shallower positive gradient than the $\bopt - \buv=0.5$ line (see Figure~\ref{fig:dusttrack}).

There are sources exhibiting bluer $\bopt$ and redder $\buv$ than the SMC models in Figure~\ref{fig:dusttrack} and than the shortest $\tau$ track in Figure~\ref{fig:passtrack}.\footnote{Extremely blue sources ($\bopt<-3.5$) are discussed in Appendix~\ref{apdx:veryblue}.}
Some of them may be explained by additional UV dust reddening to the quiescent tracks, while quiescent galaxies are generally dust-poor.
An alternative scenario is to consider steeper dust reddening laws than the SMC law.
For example, some red quasars exhibit a much steeper reddening curve than the SMC law \citep{Zafar2015}.
However, it is unclear whether such a reddening law for a special class of quasars is applicable to our sample composed of general star-forming galaxies.
On the other hand, the analysis by \citet{SunFW2026} of a distinct sample of background galaxies, whose emitted radiation propagates through the circumgalactic medium (CGM) of foreground massive galaxies in the JADES field, reveals similar or even steeper reddening curves.
However, this traces CGM dust and it is unclear whether this applies to the ISM dust for our sample galaxies.
In any case, it would be interesting to investigate these bluer $\bopt$ and redder $\buv$ galaxies in future work.

\subsection{The Vertical Trend: BH* Sequence}
\label{subsec:mixtrack}

As shown in Figure~\ref{fig:3mixtracks}, the tracks initially exhibit a steep increase in $\bopt$ with only minor variations in $\buv$ as $\eta$ increases from $0$ to $\sim 0.01$. Beyond this value of $\eta$, the trend reverses: further increases in $\eta$ induce significant reddening of $\buv$, while $\bopt$ changes only marginally. In our model, this behavior arises because increasing $\eta$ beyond this threshold effectively flattens the UV slope until it becomes fully flat ($\buv = 0$), resembling our BH* template derived from The Cliff. This flat UV continuum imposed on the BH* template is the main driver of the drastic increase in $\buv$ at high values of $\eta$ shown in Figure~\ref{fig:3mixtracks}, which reproduces the rightward curvature of our model tracks at the LRD locus.

\begin{figure}
    \centering
    \includegraphics[width=0.95\columnwidth]{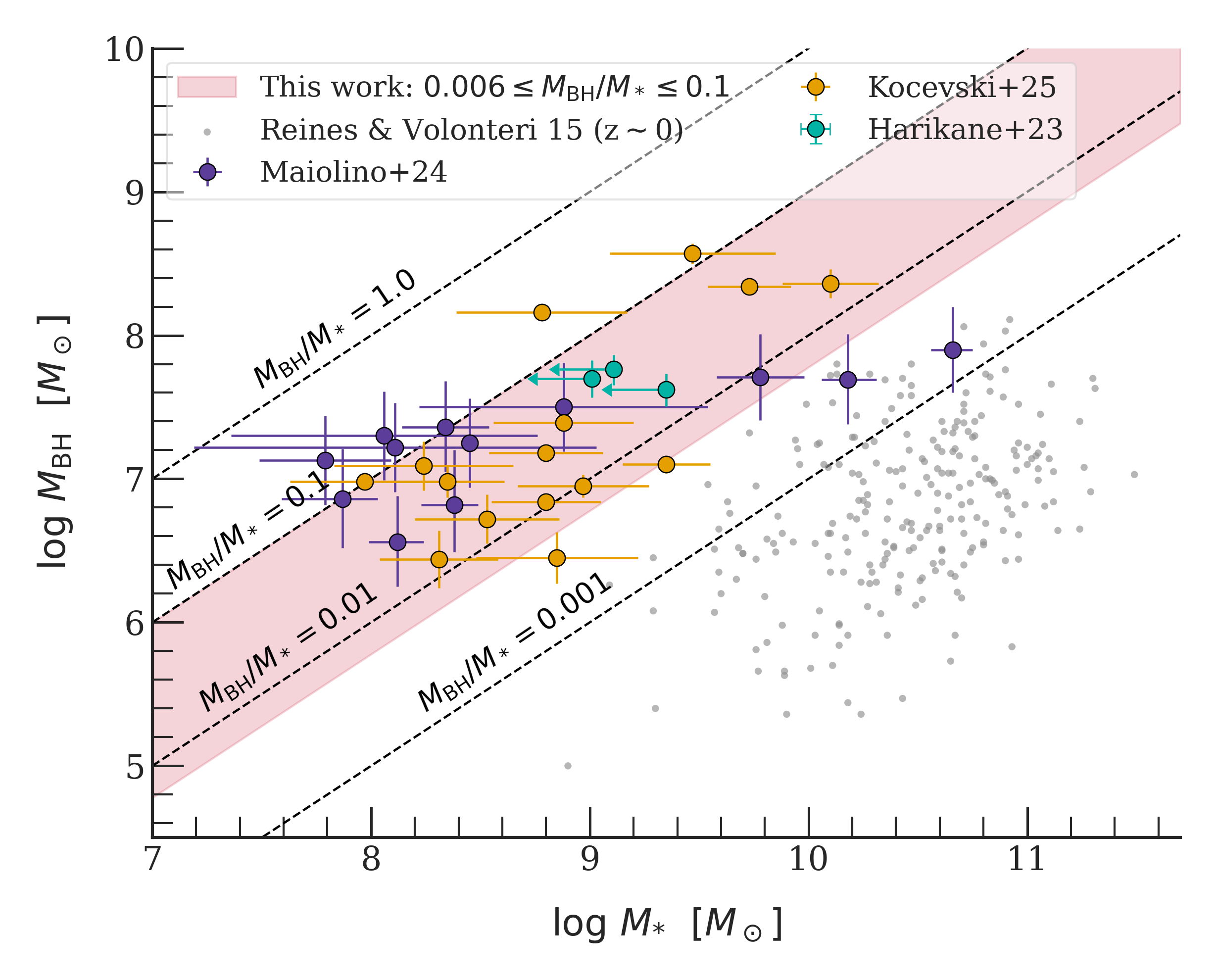} 
    \caption{The distribution of sources in the $M_{\mathrm{BH}}$ versus $M_*$ plane. The gold points denote the broad-line LRDs compiled by \citet{Kocevski2024}; the purple points are the red, compact, broad-line AGNs identified in the JADES survey \citep{Maiolino2024}; and the cyan points indicate the red, compact, broad-line AGNs discovered in the CEERS survey \citep{Harikane2023} that are best described by a pure PSF model. The gray points show the sample of local broad-line AGNs from \citet{ReinesVolonteri2015}, while the red-shaded region illustrates the range of $\eta = M_{\mathrm{BH}}/M_*$ inferred in this work.}
    \label{fig:MBH_M*}
\end{figure}

The values of $\eta$ inferred for the LRDs from Figure~\ref{fig:3mixtracks} span approximately $0.006$--$0.1$, as determined by the location at which the green track in Figure~\ref{fig:3mixtracks} crosses over into the LRD boundary. This places LRDs about two orders of magnitude above the majority of the local $M_\mathrm{BH}/M_*$ broad line AGNs of \citet{ReinesVolonteri2015}, and in agreement with the inferred black hole mass-to-stellar mass relation ($M_{\mathrm{BH}}$--$M_*$) of many LRD studies, as shown in Figure~\ref{fig:MBH_M*}.
Interestingly, even a level of $\eta=0.001$--$0.01$ already makes $\bopt$ significantly redder than the locus of star-forming galaxies. This intermediate population may offer a clue to the transition mechanism between general star-forming galaxies and LRDs. It will be interesting to investigate this population of galaxies in future work.

Several caveats and points of caution remain regarding our model.
First, fixing $L_{\lambda,\mathrm{Cliff}}/M_{\mathrm{BH,Cliff}}$ implies that every BH*-like object accretes at the same Eddington ratio as The Cliff, whereas the accretion rates of LRDs may vary considerably from object to object: the reported values range from sub-Eddington ratios \citep[e.g.,][]{Greene2024, Maiolino2024} to near or super-Eddington ratios \citep[e.g.,][]{Lambrides2024, Naidu2025, Kocevski2024}. As the BH* term should scale with the Eddington ratio, an object accreting at a different rate would correspond to a different rescaled value of $\eta$.

Second, the inferred stellar mass-to-light ratios carry their own systematic uncertainties. When modeling the SEDs of the galaxies used in the stack, the majority of the sources were best fitted with a comparatively high ionization parameter (median $\log(U)\sim -1.5$), which is somewhat unexpected compared to the previous analyses of JWST-selected galaxies at similar redshifts \citep{Reddy2023, Cleri2026}. This may be caused by line blending due to the low resolution of the PRISM data. Therefore, the inferred stellar masses may be underestimated as a part of the UV flux is attributed to the nebular continuum emission rather than the underlying stellar population.

Third, the black hole mass of The Cliff may be overestimated if the observed H$\alpha$ emission line is broadened by electron or resonant scattering in the dense envelope rather than by gravitational motion, as expected within the BH* framework \citep{Rusakov2026, Sun2026}. Together, the latter two effects imply that the inferred values of $\eta$ may be upper limits, indicating that LRDs could be less overmassive than suggested in this analysis, while object-to-object variation in the Eddington ratio would introduce a scatter in either direction.

As we adopt The Cliff as our BH* template, the optical slope saturates at its value of $\bopt\approx1.5$. As a consequence, our mixing tracks are unable to reproduce the LRDs that display redder optical slopes of $\bopt>1.5$. 
However, this is not a critical problem because there is a range of blackbody temperatures of the dense gas cocoon of BH*s \citep[$T\sim2000-7000\,\mathrm{K}$;][]{deGraaff2025b}, whereas The Cliff lies near the middle of this range \citep[$T\approx4000\,\mathrm{K}$;][]{deGraaff2025b}. A lower blackbody temperature would produce a continuum that rises more steeply across our optical fitting window, thus yielding a redder $\bopt$.

\subsection{Morphology and Intermediate Population}
\label{Morphology}

As found in Figures~\ref{fig:ER+Median_Stack_UV} and \ref{fig:ER+Median_Stack_Optical}, the average morphology of the sources along the vertical ($\bopt$) trend changes from a compact point-like shape in the LRD area (Bin 1) to an extended shape in the star-forming galaxy region (Bin 4). The two intermediate Bins 2 and 3 exhibit intermediate values of the extendedness ratio (ER) and a gradual increase toward Bin 4. 
Indeed, both the median ER values derived from the extendedness plot and the ER of the median stacked image in the UV and optical regimes exhibit a relatively strong positive correlation with increasing bin number (corresponding to a bluer $\bopt$), as quantified by Spearman correlation coefficients of $r_{\mathrm{UV}} = 0.604$ and $r_{\mathrm{optical}} = 0.687$.

However, two important caveats should be emphasized. Firstly, the stacked images are constructed from the Full sample rather than the Clean subsample, to maximize the number of objects per stack (see \S\ref{morphologyhowto}). Therefore, emission-line contamination may be present in the median-stacked images in both the UV and optical regimes. 
Although the number of the Clean sources in the optical regime is limited, the trends obtained from the Full sample and the Clean subsample appear to be in good agreement.
A continuum-extracted analysis of NIRCam sources such as that demonstrated by \citet{NL_LRDS} would be better; 
however, this is beyond the scope of this current research. Secondly, some sources exhibit extended morphologies that deviate significantly from a circular, point-like configuration. For such objects, a Gaussian model does not fully describe the true extent of the morphology of such galaxies, and the fitted ER should be interpreted primarily as a relative indicator supporting the morphological trend from compact LRDs towards more extended star-forming galaxies. Nevertheless, the positive correlation of ER with bin number is expected to remain statistically robust even in the presence of these limitations.

The intermediate population in Bins 2 and 3 connects the LRDs and star-forming galaxies smoothly in terms of morphology and the strength of the V-shaped SED.
Indeed, as can be seen in Figure~\ref{fig:contour}, the vertical sequence extends smoothly from the star-forming galaxy area to the LRD area. 
These facts strongly suggest that LRDs are not a completely distinct population from star-forming galaxies, but rather an extreme end of the continuous sequence, supporting the scenario of a linear mixing model of a host galaxy and a BH*.

The properties of the intermediate population in Bins 2 and 3 deserve explicit consideration. These sources exhibit $\bopt$ values spanning approximately from −1 to 0, thereby bridging the point-like LRDs in Bin 1 with the clearly resolved star-forming galaxies in Bin 4, and they account for $\sim 50$ objects distributed along the vertical sequence. Within the framework of our linear mixing model, such systems can be interpreted as composites in which the BH* and the host galaxy contribute comparably to the rest-frame UV–optical continuum. Specifically, the BH* component steepens $\bopt$ and introduces an unresolved nuclear source, while the host galaxy dilutes the characteristic V-shaped spectral feature and adds extended emission. This framework implies that intermediate sources should display composite morphologies, characterized by a compact core embedded within spatially resolved extended emission---a structure that remains unresolved at the resolution of our median-stacked images.

\subsection{A Unified Mixing Picture}
\label{sec:mixing_picture}

Our analysis yields three main results: (i) neither dust reddening nor quiescent stellar populations can account for the vertical sequence in the $\buv$--$\bopt$ plane; (ii) both $\bopt$ and the morphological extent vary smoothly between the star-forming galaxy and the LRD area, with no clear evidence of bimodality; and (iii) the LRD end of the sequence can possibly be described by a linear mixing of a star-forming host and a BH* component with $\eta$ ranging from $0.006$--$0.1$, consistent with previous $M_{\mathrm{BH}}$--$M_*$ estimates for LRDs (Figure~\ref{fig:MBH_M*}).

These results motivate a simple unified picture: a galaxy's position along the sequence is set by the relative contribution of the BH* component to the total continuum. The BH*-dominant phase is observed as an LRD, the host-dominant phase as an ordinary star-forming galaxy, and the intermediate population of Bins 2 and 3 as the transition between the two. Crucially, this interpretation depends only on the variation of the BH* contribution along the sequence, not on the absolute value of $\eta$.

An independent spectral analysis conducted by \citet{Merida2026} further supports the composite nature of the proposed mixing sequence. Their results indicate that pure BH* models do not adequately reproduce the spectra of the majority of LRDs: an additional stellar contribution to the optical continuum is frequently required, and the BH* solutions exhibit strong degeneracies with alternative model configurations. This behavior is expected within our framework, in which every source along the sequence comprises both stellar and BH* emission in continuously varying proportions. In this context, the stellar contribution to the optical continuum is an intrinsic property of the mixing sequence, increasing toward the star-forming end.

However, the direction of evolution along the sequence remains open. As our bins are defined by $\bopt$, our data alone cannot establish which end represents the earlier evolutionary stage, and the literature is divided: \citet{Perez2026b} infer that the strongest V-shaped phases come first and are progressively diluted by the growth of the host, while \citet{Merida2026} find that when the AGN emission is suppressed, LRDs with higher optical-to-UV ratios exhibit larger stellar contributions and are typically found at lower redshifts, which they tentatively interpret as the opposite temporal orientation. Our interpretation is agnostic on this point: both directions traverse the same mixing sequence. The morphological trend along the vertical track, together with the anti-correlation between UV size and Balmer break strength reported by \citet{Ando2026}, is consistent with either picture.

In the above discussion, we emphasize two important caveats. First, as detailed in Section~\ref{subsec:mixtrack}, the inferred $\eta$ values are subject to systematic uncertainties from the assumed mass-to-light ratios of both components: the possible overestimation of the black hole mass of The Cliff \citep{deGraaff2025a, Rusakov2026} and the probable underestimation of the host stellar masses imply that $\eta$ should be interpreted as an upper limit, while the object-to-object variation in Eddington ratio may contribute to additional scatter in either direction. Consequently, the agreement between our values and previous studies in Figure~\ref{fig:MBH_M*} demonstrates consistency under a common set of assumptions rather than an independent confirmation of overmassive black holes.

Second, $\eta$ is unlikely to remain constant in reality: the BH* luminosity is likely powered by accretion onto the disk \citep[e.g.,][]{Begelman2025, Kido2025, Rusakov2026}, so $L_{\lambda}/M_{\mathrm{BH}}$ is expected to decline steeply once rapid accretion ceases \citep[e.g.,][]{Inayoshi2025b}. The mixing fraction is therefore more plausibly governed by the accretion rate and the duty cycle of the BH* phase \citep[e.g.,][]{Inayoshi2025b, ZhangC2026} than by $M_{\mathrm{BH}}$ alone. Testing this hypothesis and determining the evolutionary direction along the sequence will require a spectral decomposition of the intermediate population and constraints on the characteristic timescales of the BH* phase, which we defer to future work.

An alternative interpretation is that the evolutionary sequence does not terminate at the star-forming population. If BH feedback induces the quenching of the host galaxy, analogous to the classical AGN-driven quenching framework, the evolutionary track would be expected to transition onto the quiescent diagonal sequence (see Section~\ref{subsec:BAGPIPESPassive} and Section~\ref{subsec:dustandpass}). In this picture, the broad-line AGNs identified by JWST in massive post-starburst galaxies at $z\approx4.7-6.4$ \citep{CarnallDust, Onoue2025}, with black hole-to-stellar mass ratios $\eta\approx0.01-0.03$ comparable to those inferred for our LRDs, are consistent with such an evolutionary pathway. Moreover, the non-negligible AGN fraction among quiescent galaxies at $1<z<5$ \citep{Ito2026} suggests that these systems may be relatively common. Nevertheless, our analysis cannot distinguish quenched post-BH* systems from galaxies that are reddened through ordinary aging or dust attenuation; therefore, this path is worth exploring more carefully in the future.

\section{Conclusions}
\label{conclusions}

This study examines the linear mixing model as a framework for interpreting the observed continuum sequence from LRDs to typical star-forming galaxies in JWST observations. Using the extensively studied LRD “The Cliff”, often regarded in the LRD literature as one of the two most compelling black hole star (BH*) candidates, as a reference object, we constructed a BH* spectral template from its observed spectrum. This template was then employed to generate synthetic linear mixing sequences designed to explore possible connections between LRDs and star-forming galaxies.

Spectroscopic and imaging data for this analysis are drawn from DJA, from which we selected 3,566 sources with high signal-to-noise spectra. We fitted the continuum of each source with independent power-law fits to the UV and optical windows of the form $F_{\lambda} \propto\lambda^{\beta}$ and derived the corresponding UV and optical slopes, denoted $\buv$ and $\bopt$, respectively. When plotting $\bopt$ against $\buv$, we identify three distinct structures in the distribution: (i) a central locus of points, (ii) an approximately diagonal sequence, and (iii) a predominantly vertical sequence.

The UV and optical spectral slope values of the sources located within the central locus are strongly correlated and exhibit a trend that is connected to the observed diagonal sequence.
The diagonal sequence can be reproduced by models including dust reddening, quiescent stellar populations, or some combination of both effects using the BAGPIPES software. Furthermore, dust-reddened star-forming models, or the quiescent galaxy tracks, or a combination of both cannot account for the observed vertical trend.

To interpret this vertical sequence, we construct synthetic spectra by linearly mixing the spectrum of the BH* template with the median stacked spectrum of star-forming galaxies. Comparing the positions of each $\eta$ along these synthetic tracks with the LRD positions in the $\buv$--$\bopt$ plane, we find that the ratio $\eta=M_{\mathrm{BH}}/M_*$ ranges around $0.006-0.1$ for LRDs, consistent with previous LRD studies. However, these inferred values may represent upper limits, largely due to the uncertainties associated with the black hole mass estimation for The Cliff and the stellar masses of the hosts, while the assumption of the common Eddington ratio for all BH* components can introduce further object-to-object scatter.

The vertical sequence is continuously populated, and no clear boundary is apparent between LRDs and star-forming galaxies in the $\buv$--$\bopt$ plane. The transition between these two populations is bridged by an intermediate class---corresponding to Bins 2 and 3, with $-1 \lesssim \bopt \lesssim 0$---which has not been studied extensively in the literature. Within our linear mixing framework, these systems are characterized by comparable contributions from the BH* component and the host galaxy to the optical continuum. We therefore propose that this region of the $\buv$--$\bopt$ plane may serve as an effective selection window for transitional candidates, whose targeted follow-up analysis would enable a direct test of the physical connection between LRDs and star-forming galaxies.

Consistent with this picture, we find that the spatial extent of our sources increases toward bluer optical slopes in both the rest-frame UV and optical regimes. This behavior is expected if the compact BH* structure, which produces an approximately blackbody-like continuum, gradually dissipates and the emission from the more extended host galaxy becomes increasingly dominant. However, our analysis constrains only the relative compactness of systems along the vertical sequence and does not determine the temporal evolution along this track, which remains an open question in the current literature.

\section*{Acknowledgements}

This work is based on observations made with the NASA/ESA/CSA James Webb Space Telescope, using the NIRSpec and NIRCam instruments. The data products presented herein were retrieved from DAWN JWST Archive (DJA). DJA is an initiative of the Cosmic Dawn Center (DAWN), funded by the Danish National Research Foundation under grant DNRF140. The full sample analyzed in this study was drawn from the following JWST programs: 1180, 1181, 1208, 1210, 1211, 1212, 1213, 1214, 1215, 1286, 1287, 1345, 1433, 1747, 2028, 2073, 2198, 2282, 2561, 2565, 2750, 2756, 2767, 3073, 3215, 4106, 4233, 4446, 4557, 5105, 5224, 6368, 6541, and 6585. This work utilized the following software packages: Astropy\,\citep{Astropy2013, Astropy2018, Astropy2022}, NumPy\,\citep{Harris2020}, Matplotlib\,\citep{Hunter2007}, SpectRes\,\citep{CarnallSpectres}, SciPy\,\citep{Virtanen2020}, and BAGPIPES\,\citep{Carnall2018}. Portions of the analysis code used in this study were developed with assistance from Claude AI (Anthropic). All AI-assisted code was reviewed and tested by the lead author, and the resulting outputs were independently verified. This work is supported by Japan Society for the Promotion of Science (JSPS) KAKENHI Grant No.~23H00131, 26K07155 and 26H02069.

\bibliographystyle{aasjournal}
\bibliography{ref}

\begin{thebibliography}{}
\expandafter\ifx\csname natexlab\endcsname\relax\def\natexlab#1{#1}\fi
\providecommand{\url}[1]{\href{#1}{#1}}
\providecommand{\dodoi}[1]{doi:~\href{http://doi.org/#1}{\nolinkurl{#1}}}
\providecommand{\doeprint}[1]{\href{http://ascl.net/#1}{\nolinkurl{http://ascl.net/#1}}}
\providecommand{\doarXiv}[1]{\href{https://arxiv.org/abs/#1}{\nolinkurl{https://arxiv.org/abs/#1}}}

\bibitem[{{Akins} {et~al.}(2025){Akins}, {Casey}, {Lambrides}, {Allen}, {Andika}, {Brinch}, {Champagne}, {Cooper}, {Ding}, {Drakos}, {Faisst}, {Finkelstein}, {Franco}, {Fujimoto}, {Gentile}, {Gillman}, {Gozaliasl}, {Harish}, {Hayward}, {Hirschmann}, {Ilbert}, {Kartaltepe}, {Kocevski}, {Koekemoer}, {Kokorev}, {Liu}, {Long}, {McCracken}, {McKinney}, {Onoue}, {Paquereau}, {Renzini}, {Rhodes}, {Robertson}, {Shuntov}, {Silverman}, {Tanaka}, {Toft}, {Trakhtenbrot}, {Valentino}, \& {Zavala}}]{Akins2025}
{Akins}, H.~B., {Casey}, C.~M., {Lambrides}, E., {et~al.} 2025, \apj, 991, 37, \dodoi{10.3847/1538-4357/ade984}

\bibitem[{{Ananna} {et~al.}(2024){Ananna}, {Bogd{\'a}n}, {Kov{\'a}cs}, {Natarajan}, \& {Hickox}}]{Ananna2024}
{Ananna}, T.~T., {Bogd{\'a}n}, {\'A}., {Kov{\'a}cs}, O.~E., {Natarajan}, P., \& {Hickox}, R.~C. 2024, \apjl, 969, L18, \dodoi{10.3847/2041-8213/ad5669}

\bibitem[{{Ando} {et~al.}(2026){Ando}, {Harikane}, {Katz}, {Inayoshi}, \& {Tanaka}}]{Ando2026}
{Ando}, M., {Harikane}, Y., {Katz}, H., {Inayoshi}, K., \& {Tanaka}, T.~S. 2026, arXiv e-prints, arXiv:2606.03522, \dodoi{10.48550/arXiv.2606.03522}

\bibitem[{{Asada} {et~al.}(2026){Asada}, {Inayoshi}, {Fei}, {Fujimoto}, \& {Willott}}]{Asada2026}
{Asada}, Y., {Inayoshi}, K., {Fei}, Q., {Fujimoto}, S., \& {Willott}, C.~J. 2026, \apjl, 1005, L22, \dodoi{10.3847/2041-8213/ae7d0a}

\bibitem[{{Astropy Collaboration} {et~al.}(2013){Astropy Collaboration}, {Robitaille}, {Tollerud}, {Greenfield}, {Droettboom}, {Bray}, {Aldcroft}, {Davis}, {Ginsburg}, {Price-Whelan}, {Kerzendorf}, {Conley}, {Crighton}, {Barbary}, {Muna}, {Ferguson}, {Grollier}, {Parikh}, {Nair}, {Unther}, {Deil}, {Woillez}, {Conseil}, {Kramer}, {Turner}, {Singer}, {Fox}, {Weaver}, {Zabalza}, {Edwards}, {Azalee Bostroem}, {Burke}, {Casey}, {Crawford}, {Dencheva}, {Ely}, {Jenness}, {Labrie}, {Lim}, {Pierfederici}, {Pontzen}, {Ptak}, {Refsdal}, {Servillat}, \& {Streicher}}]{Astropy2013}
{Astropy Collaboration}, {Robitaille}, T.~P., {Tollerud}, E.~J., {et~al.} 2013, \aap, 558, A33, \dodoi{10.1051/0004-6361/201322068}

\bibitem[{{Astropy Collaboration} {et~al.}(2018){Astropy Collaboration}, {Price-Whelan}, {Sip{\H{o}}cz}, {G{\"u}nther}, {Lim}, {Crawford}, {Conseil}, {Shupe}, {Craig}, {Dencheva}, {Ginsburg}, {VanderPlas}, {Bradley}, {P{\'e}rez-Su{\'a}rez}, {de Val-Borro}, {Aldcroft}, {Cruz}, {Robitaille}, {Tollerud}, {Ardelean}, {Babej}, {Bach}, {Bachetti}, {Bakanov}, {Bamford}, {Barentsen}, {Barmby}, {Baumbach}, {Berry}, {Biscani}, {Boquien}, {Bostroem}, {Bouma}, {Brammer}, {Bray}, {Breytenbach}, {Buddelmeijer}, {Burke}, {Calderone}, {Cano Rodr{\'\i}guez}, {Cara}, {Cardoso}, {Cheedella}, {Copin}, {Corrales}, {Crichton}, {D'Avella}, {Deil}, {Depagne}, {Dietrich}, {Donath}, {Droettboom}, {Earl}, {Erben}, {Fabbro}, {Ferreira}, {Finethy}, {Fox}, {Garrison}, {Gibbons}, {Goldstein}, {Gommers}, {Greco}, {Greenfield}, {Groener}, {Grollier}, {Hagen}, {Hirst}, {Homeier}, {Horton}, {Hosseinzadeh}, {Hu}, {Hunkeler}, {Ivezi{\'c}}, {Jain}, {Jenness}, {Kanarek}, {Kendrew}, {Kern}, {Kerzendorf}, {Khvalko}, {King}, {Kirkby}, {Kulkarni},
  {Kumar}, {Lee}, {Lenz}, {Littlefair}, {Ma}, {Macleod}, {Mastropietro}, {McCully}, {Montagnac}, {Morris}, {Mueller}, {Mumford}, {Muna}, {Murphy}, {Nelson}, {Nguyen}, {Ninan}, {N{\"o}the}, {Ogaz}, {Oh}, {Parejko}, {Parley}, {Pascual}, {Patil}, {Patil}, {Plunkett}, {Prochaska}, {Rastogi}, {Reddy Janga}, {Sabater}, {Sakurikar}, {Seifert}, {Sherbert}, {Sherwood-Taylor}, {Shih}, {Sick}, {Silbiger}, {Singanamalla}, {Singer}, {Sladen}, {Sooley}, {Sornarajah}, {Streicher}, {Teuben}, {Thomas}, {Tremblay}, {Turner}, {Terr{\'o}n}, {van Kerkwijk}, {de la Vega}, {Watkins}, {Weaver}, {Whitmore}, {Woillez}, {Zabalza}, \& {Astropy Contributors}}]{Astropy2018}
{Astropy Collaboration}, {Price-Whelan}, A.~M., {Sip{\H{o}}cz}, B.~M., {et~al.} 2018, \aj, 156, 123, \dodoi{10.3847/1538-3881/aabc4f}

\bibitem[{{Astropy Collaboration} {et~al.}(2022){Astropy Collaboration}, {Price-Whelan}, {Lim}, {Earl}, {Starkman}, {Bradley}, {Shupe}, {Patil}, {Corrales}, {Brasseur}, {N{\"o}the}, {Donath}, {Tollerud}, {Morris}, {Ginsburg}, {Vaher}, {Weaver}, {Tocknell}, {Jamieson}, {van Kerkwijk}, {Robitaille}, {Merry}, {Bachetti}, {G{\"u}nther}, {Aldcroft}, {Alvarado-Montes}, {Archibald}, {B{\'o}di}, {Bapat}, {Barentsen}, {Baz{\'a}n}, {Biswas}, {Boquien}, {Burke}, {Cara}, {Cara}, {Conroy}, {Conseil}, {Craig}, {Cross}, {Cruz}, {D'Eugenio}, {Dencheva}, {Devillepoix}, {Dietrich}, {Eigenbrot}, {Erben}, {Ferreira}, {Foreman-Mackey}, {Fox}, {Freij}, {Garg}, {Geda}, {Glattly}, {Gondhalekar}, {Gordon}, {Grant}, {Greenfield}, {Groener}, {Guest}, {Gurovich}, {Handberg}, {Hart}, {Hatfield-Dodds}, {Homeier}, {Hosseinzadeh}, {Jenness}, {Jones}, {Joseph}, {Kalmbach}, {Karamehmetoglu}, {Ka{\l}uszy{\'n}ski}, {Kelley}, {Kern}, {Kerzendorf}, {Koch}, {Kulumani}, {Lee}, {Ly}, {Ma}, {MacBride}, {Maljaars}, {Muna}, {Murphy}, {Norman},
  {O'Steen}, {Oman}, {Pacifici}, {Pascual}, {Pascual-Granado}, {Patil}, {Perren}, {Pickering}, {Rastogi}, {Roulston}, {Ryan}, {Rykoff}, {Sabater}, {Sakurikar}, {Salgado}, {Sanghi}, {Saunders}, {Savchenko}, {Schwardt}, {Seifert-Eckert}, {Shih}, {Jain}, {Shukla}, {Sick}, {Simpson}, {Singanamalla}, {Singer}, {Singhal}, {Sinha}, {Sip{\H{o}}cz}, {Spitler}, {Stansby}, {Streicher}, {{\v{S}}umak}, {Swinbank}, {Taranu}, {Tewary}, {Tremblay}, {de Val-Borro}, {Van Kooten}, {Vasovi{\'c}}, {Verma}, {de Miranda Cardoso}, {Williams}, {Wilson}, {Winkel}, {Wood-Vasey}, {Xue}, {Yoachim}, {Zhang}, {Zonca}, \& {Astropy Project Contributors}}]{Astropy2022}
{Astropy Collaboration}, {Price-Whelan}, A.~M., {Lim}, P.~L., {et~al.} 2022, \apj, 935, 167, \dodoi{10.3847/1538-4357/ac7c74}

\bibitem[{{Baggen} {et~al.}(2024){Baggen}, {van Dokkum}, {Brammer}, {de Graaff}, {Franx}, {Greene}, {Labb{\'e}}, {Leja}, {Maseda}, {Nelson}, {Rix}, {Wang}, \& {Weibel}}]{Baggen2024}
{Baggen}, J. F.~W., {van Dokkum}, P., {Brammer}, G., {et~al.} 2024, \apjl, 977, L13, \dodoi{10.3847/2041-8213/ad90b8}

\bibitem[{{Barro} {et~al.}(2024){Barro}, {P{\'e}rez-Gonz{\'a}lez}, {Kocevski}, {McGrath}, {Trump}, {Simons}, {Somerville}, {Yung}, {Arrabal Haro}, {Akins}, {Bagley}, {Cleri}, {Costantin}, {Davis}, {Dickinson}, {Finkelstein}, {Giavalisco}, {G{\'o}mez-Guijarro}, {Hathi}, {Hirschmann}, {Holwerda}, {Huertas-Company}, {Kartaltepe}, {Koekemoer}, {Lucas}, {Papovich}, {Pirzkal}, {Seill{\'e}}, {Tacchella}, {Wuyts}, {Wilkins}, {de la Vega}, {Yang}, \& {Zavala}}]{Barro2024}
{Barro}, G., {P{\'e}rez-Gonz{\'a}lez}, P.~G., {Kocevski}, D.~D., {et~al.} 2024, \apj, 963, 128, \dodoi{10.3847/1538-4357/ad167e}

\bibitem[{{Barro} {et~al.}(2026){Barro}, {P{\'e}rez-Gonz{\'a}lez}, {Kocevski}, {Trump}, {Dickinson}, {Arrabal Haro}, {Brooks}, {Donnan}, {Dunlop}, {Finkelstein}, {Franco}, {Gandolfi}, {Giavalisco}, {Grogin}, {Hirschmann}, {Kartaltepe}, {Koekemoer}, {Larson}, {Leung}, {Lucas}, {McGrath}, {Papovich}, {P{\'e}rez-D{\'\i}az}, {Somerville}, {Taylor}, {Taylor}, {Tripodi}, {Yung}, \& {Wang}}]{Barro2026}
{Barro}, G., {P{\'e}rez-Gonz{\'a}lez}, P.~G., {Kocevski}, D., {et~al.} 2026, \apj, 1003, 96, \dodoi{10.3847/1538-4357/ae5d2c}

\bibitem[{{Begelman} \& {Dexter}(2026)}]{Begelman2025}
{Begelman}, M.~C., \& {Dexter}, J. 2026, \apj, 996, 48, \dodoi{10.3847/1538-4357/ae274a}

\bibitem[{{Billand} {et~al.}(2026){Billand}, {Elbaz}, {Franco}, {Gentile}, {Daddi}, {Giavalisco}, {Kocevski}, {Lewis}, {Magnelli}, {Sangalli}, \& {Tarrasse}}]{Billand2026}
{Billand}, J.-B., {Elbaz}, D., {Franco}, M., {et~al.} 2026, arXiv e-prints, arXiv:2604.11677, \dodoi{10.48550/arXiv.2604.11677}

\bibitem[{{Brammer}(2023{\natexlab{a}})}]{Brammer2023}
{Brammer}, G. 2023{\natexlab{a}}, {msaexp: NIRSpec analysis tools},  Zenodo, \dodoi{10.5281/zenodo.8319596}

\bibitem[{{Brammer}(2023{\natexlab{b}})}]{Brammer2023b}
---. 2023{\natexlab{b}}, {grizli},  Zenodo, \dodoi{10.5281/zenodo.8370018}

\bibitem[{{Brammer} \& {Valentino}(2025)}]{BrammerValentino2025}
{Brammer}, G., \& {Valentino}, F. 2025, {The DAWN JWST Archive: Compilation of Public NIRSpec Spectra},  Zenodo, \dodoi{10.5281/zenodo.15472354}

\bibitem[{{Brazzini} {et~al.}(2026){Brazzini}, {D'Eugenio}, {Maiolino}, {Lyu}, {DeCoursey}, {{\"U}bler}, {Ji}, {Juod{\v{z}}balis}, {Scholtz}, {Jones}, {Hainline}, {Dalla Bont{\`a}}, {{\'e}rez-Gonz{\'a}lez}, {Geris}, {Harshan}, {Feruglio}, {Bischetti}, {Mazzolari}, {Rieke}, {Alberts}, {Trefoloni}, {Carniani}, {Parlanti}, {Marconi}, {Risaliti}, {Ramos Almeida}, {Rinaldi}, {Perna}, {Zamora}, {Lamperti}, {Venturi}, {Cresci}, {Bunker}, \& {Ivey}}]{Brazzini2026}
{Brazzini}, M., {D'Eugenio}, F., {Maiolino}, R., {et~al.} 2026, arXiv e-prints, arXiv:2601.22214, \dodoi{10.48550/arXiv.2601.22214}

\bibitem[{{Byler} {et~al.}(2017){Byler}, {Dalcanton}, {Conroy}, \& {Johnson}}]{Byler2017}
{Byler}, N., {Dalcanton}, J.~J., {Conroy}, C., \& {Johnson}, B.~D. 2017, \apj, 840, 44, \dodoi{10.3847/1538-4357/aa6c66}

\bibitem[{{Calzetti} {et~al.}(2000){Calzetti}, {Armus}, {Bohlin}, {Kinney}, {Koornneef}, \& {Storchi-Bergmann}}]{Calzetti2000}
{Calzetti}, D., {Armus}, L., {Bohlin}, R.~C., {et~al.} 2000, \apj, 533, 682, \dodoi{10.1086/308692}

\bibitem[{{Carnall}(2017)}]{CarnallSpectres}
{Carnall}, A.~C. 2017, arXiv e-prints, arXiv:1705.05165, \dodoi{10.48550/arXiv.1705.05165}

\bibitem[{{Carnall} {et~al.}(2018){Carnall}, {McLure}, {Dunlop}, \& {Dav{\'e}}}]{Carnall2018}
{Carnall}, A.~C., {McLure}, R.~J., {Dunlop}, J.~S., \& {Dav{\'e}}, R. 2018, \mnras, 480, 4379, \dodoi{10.1093/mnras/sty2169}

\bibitem[{{Carnall} {et~al.}(2023){Carnall}, {McLure}, {Dunlop}, {McLeod}, {Wild}, {Cullen}, {Magee}, {Begley}, {Cimatti}, {Donnan}, {Hamadouche}, {Jewell}, \& {Walker}}]{CarnallDust}
{Carnall}, A.~C., {McLure}, R.~J., {Dunlop}, J.~S., {et~al.} 2023, \nat, 619, 716, \dodoi{10.1038/s41586-023-06158-6}

\bibitem[{{Carnall} {et~al.}(2024){Carnall}, {Cullen}, {McLure}, {McLeod}, {Begley}, {Donnan}, {Dunlop}, {Shapley}, {Rowlands}, {Almaini}, {Arellano-C{\'o}rdova}, {Barrufet}, {Cimatti}, {Ellis}, {Grogin}, {Hamadouche}, {Illingworth}, {Koekemoer}, {Leung}, {Lovell}, {P{\'e}rez-Gonz{\'a}lez}, {Santini}, {Stanton}, \& {Wild}}]{CarnallMetal}
{Carnall}, A.~C., {Cullen}, F., {McLure}, R.~J., {et~al.} 2024, \mnras, 534, 325, \dodoi{10.1093/mnras/stae2092}

\bibitem[{{Casey} {et~al.}(2024){Casey}, {Akins}, {Kokorev}, {McKinney}, {Cooper}, {Long}, {Franco}, \& {Manning}}]{Casey2024}
{Casey}, C.~M., {Akins}, H.~B., {Kokorev}, V., {et~al.} 2024, \apjl, 975, L4, \dodoi{10.3847/2041-8213/ad7ba7}

\bibitem[{{Chac{\'o}n} \& {Duong}(2018)}]{ChaconDoung2018}
{Chac{\'o}n}, J.~E., \& {Duong}, T. 2018, {Multivariate Kernel Smoothing and Its Applications} (CRC Press), \dodoi{10.1201/9780429485572}

\bibitem[{{Chen} {et~al.}(2025){Chen}, {Ho}, {Li}, \& {Zhuang}}]{Chen2025}
{Chen}, C.-H., {Ho}, L.~C., {Li}, R., \& {Zhuang}, M.-Y. 2025, \apj, 983, 60, \dodoi{10.3847/1538-4357/ada93a}

\bibitem[{{Chen} {et~al.}(2026){Chen}, {Ichikawa}, {Akiyama}, {Inayoshi}, {Inoue}, {Onoue}, {Toba}, {Zavala}, {Bakx}, {Kawaguchi}, {Lee}, {Matsumoto}, \& {Vijarnwannaluk}}]{Chen2026}
{Chen}, X., {Ichikawa}, K., {Akiyama}, M., {et~al.} 2026, \apj, 999, 30, \dodoi{10.3847/1538-4357/ae3683}

\bibitem[{{Cleri} {et~al.}(2026){Cleri}, {Lewis}, {Leja}, {Helton}, {Burnham}, {Curtis}, {de Graaff}, {Hirschmann}, {Katz}, {Maseda}, {McConachie}, {Plat}, \& {Scharre}}]{Cleri2026}
{Cleri}, N.~J., {Lewis}, Z.~J., {Leja}, J., {et~al.} 2026, arXiv e-prints, arXiv:2605.30410, \dodoi{10.48550/arXiv.2605.30410}

\bibitem[{{Curti} {et~al.}(2024){Curti}, {Maiolino}, {Curtis-Lake}, {Chevallard}, {Carniani}, {D'Eugenio}, {Looser}, {Scholtz}, {Charlot}, {Cameron}, {{\"U}bler}, {Witstok}, {Boyett}, {Laseter}, {Sandles}, {Arribas}, {Bunker}, {Giardino}, {Maseda}, {Rawle}, {Rodr{\'\i}guez Del Pino}, {Smit}, {Willott}, {Eisenstein}, {Hausen}, {Johnson}, {Rieke}, {Robertson}, {Tacchella}, {Williams}, {Willmer}, {Baker}, {Bhatawdekar}, {Egami}, {Helton}, {Ji}, {Kumari}, {Perna}, {Shivaei}, \& {Sun}}]{Curti2024}
{Curti}, M., {Maiolino}, R., {Curtis-Lake}, E., {et~al.} 2024, \aap, 684, A75, \dodoi{10.1051/0004-6361/202346698}

\bibitem[{{de Graaff} {et~al.}(2025{\natexlab{a}}){de Graaff}, {Rix}, {Naidu}, {Labb{\'e}}, {Wang}, {Leja}, {Matthee}, {Katz}, {Greene}, {Hviding}, {Baggen}, {Bezanson}, {Boogaard}, {Brammer}, {Dayal}, {van Dokkum}, {Goulding}, {Hirschmann}, {Maseda}, {McConachie}, {Miller}, {Nelson}, {Oesch}, {Setton}, {Shivaei}, {Weibel}, {Whitaker}, \& {Williams}}]{deGraaff2025a}
{de Graaff}, A., {Rix}, H.-W., {Naidu}, R.~P., {et~al.} 2025{\natexlab{a}}, \aap, 701, A168, \dodoi{10.1051/0004-6361/202554681}

\bibitem[{{de Graaff} {et~al.}(2025{\natexlab{b}}){de Graaff}, {Brammer}, {Weibel}, {Lewis}, {Maseda}, {Oesch}, {Bezanson}, {Boogaard}, {Cleri}, {Cooper}, {Gottumukkala}, {Greene}, {Hirschmann}, {Hviding}, {Katz}, {Labb{\'e}}, {Leja}, {Matthee}, {McConachie}, {Miller}, {Naidu}, {Price}, {Rix}, {Setton}, {Suess}, {Wang}, {Whitaker}, \& {Williams}}]{RUBIES}
{de Graaff}, A., {Brammer}, G., {Weibel}, A., {et~al.} 2025{\natexlab{b}}, \aap, 697, A189, \dodoi{10.1051/0004-6361/202452186}

\bibitem[{{de Graaff} {et~al.}(2025{\natexlab{c}}){de Graaff}, {Setton}, {Brammer}, {Cutler}, {Suess}, {Labb{\'e}}, {Leja}, {Weibel}, {Maseda}, {Whitaker}, {Bezanson}, {Boogaard}, {Cleri}, {De Lucia}, {Franx}, {Greene}, {Hirschmann}, {Matthee}, {McConachie}, {Naidu}, {Oesch}, {Price}, {Rix}, {Valentino}, {Wang}, \& {Williams}}]{deGraaffDust}
{de Graaff}, A., {Setton}, D.~J., {Brammer}, G., {et~al.} 2025{\natexlab{c}}, Nature Astronomy, 9, 280, \dodoi{10.1038/s41550-024-02424-3}

\bibitem[{{de Graaff} {et~al.}(2026){de Graaff}, {Hviding}, {Naidu}, {Greene}, {Miller}, {Leja}, {Matthee}, {Brammer}, {Katz}, {Bezanson}, {Boogaard}, {Bose}, {Chisholm}, {Cleri}, {Dayal}, {Feldmann}, {Fudamoto}, {Fujimoto}, {Furtak}, {Glazebrook}, {Gottumukkala}, {Heintz}, {Kokorev}, {Labbe}, {Maseda}, {McConachie}, {Nanayakkara}, {Nelson}, {Nowaczyk}, {Oesch}, {Rix}, {Setton}, {Torralba}, {Walter}, {Wang}, {Weibel}, \& {van der Wel}}]{deGraaff2025b}
{de Graaff}, A., {Hviding}, R.~E., {Naidu}, R.~P., {et~al.} 2026, \mnras, stag1567, \dodoi{10.1093/mnras/stag1567}

\bibitem[{{D'Eugenio} {et~al.}(2025){D'Eugenio}, {Nelson}, {Ji}, {Baggen}, {Greene}, {Labb{\'e}}, {Pezzulli}, {Brown}, {Maiolino}, {Matthee}, {Terlevich}, {Terlevich}, {Torralba}, \& {Carniani}}]{DEugenio2025}
{D'Eugenio}, F., {Nelson}, E., {Ji}, X., {et~al.} 2025, arXiv e-prints, arXiv:2510.00101, \dodoi{10.48550/arXiv.2510.00101}

\bibitem[{{Dottorini} {et~al.}(2025){Dottorini}, {Calabr{\`o}}, {Pentericci}, {Mascia}, {Llerena}, {Napolitano}, {Santini}, {Roberts-Borsani}, {Castellano}, {Amorin}, {Dickinson}, {Fontana}, {Hathi}, {Hirschmann}, {Koekemoer}, {Lucas}, {Merlin}, {Morales}, {Pacucci}, {Wilkins}, {Arrabal Haro}, {Bagley}, {Finkelstein}, {Kartaltepe}, {Papovich}, \& {Pirzkal}}]{UVSlope}
{Dottorini}, D., {Calabr{\`o}}, A., {Pentericci}, L., {et~al.} 2025, \aap, 698, A234, \dodoi{10.1051/0004-6361/202453267}

\bibitem[{{Gardner} {et~al.}(2023){Gardner}, {Mather}, {Abbott}, {Abell}, {Abernathy}, {Abney}, {Abraham}, {Abraham}, {Abul-Huda}, {Acton}, {Adams}, {Adams}, {Adler}, {Adriaensen}, {Aguilar}, {Ahmed}, {Ahmed}, {Ahmed}, {Albat}, {Albert}, {Alberts}, {Aldridge}, {Allen}, {Allen}, {Altenburg}, {Altunc}, {Alvarez}, {{\'A}lvarez-M{\'a}rquez}, {Alves de Oliveira}, {Ambrose}, {Anandakrishnan}, {Andersen}, {Anderson}, {Anderson}, {Anderson}, {Anderson}, {Aprea}, {Archer}, {Arenberg}, {Argyriou}, {Arribas}, {Artigau}, {Arvai}, {Atcheson}, {Atkinson}, {Averbukh}, {Aymergen}, {Bacinski}, {Baggett}, {Bagnasco}, {Baker}, {Balzano}, {Banks}, {Baran}, {Barker}, {Barrett}, {Barringer}, {Barto}, {Bast}, {Baudoz}, {Baum}, {Beatty}, {Beaulieu}, {Bechtold}, {Beck}, {Beddard}, {Beichman}, {Bellagama}, {Bely}, {Berger}, {Bergeron}, {Bernier}, {Bertch}, {Beskow}, {Betz}, {Biagetti}, {Birkmann}, {Bjorklund}, {Blackwood}, {Blazek}, {Blossfeld}, {Bluth}, {Boccaletti}, {Boegner}, {Bohlin}, {Boia}, {B{\"o}ker}, {Bonaventura}, {Bond},
  {Bosley}, {Boucarut}, {Bouchet}, {Bouwman}, {Bower}, {Bowers}, {Bowers}, {Boyce}, {Boyer}, {Boyer}, {Boyer}, {Boyer}, {Bradley}, {Brady}, {Brandl}, {Brannen}, {Breda}, {Bremmer}, {Brennan}, {Bresnahan}, {Bright}, {Broiles}, {Bromenschenkel}, {Brooks}, {Brooks}, {Brown}, {Brown}, {Brown}, {Bruce}, {Bryson}, {Bujanda}, {Bullock}, {Bunker}, {Bureo}, {Burt}, {Bush}, {Bushouse}, {Bussman}, {Cabaud}, {Cale}, {Calhoon}, {Calvani}, {Canipe}, {Caputo}, {Cara}, {Carey}, {Case}, {Cesari}, {Cetorelli}, {Chance}, {Chandler}, {Chaney}, {Chapman}, {Charlot}, {Chayer}, {Cheezum}, {Chen}, {Chen}, {Cherinka}, {Chichester}, {Chilton}, {Chittiraibalan}, {Clampin}, {Clark}, {Clark}, {Clark}, {Claybrooks}, {Cleveland}, {Cohen}, {Cohen}, {Col{\'o}n}, {Coleman}, {Colina}, {Comber}, {Comeau}, {Comer}, {Conde Reis}, {Connolly}, {Conroy}, {Contos}, {Contreras}, {Cook}, {Cooper}, {Cooper}, {Correia}, {Correnti}, {Cossou}, {Costanza}, {Coulais}, {Cox}, {Coyle}, {Cracraft}, {Crew}, {Curtis}, {Cusveller}, {Da Costa Maciel}, {Dailey},
  {Daugeron}, {Davidson}, {Davies}, {Davis}, {Davis}, {Day}, {de Chambure}, {de Jong}, {De Marchi}, {Dean}, {Decker}, {Delisa}, {Dell}, \& {Dellagatta}}]{JWST}
{Gardner}, J.~P., {Mather}, J.~C., {Abbott}, R., {et~al.} 2023, \pasp, 135, 068001, \dodoi{10.1088/1538-3873/acd1b5}

\bibitem[{{Greene} \& {Ho}(2005)}]{GreeneHo2005}
{Greene}, J.~E., \& {Ho}, L.~C. 2005, \apj, 630, 122, \dodoi{10.1086/431897}

\bibitem[{{Greene} {et~al.}(2024){Greene}, {Labbe}, {Goulding}, {Furtak}, {Chemerynska}, {Kokorev}, {Dayal}, {Volonteri}, {Williams}, {Wang}, {Setton}, {Burgasser}, {Bezanson}, {Atek}, {Brammer}, {Cutler}, {Feldmann}, {Fujimoto}, {Glazebrook}, {de Graaff}, {Khullar}, {Leja}, {Marchesini}, {Maseda}, {Matthee}, {Miller}, {Naidu}, {Nanayakkara}, {Oesch}, {Pan}, {Papovich}, {Price}, {van Dokkum}, {Weaver}, {Whitaker}, \& {Zitrin}}]{Greene2024}
{Greene}, J.~E., {Labbe}, I., {Goulding}, A.~D., {et~al.} 2024, \apj, 964, 39, \dodoi{10.3847/1538-4357/ad1e5f}

\bibitem[{{Harikane} {et~al.}(2023){Harikane}, {Zhang}, {Nakajima}, {Ouchi}, {Isobe}, {Ono}, {Hatano}, {Xu}, \& {Umeda}}]{Harikane2023}
{Harikane}, Y., {Zhang}, Y., {Nakajima}, K., {et~al.} 2023, \apj, 959, 39, \dodoi{10.3847/1538-4357/ad029e}

\bibitem[{{Harris} {et~al.}(2020){Harris}, {Millman}, {van der Walt}, {Gommers}, {Virtanen}, {Cournapeau}, {Wieser}, {Taylor}, {Berg}, {Smith}, {Kern}, {Picus}, {Hoyer}, {van Kerkwijk}, {Brett}, {Haldane}, {del R{\'\i}o}, {Wiebe}, {Peterson}, {G{\'e}rard-Marchant}, {Sheppard}, {Reddy}, {Weckesser}, {Abbasi}, {Gohlke}, \& {Oliphant}}]{Harris2020}
{Harris}, C.~R., {Millman}, K.~J., {van der Walt}, S.~J., {et~al.} 2020, \nat, 585, 357, \dodoi{10.1038/s41586-020-2649-2}

\bibitem[{{Heintz} {et~al.}(2024){Heintz}, {Watson}, {Brammer}, {Vejlgaard}, {Hutter}, {Strait}, {Matthee}, {Oesch}, {Jakobsson}, {Tanvir}, {Laursen}, {Naidu}, {Mason}, {Killi}, {Jung}, {Hsiao}, {Abdurro'uf}, {Coe}, {Arrabal Haro}, {Finkelstein}, \& {Toft}}]{Heintz2024}
{Heintz}, K.~E., {Watson}, D., {Brammer}, G., {et~al.} 2024, Science, 384, 890, \dodoi{10.1126/science.adj0343}

\bibitem[{{Hogg}(1999)}]{Hogg2000}
{Hogg}, D.~W. 1999, arXiv e-prints, arXiv:astro-ph/9905116, \dodoi{10.48550/arXiv.astro-ph/9905116}

\bibitem[{{Hopkins} {et~al.}(2004){Hopkins}, {Strauss}, {Hall}, {Richards}, {Cooper}, {Schneider}, {Vanden Berk}, {Jester}, {Brinkmann}, \& {Szokoly}}]{Hopkins2004}
{Hopkins}, P.~F., {Strauss}, M.~A., {Hall}, P.~B., {et~al.} 2004, \aj, 128, 1112, \dodoi{10.1086/423291}

\bibitem[{{Hunter}(2007)}]{Hunter2007}
{Hunter}, J.~D. 2007, Computing in Science and Engineering, 9, 90, \dodoi{10.1109/MCSE.2007.55}

\bibitem[{{Hviding} {et~al.}(2025){Hviding}, {de Graaff}, {Miller}, {Setton}, {Greene}, {Labb{\'e}}, {Brammer}, {Bezanson}, {Boogaard}, {Cleri}, {Leja}, {Maseda}, {McConachie}, {Matthee}, {Naidu}, {Oesch}, {Wang}, {Whitaker}, \& {Williams}}]{HvidingRUBIES}
{Hviding}, R.~E., {de Graaff}, A., {Miller}, T.~B., {et~al.} 2025, \aap, 702, A57, \dodoi{10.1051/0004-6361/202555816}

\bibitem[{{Inayoshi}(2025)}]{Inayoshi2025b}
{Inayoshi}, K. 2025, \apjl, 988, L22, \dodoi{10.3847/2041-8213/adea66}

\bibitem[{{Inayoshi} \& {Maiolino}(2025)}]{Inayoshi2024}
{Inayoshi}, K., \& {Maiolino}, R. 2025, \apjl, 980, L27, \dodoi{10.3847/2041-8213/adaebd}

\bibitem[{{Inayoshi} {et~al.}(2026){Inayoshi}, {Murase}, \& {Kashiyama}}]{Inayoshi2026}
{Inayoshi}, K., {Murase}, K., \& {Kashiyama}, K. 2026, \apj, 1000, 90, \dodoi{10.3847/1538-4357/ae42ce}

\bibitem[{{Ishikawa} {et~al.}(2026){Ishikawa}, {Eilers}, {Naidu}, {Matthee}, {Bordoloi}, {Chisholm}, {Greene}, {Ma}, {Oesch}, {Sun}, {Torralba}, {Weaver}, {Wuyts}, \& {Xiao}}]{Ishikawa2026}
{Ishikawa}, Y., {Eilers}, A.-C., {Naidu}, R.~P., {et~al.} 2026, arXiv e-prints, arXiv:2607.09647, \dodoi{10.48550/arXiv.2607.09647}

\bibitem[{{Ito} {et~al.}(2026){Ito}, {Valentino}, {Brammer}, {Hamadouche}, {Whitaker}, {Kokorev}, {Zhu}, {Kakimoto}, {Wu}, {Antwi-Danso}, {Baker}, {Ceverino}, {Faisst}, {Farcy}, {Fujimoto}, {Gallazzi}, {Gillman}, {Gottumukkala}, {Heintz}, {Hirschmann}, {Jespersen}, {Kubo}, {Lee}, {Magdis}, {Onodera}, {Shimakawa}, {Tanaka}, {Toft}, \& {Weaver}}]{Ito2026}
{Ito}, K., {Valentino}, F., {Brammer}, G., {et~al.} 2026, \aap, 710, A364, \dodoi{10.1051/0004-6361/202556137}

\bibitem[{{Jakobsen} {et~al.}(2022)}]{NIRSpec}
{Jakobsen}, P., {et~al.} 2022, \aap, 661, A81, \dodoi{10.1051/0004-6361/202142663}

\bibitem[{{Ji} {et~al.}(2025){Ji}, {Maiolino}, {{\"U}bler}, {Scholtz}, {D'Eugenio}, {Sun}, {Perna}, {Turner}, {Carniani}, {Arribas}, {Bennett}, {Bunker}, {Charlot}, {Cresci}, {Curti}, {Egami}, {Fabian}, {Inayoshi}, {Isobe}, {Jones}, {Juod{\v{z}}balis}, {Kumari}, {Lyu}, {Mazzolari}, {Parlanti}, {Robertson}, {Rodr{\'\i}guez Del Pino}, {Schneider}, {Sijacki}, {Tacchella}, {Trinca}, {Valiante}, {Venturi}, {Volonteri}, {Willott}, {Witten}, \& {Witstok}}]{BlackThunder2}
{Ji}, X., {Maiolino}, R., {{\"U}bler}, H., {et~al.} 2025, \mnras, 544, 3900, \dodoi{10.1093/mnras/staf1867}

\bibitem[{{Ji} {et~al.}(2026{\natexlab{a}}){Ji}, {D'Eugenio}, {Juod{\v{z}}balis}, {Walton}, {Fabian}, {Maiolino}, {Ramos Almeida}, {Acosta Pulido}, {Belokurov}, {Isobe}, {Jones}, {Maraston}, {Scholtz}, {Simmonds}, {Tacchella}, {Terlevich}, \& {Terlevich}}]{Ji2026b}
{Ji}, X., {D'Eugenio}, F., {Juod{\v{z}}balis}, I., {et~al.} 2026{\natexlab{a}}, \mnras, 545, 1, \dodoi{10.1093/mnras/staf2235}

\bibitem[{{Ji} {et~al.}(2026{\natexlab{b}}){Ji}, {Sun}, {Giavalisco}, {Zhu}, {Rieke}, {Williams}, {Maseda}, {Lyu}, {Rieke}, \& {Tacchella}}]{Ji2026}
{Ji}, Z., {Sun}, Y., {Giavalisco}, M., {et~al.} 2026{\natexlab{b}}, arXiv e-prints, arXiv:2606.18342, \dodoi{10.48550/arXiv.2606.18342}

\bibitem[{{Kido} {et~al.}(2025){Kido}, {Ioka}, {Hotokezaka}, {Inayoshi}, \& {Irwin}}]{Kido2025}
{Kido}, D., {Ioka}, K., {Hotokezaka}, K., {Inayoshi}, K., \& {Irwin}, C.~M. 2025, \mnras, 544, 3407, \dodoi{10.1093/mnras/staf1898}

\bibitem[{{Killi} {et~al.}(2024){Killi}, {Watson}, {Brammer}, {McPartland}, {Antwi-Danso}, {Newshore}, {Coe}, {Allen}, {Fynbo}, {Gould}, {Heintz}, {Rusakov}, \& {Vejlgaard}}]{Killi2024}
{Killi}, M., {Watson}, D., {Brammer}, G., {et~al.} 2024, \aap, 691, A52, \dodoi{10.1051/0004-6361/202348857}

\bibitem[{{Kocevski} {et~al.}(2023){Kocevski}, {Onoue}, {Inayoshi}, {Trump}, {Arrabal Haro}, {Grazian}, {Dickinson}, {Finkelstein}, {Kartaltepe}, {Hirschmann}, {Aird}, {Holwerda}, {Fujimoto}, {Juneau}, {Amor{\'\i}n}, {Backhaus}, {Bagley}, {Barro}, {Bell}, {Bisigello}, {Calabr{\`o}}, {Cleri}, {Cooper}, {Ding}, {Grogin}, {Ho}, {Hutchison}, {Inoue}, {Jiang}, {Jones}, {Koekemoer}, {Li}, {Li}, {McGrath}, {Molina}, {Papovich}, {P{\'e}rez-Gonz{\'a}lez}, {Pirzkal}, {Wilkins}, {Yang}, \& {Yung}}]{Kocevski2023}
{Kocevski}, D.~D., {Onoue}, M., {Inayoshi}, K., {et~al.} 2023, \apjl, 954, L4, \dodoi{10.3847/2041-8213/ace5a0}

\bibitem[{{Kocevski} {et~al.}(2025){Kocevski}, {Finkelstein}, {Barro}, {Taylor}, {Calabr{\`o}}, {Laloux}, {Buchner}, {Trump}, {Leung}, {Yang}, {Dickinson}, {P{\'e}rez-Gonz{\'a}lez}, {Pacucci}, {Inayoshi}, {Somerville}, {McGrath}, {Akins}, {Bagley}, {Bowler}, {Bisigello}, {Carnall}, {Casey}, {Cheng}, {Cleri}, {Costantin}, {Cullen}, {Davis}, {Donnan}, {Dunlop}, {Ellis}, {Ferguson}, {Fujimoto}, {Fontana}, {Giavalisco}, {Grazian}, {Grogin}, {Hathi}, {Hirschmann}, {Huertas-Company}, {Holwerda}, {Illingworth}, {Juneau}, {Kartaltepe}, {Koekemoer}, {Li}, {Lucas}, {Magee}, {Mason}, {McLeod}, {McLure}, {Napolitano}, {Papovich}, {Pirzkal}, {Rodighiero}, {Santini}, {Wilkins}, \& {Yung}}]{Kocevski2024}
{Kocevski}, D.~D., {Finkelstein}, S.~L., {Barro}, G., {et~al.} 2025, \apj, 986, 126, \dodoi{10.3847/1538-4357/adbc7d}

\bibitem[{{Kokorev} {et~al.}(2024){Kokorev}, {Caputi}, {Greene}, {Dayal}, {Trebitsch}, {Cutler}, {Fujimoto}, {Labb{\'e}}, {Miller}, {Iani}, {Navarro-Carrera}, \& {Rinaldi}}]{Kokorev2024}
{Kokorev}, V., {Caputi}, K.~I., {Greene}, J.~E., {et~al.} 2024, \apj, 968, 38, \dodoi{10.3847/1538-4357/ad4265}

\bibitem[{{Kutner}(1987)}]{RayleighJeans}
{Kutner}, M.~L. 1987, {Astronomy: A Physical Perspective} (John Wiley \& Sons, Ltd)

\bibitem[{{Labb{\'e}} {et~al.}(2023){Labb{\'e}}, {van Dokkum}, {Nelson}, {Bezanson}, {Suess}, {Leja}, {Brammer}, {Whitaker}, {Mathews}, {Stefanon}, \& {Wang}}]{Labbe2023}
{Labb{\'e}}, I., {van Dokkum}, P., {Nelson}, E., {et~al.} 2023, \nat, 616, 266, \dodoi{10.1038/s41586-023-05786-2}

\bibitem[{{Labbe} {et~al.}(2024){Labbe}, {Greene}, {Matthee}, {Treiber}, {Kokorev}, {Miller}, {Kramarenko}, {Setton}, {Ma}, {Goulding}, {Bezanson}, {Naidu}, {Williams}, {Atek}, {Brammer}, {Cutler}, {Chemerynska}, {Cloonan}, {Dayal}, {de Graaff}, {Fudamoto}, {Fujimoto}, {Furtak}, {Glazebrook}, {Heintz}, {Leja}, {Marchesini}, {Nanayakkara}, {Nelson}, {Oesch}, {Pan}, {Price}, {Shivaei}, {Sobral}, {Suess}, {van Dokkum}, {Wang}, {Weaver}, {Whitaker}, \& {Zitrin}}]{Labbe2024b}
{Labbe}, I., {Greene}, J.~E., {Matthee}, J., {et~al.} 2024, arXiv e-prints, arXiv:2412.04557, \dodoi{10.48550/arXiv.2412.04557}

\bibitem[{{Labbe} {et~al.}(2025){Labbe}, {Greene}, {Bezanson}, {Fujimoto}, {Furtak}, {Goulding}, {Matthee}, {Naidu}, {Oesch}, {Atek}, {Brammer}, {Chemerynska}, {Coe}, {Cutler}, {Dayal}, {Feldmann}, {Franx}, {Glazebrook}, {Leja}, {Maseda}, {Marchesini}, {Nanayakkara}, {Nelson}, {Pan}, {Papovich}, {Price}, {Suess}, {Wang}, {Weaver}, {Whitaker}, {Williams}, \& {Zitrin}}]{Labbe2025}
{Labbe}, I., {Greene}, J.~E., {Bezanson}, R., {et~al.} 2025, \apj, 978, 92, \dodoi{10.3847/1538-4357/ad3551}

\bibitem[{{Lambrides} {et~al.}(2026){Lambrides}, {Larson}, {Garofali}, {Ptak}, {Chiaberge}, {Long}, {Hutchison}, {Norman}, {McKinney}, {Akins}, {Berg}, {Chisholm}, {Civano}, {Cloonan}, {Endsley}, {Faisst}, {Gilli}, {Gillman}, {Hirschmann}, {Kartaltepe}, {Kocevski}, {Kokorev}, {Pacucci}, {Richardson}, {Stiavelli}, \& {Whalen}}]{Lambrides2024}
{Lambrides}, E., {Larson}, R.~L., {Garofali}, K., {et~al.} 2026, Nature Astronomy, 10, 868, \dodoi{10.1038/s41550-026-02813-w}

\bibitem[{{Lin} {et~al.}(2024){Lin}, {Wang}, {Fan}, {Cai}, {Champagne}, {Sun}, {Volonteri}, {Yang}, {Hennawi}, {Ba{\~n}ados}, {Barth}, {Eilers}, {Farina}, {Liu}, {Jin}, {Jun}, {Lupi}, {Kakiichi}, {Mazzucchelli}, {Onoue}, {Pan}, {Pizzati}, {Rojas-Ruiz}, {Schindler}, {Trakhtenbrot}, {Shen}, {Trebitsch}, {Zhuang}, {Endsley}, {Meyer}, {Li}, {Li}, {Pudoka}, {Tee}, {Wu}, \& {Zhang}}]{Lin2024}
{Lin}, X., {Wang}, F., {Fan}, X., {et~al.} 2024, \apj, 974, 147, \dodoi{10.3847/1538-4357/ad6565}

\bibitem[{{Lin} {et~al.}(2026){Lin}, {Fan}, {Cai}, {Bian}, {Liu}, {Sun}, {Ma}, {Greene}, {Strauss}, {Green}, {Lyu}, {Champagne}, {Goulding}, {Inayoshi}, {Jin}, {Leung}, {Li}, {Liu}, {Liu}, {Mao}, {Pudoka}, {Tee}, {Wang}, {Wang}, {Wu}, {Yang}, {Zhang}, \& {Zhu}}]{Lin2026}
{Lin}, X., {Fan}, X., {Cai}, Z., {et~al.} 2026, \apj, 997, 364, \dodoi{10.3847/1538-4357/ae2bdf}

\bibitem[{{Liu} {et~al.}(2025){Liu}, {Jiang}, {Quataert}, {Greene}, \& {Ma}}]{Liu2025}
{Liu}, H., {Jiang}, Y.-F., {Quataert}, E., {Greene}, J.~E., \& {Ma}, Y. 2025, \apj, 994, 113, \dodoi{10.3847/1538-4357/ae0c19}

\bibitem[{{Ma} {et~al.}(2025){Ma}, {Greene}, {Setton}, {Volonteri}, {Leja}, {Wang}, {Bezanson}, {Brammer}, {Cutler}, {Dayal}, {van Dokkum}, {Furtak}, {Glazebrook}, {Goulding}, {de Graaff}, {Kokorev}, {Labbe}, {Pan}, {Price}, {Weaver}, {Williams}, {Whitaker}, \& {Zitrin}}]{Ma2025}
{Ma}, Y., {Greene}, J.~E., {Setton}, D.~J., {et~al.} 2025, \apj, 981, 191, \dodoi{10.3847/1538-4357/ada613}

\bibitem[{{Madau} \& {Maiolino}(2026)}]{MadauMaio2026}
{Madau}, P., \& {Maiolino}, R. 2026, arXiv e-prints, arXiv:2602.22386, \dodoi{10.48550/arXiv.2602.22386}

\bibitem[{{Maiolino} {et~al.}(2024){Maiolino}, {Scholtz}, {Curtis-Lake}, {Carniani}, {Baker}, {de Graaff}, {Tacchella}, {{\"U}bler}, {D'Eugenio}, {Witstok}, {Curti}, {Arribas}, {Bunker}, {Charlot}, {Chevallard}, {Eisenstein}, {Egami}, {Ji}, {Jones}, {Lyu}, {Rawle}, {Robertson}, {Rujopakarn}, {Perna}, {Sun}, {Venturi}, {Williams}, \& {Willott}}]{Maiolino2024}
{Maiolino}, R., {Scholtz}, J., {Curtis-Lake}, E., {et~al.} 2024, \aap, 691, A145, \dodoi{10.1051/0004-6361/202347640}

\bibitem[{{Matthee} {et~al.}(2024){Matthee}, {Naidu}, {Brammer}, {Chisholm}, {Eilers}, {Goulding}, {Greene}, {Kashino}, {Labbe}, {Lilly}, {Mackenzie}, {Oesch}, {Weibel}, {Wuyts}, {Xiao}, {Bordoloi}, {Bouwens}, {van Dokkum}, {Illingworth}, {Kramarenko}, {Maseda}, {Mason}, {Meyer}, {Nelson}, {Reddy}, {Shivaei}, {Simcoe}, \& {Yue}}]{MattheeLRDPioneer}
{Matthee}, J., {Naidu}, R.~P., {Brammer}, G., {et~al.} 2024, \apj, 963, 129, \dodoi{10.3847/1538-4357/ad2345}

\bibitem[{{M{\'e}rida} {et~al.}(2026){M{\'e}rida}, {Sawicki}, {Willott}, {Gaspar}, \& {Iyer}}]{Merida2026}
{M{\'e}rida}, R.~M., {Sawicki}, M., {Willott}, C.~J., {Gaspar}, G., \& {Iyer}, K.~G. 2026, arXiv e-prints, arXiv:2606.12355, \dodoi{10.48550/arXiv.2606.12355}

\bibitem[{{Naidu} {et~al.}(2026){Naidu}, {Matthee}, {Katz}, {de Graaff}, {Oesch}, {Smith}, {Greene}, {Brammer}, {Weibel}, {Hviding}, {Chisholm}, {Labb{\'e}}, {Simcoe}, {Witten}, {Sun}, {Atek}, {Baggen}, {Belli}, {Bezanson}, {Boogaard}, {Bose}, {Bouwens}, {Covelo-Paz}, {Dayal}, {Fudamoto}, {Furtak}, {Giovinazzo}, {Goulding}, {Gronke}, {Heintz}, {Hirschmann}, {Illingworth}, {Inoue}, {Johnson}, {Leja}, {Leonova}, {McConachie}, {Maseda}, {Natarajan}, {Nelson}, {Setton}, {Shivaei}, {Sobral}, {Stefanon}, {Tacchella}, {Toft}, {Torralba}, {van Dokkum}, {van der Wel}, {Volonteri}, {Walter}, {Wang}, {Watson}, \& {Whitaker}}]{Naidu2025}
{Naidu}, R.~P., {Matthee}, J., {Katz}, H., {et~al.} 2026, \nat, 656, 329, \dodoi{10.1038/s41586-026-10846-4}

\bibitem[{{Noboriguchi} {et~al.}(2023){Noboriguchi}, {Inoue}, {Nagao}, {Toba}, \& {Misawa}}]{Noboriguchi2023}
{Noboriguchi}, A., {Inoue}, A.~K., {Nagao}, T., {Toba}, Y., \& {Misawa}, T. 2023, \apjl, 959, L14, \dodoi{10.3847/2041-8213/ad0e00}

\bibitem[{{Oesch} {et~al.}(2023){Oesch}, {Brammer}, {Naidu}, {Bouwens}, {Chisholm}, {Illingworth}, {Matthee}, {Nelson}, {Qin}, {Reddy}, {Shapley}, {Shivaei}, {van Dokkum}, {Weibel}, {Whitaker}, {Wuyts}, {Covelo-Paz}, {Endsley}, {Fudamoto}, {Giovinazzo}, {Herard-Demanche}, {Kerutt}, {Kramarenko}, {Labbe}, {Leonova}, {Lin}, {Magee}, {Marchesini}, {Maseda}, {Mason}, {Matharu}, {Meyer}, {Neufeld}, {Prieto Lyon}, {Schaerer}, {Sharma}, {Shuntov}, {Smit}, {Stefanon}, {Wyithe}, \& {Xiao}}]{Oesch2023}
{Oesch}, P.~A., {Brammer}, G., {Naidu}, R.~P., {et~al.} 2023, \mnras, 525, 2864, \dodoi{10.1093/mnras/stad2411}

\bibitem[{{Oke} \& {Gunn}(1983)}]{OkeGunn}
{Oke}, J.~B., \& {Gunn}, J.~E. 1983, \apj, 266, 713, \dodoi{10.1086/160817}

\bibitem[{{Onoue} {et~al.}(2025){Onoue}, {Ding}, {Silverman}, {Matsuoka}, {Izumi}, {Strauss}, {Ward}, {Phillips}, {Ito}, {Andika}, {Aoki}, {Arita}, {Baba}, {Bieri}, {Bosman}, {Eilers}, {Fujimoto}, {Habouzit}, {Haiman}, {Imanishi}, {Inayoshi}, {Iwasawa}, {Jahnke}, {Kashikawa}, {Kawaguchi}, {Kohno}, {Lee}, {Li}, {Lupi}, {Lyu}, {Nagao}, {Overzier}, {Schindler}, {Schramm}, {Scoggins}, {Shimasaku}, {Toba}, {Trakhtenbrot}, {Trebitsch}, {Treu}, {Umehata}, {Venemans}, {Vestergaard}, {Volonteri}, {Walter}, {Wang}, {Yang}, \& {Zhang}}]{Onoue2025}
{Onoue}, M., {Ding}, X., {Silverman}, J.~D., {et~al.} 2025, Nature Astronomy, 9, 1541, \dodoi{10.1038/s41550-025-02628-1}

\bibitem[{{Pei}(1992)}]{Pei1992}
{Pei}, Y.~C. 1992, \apj, 395, 130, \dodoi{10.1086/171637}

\bibitem[{{P{\'e}rez-Gonz{\'a}lez} {et~al.}(2024){P{\'e}rez-Gonz{\'a}lez}, {Barro}, {Rieke}, {Lyu}, {Rieke}, {Alberts}, {Williams}, {Hainline}, {Sun}, {Pusk{\'a}s}, {Annunziatella}, {Baker}, {Bunker}, {Egami}, {Ji}, {Johnson}, {Robertson}, {Rodr{\'\i}guez Del Pino}, {Rujopakarn}, {Shivaei}, {Tacchella}, {Willmer}, \& {Willott}}]{Perez-Gonzalez2024}
{P{\'e}rez-Gonz{\'a}lez}, P.~G., {Barro}, G., {Rieke}, G.~H., {et~al.} 2024, \apj, 968, 4, \dodoi{10.3847/1538-4357/ad38bb}

\bibitem[{{P{\'e}rez-Gonz{\'a}lez} {et~al.}(2026){P{\'e}rez-Gonz{\'a}lez}, {Barro}, {Carniani}, {D'Eugenio}, {Rieke}, {Tripodi}, {Bunker}, {Ji}, {Marques-Chaves}, {Schaerer}, {Venturi}, {Ar{\'e}valo-Gonz{\'a}lez}, {Arribas}, {Rinaldi}, {Rodr{\'\i}guez Del Pino}, {Witstok}, {Bhatawdekar}, {Boogaard}, {Charlot}, {Chevallard}, {Costantin}, {Curti}, {Curtis-Lake}, {Daddi}, {Davis}, {Dickinson}, {Donnan}, {Donnan}, {Dunlop}, {Eisenstein}, {Ferguson}, {Fern{\'a}ndez Aranda}, {Finkelstein}, {Fujimoto}, {Gandolfi}, {Giavalisco}, {Grogin}, {Hamed}, {Hirschmann}, {Kartaltepe}, {Kocevski}, {Koekemoer}, {Leung}, {Lofaro}, {Lucas}, {McLeod}, {Melinder}, {{\"O}stlin}, {Papovich}, {Pentericci}, {P{\'e}rez-D{\'\i}az}, {Rieke}, {Scholtz}, {Somerville}, {Stanton}, {Stevenson}, {Shivaei}, {Tacchella}, {Trump}, {{\"U}bler}, {Wang}, {Williams}, {Willmer}, {Yung}, \& {Zhu}}]{Perez2026b}
{P{\'e}rez-Gonz{\'a}lez}, P.~G., {Barro}, G., {Carniani}, S., {et~al.} 2026, arXiv e-prints, arXiv:2602.20247, \dodoi{10.48550/arXiv.2602.20247}

\bibitem[{{Perger} {et~al.}(2025){Perger}, {Fogasy}, {Frey}, \& {Gab{\'a}nyi}}]{Perger2025}
{Perger}, K., {Fogasy}, J., {Frey}, S., \& {Gab{\'a}nyi}, K.~{\'E}. 2025, \aap, 693, L2, \dodoi{10.1051/0004-6361/202452422}

\bibitem[{{Planck Collaboration} {et~al.}(2020){Planck Collaboration}, {Aghanim}, {Akrami}, {Ashdown}, {Aumont}, {Baccigalupi}, {Ballardini}, {Banday}, {Barreiro}, {Bartolo}, {Basak}, {Battye}, {Benabed}, {Bernard}, {Bersanelli}, {Bielewicz}, {Bock}, {Bond}, {Borrill}, {Bouchet}, {Boulanger}, {Bucher}, {Burigana}, {Butler}, {Calabrese}, {Cardoso}, {Carron}, {Challinor}, {Chiang}, {Chluba}, {Colombo}, {Combet}, {Contreras}, {Crill}, {Cuttaia}, {de Bernardis}, {de Zotti}, {Delabrouille}, {Delouis}, {Di Valentino}, {Diego}, {Dor{\'e}}, {Douspis}, {Ducout}, {Dupac}, {Dusini}, {Efstathiou}, {Elsner}, {En{\ss}lin}, {Eriksen}, {Fantaye}, {Farhang}, {Fergusson}, {Fernandez-Cobos}, {Finelli}, {Forastieri}, {Frailis}, {Fraisse}, {Franceschi}, {Frolov}, {Galeotta}, {Galli}, {Ganga}, {G{\'e}nova-Santos}, {Gerbino}, {Ghosh}, {Gonz{\'a}lez-Nuevo}, {G{\'o}rski}, {Gratton}, {Gruppuso}, {Gudmundsson}, {Hamann}, {Handley}, {Hansen}, {Herranz}, {Hildebrandt}, {Hivon}, {Huang}, {Jaffe}, {Jones}, {Karakci}, {Keih{\"a}nen},
  {Keskitalo}, {Kiiveri}, {Kim}, {Kisner}, {Knox}, {Krachmalnicoff}, {Kunz}, {Kurki-Suonio}, {Lagache}, {Lamarre}, {Lasenby}, {Lattanzi}, {Lawrence}, {Le Jeune}, {Lemos}, {Lesgourgues}, {Levrier}, {Lewis}, {Liguori}, {Lilje}, {Lilley}, {Lindholm}, {L{\'o}pez-Caniego}, {Lubin}, {Ma}, {Mac{\'\i}as-P{\'e}rez}, {Maggio}, {Maino}, {Mandolesi}, {Mangilli}, {Marcos-Caballero}, {Maris}, {Martin}, {Martinelli}, {Mart{\'\i}nez-Gonz{\'a}lez}, {Matarrese}, {Mauri}, {McEwen}, {Meinhold}, {Melchiorri}, {Mennella}, {Migliaccio}, {Millea}, {Mitra}, {Miville-Desch{\^e}nes}, {Molinari}, {Montier}, {Morgante}, {Moss}, {Natoli}, {N{\o}rgaard-Nielsen}, {Pagano}, {Paoletti}, {Partridge}, {Patanchon}, {Peiris}, {Perrotta}, {Pettorino}, {Piacentini}, {Polastri}, {Polenta}, {Puget}, {Rachen}, {Reinecke}, {Remazeilles}, {Renzi}, {Rocha}, {Rosset}, {Roudier}, {Rubi{\~n}o-Mart{\'\i}n}, {Ruiz-Granados}, {Salvati}, {Sandri}, {Savelainen}, {Scott}, {Shellard}, {Sirignano}, {Sirri}, {Spencer}, {Sunyaev}, {Suur-Uski}, {Tauber}, {Tavagnacco},
  {Tenti}, {Toffolatti}, {Tomasi}, {Trombetti}, {Valenziano}, {Valiviita}, {Van Tent}, {Vibert}, {Vielva}, {Villa}, {Vittorio}, {Wandelt}, {Wehus}, {White}, {White}, {Zacchei}, \& {Zonca}}]{Planck2018}
{Planck Collaboration}, {Aghanim}, N., {Akrami}, Y., {et~al.} 2020, \aap, 641, A6, \dodoi{10.1051/0004-6361/201833910}

\bibitem[{{Pollock} {et~al.}(2026){Pollock}, {Gottumukkala}, {Heintz}, {Brammer}, {Roberts-Borsani}, {Oesch}, {Witstok}, {Arellano-C{\'o}rdova}, {Cullen}, {Scholte}, {Terp}, {Rowland}, {Sneppen}, {Ito}, {Valentino}, {Matthee}, {Watson}, \& {Toft}}]{Pollock2026}
{Pollock}, C.~L., {Gottumukkala}, R., {Heintz}, K.~E., {et~al.} 2026, \aap, 708, A203, \dodoi{10.1051/0004-6361/202556032}

\bibitem[{{Reddy} {et~al.}(2023){Reddy}, {Topping}, {Sanders}, {Shapley}, \& {Brammer}}]{Reddy2023}
{Reddy}, N.~A., {Topping}, M.~W., {Sanders}, R.~L., {Shapley}, A.~E., \& {Brammer}, G. 2023, \apj, 952, 167, \dodoi{10.3847/1538-4357/acd754}

\bibitem[{{Reines} \& {Volonteri}(2015)}]{ReinesVolonteri2015}
{Reines}, A.~E., \& {Volonteri}, M. 2015, \apj, 813, 82, \dodoi{10.1088/0004-637X/813/2/82}

\bibitem[{{Rieke} {et~al.}(2023){Rieke}, {Kelly}, {Misselt}, {Stansberry}, {Boyer}, {Beatty}, {Egami}, {Florian}, {Greene}, {Hainline}, {Leisenring}, {Roellig}, {Schlawin}, {Sun}, {Tinnin}, {Williams}, {Willmer}, {Wilson}, {Clark}, {Rohrbach}, {Brooks}, {Canipe}, {Correnti}, {DiFelice}, {Gennaro}, {Girard}, {Hartig}, {Hilbert}, {Koekemoer}, {Nikolov}, {Pirzkal}, {Rest}, {Robberto}, {Sunnquist}, {Telfer}, {Wu}, {Ferry}, {Lewis}, {Baum}, {Beichman}, {Doyon}, {Dressler}, {Eisenstein}, {Ferrarese}, {Hodapp}, {Horner}, {Jaffe}, {Johnstone}, {Krist}, {Martin}, {McCarthy}, {Meyer}, {Rieke}, {Trauger}, \& {Young}}]{Rieke2023}
{Rieke}, M.~J., {Kelly}, D.~M., {Misselt}, K., {et~al.} 2023, \pasp, 135, 028001, \dodoi{10.1088/1538-3873/acac53}

\bibitem[{{Rinaldi} {et~al.}(2025){Rinaldi}, {Bonaventura}, {Rieke}, {Alberts}, {Caputi}, {Baker}, {Baum}, {Bhatawdekar}, {Bunker}, {Carniani}, {Curtis-Lake}, {D'Eugenio}, {Egami}, {Ji}, {Johnson}, {Hainline}, {Helton}, {Lin}, {Lyu}, {Ma}, {Maiolino}, {P{\'e}rez-Gonz{\'a}lez}, {Rieke}, {Robertson}, {Shivaei}, {Stone}, {Sun}, {Tacchella}, {{\"U}bler}, {Williams}, {Willmer}, {Willott}, {Zhang}, \& {Zhu}}]{Rinialdi2025}
{Rinaldi}, P., {Bonaventura}, N., {Rieke}, G.~H., {et~al.} 2025, \apj, 992, 71, \dodoi{10.3847/1538-4357/adfa10}

\bibitem[{{Ronayne} {et~al.}(2026){Ronayne}, {Papovich}, {Kirkpatrick}, {Backhaus}, {Cullen}, {Shen}, {Bagley}, {Barro}, {Finkelstein}, {Hamblin}, {Kartaltepe}, {Kocevski}, {Koekemoer}, {Lambrides}, {Pacucci}, \& {Yang}}]{Ronayne2025}
{Ronayne}, K., {Papovich}, C., {Kirkpatrick}, A., {et~al.} 2026, \apj, 1003, 234, \dodoi{10.3847/1538-4357/ae5e6b}

\bibitem[{{Rusakov} {et~al.}(2026){Rusakov}, {Watson}, {Nikopoulos}, {Brammer}, {Gottumukkala}, {Harvey}, {Heintz}, {Damgaard}, {Sim}, {Sneppen}, {Vijayan}, {Adams}, {Austin}, {Conselice}, {Goolsby}, {Toft}, \& {Witstok}}]{Rusakov2026}
{Rusakov}, V., {Watson}, D., {Nikopoulos}, G.~P., {et~al.} 2026, \nat, 649, 574, \dodoi{10.1038/s41586-025-09900-4}

\bibitem[{{Salim} \& {Narayanan}(2020)}]{Salim2020}
{Salim}, S., \& {Narayanan}, D. 2020, \araa, 58, 529, \dodoi{10.1146/annurev-astro-032620-021933}

\bibitem[{{Scholtz} {et~al.}(2026){Scholtz}, {D'Eugenio}, {Maiolino}, {Brazzini}, {{\"U}bler}, {Ji}, {Perna}, {Sun}, {Brocchi}, {Carniani}, {Cresci}, {Ivey}, {Juod{\v{z}}balis}, {Marconi}, {Mazzolari}, {Risaliti}, \& {Trefoloni}}]{Scholtz2026}
{Scholtz}, J., {D'Eugenio}, F., {Maiolino}, R., {et~al.} 2026, arXiv e-prints, arXiv:2603.22277, \dodoi{10.48550/arXiv.2603.22277}

\bibitem[{{Setton} {et~al.}(2025{\natexlab{a}}){Setton}, {Greene}, {de Graaff}, {Ma}, {Leja}, {Matthee}, {Bezanson}, {Boogaard}, {Cleri}, {Katz}, {Labbe}, {Maseda}, {McConachie}, {Miller}, {Price}, {Suess}, {van Dokkum}, {Wang}, {Weibel}, {Whitaker}, \& {Williams}}]{Setton2025}
{Setton}, D.~J., {Greene}, J.~E., {de Graaff}, A., {et~al.} 2025{\natexlab{a}}, \apj, 995, 118, \dodoi{10.3847/1538-4357/ae1500}

\bibitem[{{Setton} {et~al.}(2025{\natexlab{b}}){Setton}, {Greene}, {Spilker}, {Williams}, {Labb{\'e}}, {Ma}, {Wang}, {Whitaker}, {Leja}, {de Graaff}, {Alberts}, {Bezanson}, {Boogaard}, {Brammer}, {Cutler}, {Cleri}, {Cooper}, {Dayal}, {Fujimoto}, {Furtak}, {Goulding}, {Hirschmann}, {Kokorev}, {Maseda}, {McConachie}, {Matthee}, {Miller}, {Naidu}, {Oesch}, {Pan}, {Price}, {Suess}, {Weaver}, {Xiao}, {Zhang}, \& {Zitrin}}]{Setton2025b}
{Setton}, D.~J., {Greene}, J.~E., {Spilker}, J.~S., {et~al.} 2025{\natexlab{b}}, \apjl, 991, L10, \dodoi{10.3847/2041-8213/ade78b}

\bibitem[{{Silverman}(1986)}]{Silverman1986}
{Silverman}, B.~W. 1986, {Density Estimation for Statistics and Data Analysis} (Chapman \& Hall), \dodoi{10.1201/9781315140919}

\bibitem[{{Sun} {et~al.}(2026{\natexlab{a}}){Sun}, {Eisenstein}, {D'Eugenio}, {Hainline}, {Helton}, {Johnson}, {Lin}, {Rieke}, {Robertson}, {Tacchella}, {Bunker}, {Chevallard}, {Curtis-Lake}, {Egami}, {Hausen}, {Ji}, {Lyu}, {Maiolino}, {Rinaldi}, {Sun}, {Trussler}, {Williams}, {Willmer}, {Witstok}, {Wu}, \& {Zhu}}]{SunFW2026}
{Sun}, F., {Eisenstein}, D.~J., {D'Eugenio}, F., {et~al.} 2026{\natexlab{a}}, arXiv e-prints, arXiv:2601.15961, \dodoi{10.48550/arXiv.2601.15961}

\bibitem[{{Sun} {et~al.}(2026{\natexlab{b}}){Sun}, {Naidu}, {Matthee}, {de Graaff}, {Chisholm}, {Greene}, {Oesch}, {Torralba}, {Hviding}, {Brammer}, {Simcoe}, {Bose}, {Bouwens}, {Dayal}, {Eilers}, {Fei}, {Furtak}, {Gottumukkala}, {Goulding}, {Heintz}, {Hirschmann}, {Kokorev}, {Leja}, {Liu}, {Natarajan}, {Santarelli}, {Setton}, {Smith}, {Tacchella}, {Volonteri}, {Walter}, {Weibel}, \& {Williams}}]{Sun2026}
{Sun}, W.~Q., {Naidu}, R.~P., {Matthee}, J., {et~al.} 2026{\natexlab{b}}, The Open Journal of Astrophysics, 9, \dodoi{10.33232/001c.162505}

\bibitem[{{Tanaka} {et~al.}(2025){Tanaka}, {Akins}, {Harikane}, {Silverman}, {Casey}, {Inayoshi}, {Schindler}, {Shimasaku}, {Kocevski}, {Onoue}, {Faisst}, {Robertson}, {Kokorev}, {Shuntov}, {Koekemoer}, {Franco}, {Egami}, {Liu}, {Taylor}, {Kartaltepe}, {Bosman}, {Champagne}, {Kakiichi}, {Harish}, {Zhang}, {Newman}, {Kakkad}, {Fei}, {Fujimoto}, {Li}, {Finkelstein}, {Li}, {Lambrides}, {Sommovigo}, {Zavala}, {Ito}, {Liu}, {Treister}, {Aravena}, {Gozaliasl}, {Zhang}, {Hatamnia}, {Umeda}, {Inoue}, {Yang}, {Ando}, {Arita}, {Ding}, {Matsui}, {Shibanuma}, {Magdis}, {Zhuang}, {Fan}, {Li}, {Liu}, {Lyu}, {Rhodes}, {Toft}, {Wang}, {Zou}, {Arango-Toro}, {Battisti}, {Gillman}, {Khostovan}, {Long}, {Mobasher}, \& {Sanders}}]{Tanaka2025}
{Tanaka}, T.~S., {Akins}, H.~B., {Harikane}, Y., {et~al.} 2025, \apj, 995, 21, \dodoi{10.3847/1538-4357/ae145f}

\bibitem[{{Taylor} {et~al.}(2025){Taylor}, {Kokorev}, {Kocevski}, {Akins}, {Cullen}, {Dickinson}, {Finkelstein}, {Arrabal Haro}, {Bromm}, {Giavalisco}, {Inayoshi}, {Juneau}, {Leung}, {P{\'e}rez-Gonz{\'a}lez}, {Somerville}, {Trump}, {Amor{\'\i}n}, {Barro}, {Burgarella}, {Brooks}, {Carnall}, {Casey}, {Cheng}, {Chisholm}, {Chworowsky}, {Davis}, {Donnan}, {Dunlop}, {Ellis}, {Fern{\'a}ndez}, {Fujimoto}, {Grogin}, {Gupta}, {Hathi}, {Jung}, {Hirschmann}, {Kartaltepe}, {Koekemoer}, {Larson}, {Leung}, {Llerena}, {Lucas}, {McLeod}, {McLure}, {Napolitano}, {Papovich}, {Stanton}, {Tripodi}, {Wang}, {Wilkins}, {Yung}, \& {Zavala}}]{Taylor2025b}
{Taylor}, A.~J., {Kokorev}, V., {Kocevski}, D.~D., {et~al.} 2025, \apjl, 989, L7, \dodoi{10.3847/2041-8213/ade789}

\bibitem[{{Torralba} {et~al.}(2026{\natexlab{a}}){Torralba}, {Matthee}, {Pezzulli}, {Urrutia}, {Gronke}, {Mascia}, {D'Eugenio}, {Di Cesare}, {Eilers}, {Greene}, {Iani}, {Ishikawa}, {Mackenzie}, {Naidu}, {Navarrete}, \& {Kotiwale}}]{Torralba2026b}
{Torralba}, A., {Matthee}, J., {Pezzulli}, G., {et~al.} 2026{\natexlab{a}}, \aap, 705, A147, \dodoi{10.1051/0004-6361/202555596}

\bibitem[{{Torralba} {et~al.}(2026{\natexlab{b}}){Torralba}, {Matthee}, {Pezzulli}, {Naidu}, {Ishikawa}, {Brammer}, {Chang}, {Chisholm}, {de Graaff}, {D'Eugenio}, {Di Cesare}, {Eilers}, {Greene}, {Gronke}, {Iani}, {Kokorev}, {Kotiwale}, {Kramarenko}, {Ma}, {Mascia}, {Navarrete}, {Nelson}, {Oesch}, {Simcoe}, \& {Wuyts}}]{Torralba2026}
---. 2026{\natexlab{b}}, \aap, 707, A75, \dodoi{10.1051/0004-6361/202557537}

\bibitem[{{Tripodi} {et~al.}(2025){Tripodi}, {Brada{\v{c}}}, {D'Eugenio}, {Martis}, {Rihtar{\v{s}}i{\v{c}}}, {Willott}, {Pentericci}, {Moreschini}, {Markevitch}, {Asada}, {Calabr{\'o}}, {Desprez}, {Felicioni}, {Gaspar}, {Gonzalez}, {Harshan}, {Ji}, {Jude{\v{z}}}, {Lemaux}, {Marconi}, {Markov}, {Merida}, {Napolitano}, {Noirot}, {Parente}, {Peter}, {Robbins}, {Robertson}, {Sarrouh}, \& {Sawicki}}]{Tripodi2025}
{Tripodi}, R., {Brada{\v{c}}}, M., {D'Eugenio}, F., {et~al.} 2025, \apjl, 994, L6, \dodoi{10.3847/2041-8213/ae13a9}

\bibitem[{{Valentino} {et~al.}(2023){Valentino}, {Brammer}, {Gould}, {Kokorev}, {Fujimoto}, {Jespersen}, {Vijayan}, {Weaver}, {Ito}, {Tanaka}, {Ilbert}, {Magdis}, {Whitaker}, {Faisst}, {Gallazzi}, {Gillman}, {Gim{\'e}nez-Arteaga}, {G{\'o}mez-Guijarro}, {Kubo}, {Heintz}, {Hirschmann}, {Oesch}, {Onodera}, {Rizzo}, {Lee}, {Strait}, \& {Toft}}]{Valentino2023NIRCam}
{Valentino}, F., {Brammer}, G., {Gould}, K. M.~L., {et~al.} 2023, \apj, 947, 20, \dodoi{10.3847/1538-4357/acbefa}

\bibitem[{{Virtanen} {et~al.}(2020){Virtanen}, {Gommers}, {Oliphant}, {Haberland}, {Reddy}, {Cournapeau}, {Burovski}, {Peterson}, {Weckesser}, {Bright}, {van der Walt}, {Brett}, {Wilson}, {Millman}, {Mayorov}, {Nelson}, {Jones}, {Kern}, {Larson}, {Carey}, {Polat}, {Feng}, {Moore}, {VanderPlas}, {Laxalde}, {Perktold}, {Cimrman}, {Henriksen}, {Quintero}, {Harris}, {Archibald}, {Ribeiro}, {Pedregosa}, {van Mulbregt}, \& {SciPy 1. 0 Contributors}}]{Virtanen2020}
{Virtanen}, P., {Gommers}, R., {Oliphant}, T.~E., {et~al.} 2020, Nature Methods, 17, 261, \dodoi{10.1038/s41592-019-0686-2}

\bibitem[{{Wang} {et~al.}(2024){Wang}, {Leja}, {de Graaff}, {Brammer}, {Weibel}, {van Dokkum}, {Baggen}, {Suess}, {Greene}, {Bezanson}, {Cleri}, {Hirschmann}, {Labb{\'e}}, {Matthee}, {McConachie}, {Naidu}, {Nelson}, {Oesch}, {Setton}, \& {Williams}}]{Wang2024}
{Wang}, B., {Leja}, J., {de Graaff}, A., {et~al.} 2024, \apjl, 969, L13, \dodoi{10.3847/2041-8213/ad55f7}

\bibitem[{{Weibel} {et~al.}(2026){Weibel}, {Naidu}, {Oesch}, {de Graaff}, {Hviding}, {Liu}, {Matthee}, {Williams}, {Brammer}, {Covelo Paz}, {Greene}, {Jespersen}, {Ji}, {Maseda}, {Setton}, {Sun}, {Torralba}, {Witten}, \& {Xiao}}]{Weibel2026}
{Weibel}, A., {Naidu}, R.~P., {Oesch}, P.~A., {et~al.} 2026, arXiv e-prints, arXiv:2606.17271, \dodoi{10.48550/arXiv.2606.17271}

\bibitem[{{Williams} {et~al.}(2024){Williams}, {Alberts}, {Ji}, {Hainline}, {Lyu}, {Rieke}, {Endsley}, {Suess}, {Sun}, {Johnson}, {Florian}, {Shivaei}, {Rujopakarn}, {Baker}, {Bhatawdekar}, {Boyett}, {Bunker}, {Cameron}, {Carniani}, {Charlot}, {Curtis-Lake}, {DeCoursey}, {de Graaff}, {Egami}, {Eisenstein}, {Gibson}, {Hausen}, {Helton}, {Maiolino}, {Maseda}, {Nelson}, {P{\'e}rez-Gonz{\'a}lez}, {Rieke}, {Robertson}, {Saxena}, {Tacchella}, {Willmer}, \& {Willott}}]{Williams2024}
{Williams}, C.~C., {Alberts}, S., {Ji}, Z., {et~al.} 2024, \apj, 968, 34, \dodoi{10.3847/1538-4357/ad3f17}

\bibitem[{{Yue} {et~al.}(2024){Yue}, {Eilers}, {Ananna}, {Panagiotou}, {Kara}, \& {Miyaji}}]{Yue2024}
{Yue}, M., {Eilers}, A.-C., {Ananna}, T.~T., {et~al.} 2024, \apjl, 974, L26, \dodoi{10.3847/2041-8213/ad7eba}

\bibitem[{{Zafar} {et~al.}(2015){Zafar}, {M{\o}ller}, {Watson}, {Fynbo}, {Krogager}, {Zafar}, {Saturni}, {Geier}, \& {Venemans}}]{Zafar2015}
{Zafar}, T., {M{\o}ller}, P., {Watson}, D., {et~al.} 2015, \aap, 584, A100, \dodoi{10.1051/0004-6361/201526570}

\bibitem[{{Zhang} {et~al.}(2026{\natexlab{a}}){Zhang}, {Zhang}, {Wu}, {Ho}, \& {Wang}}]{ZhangC2026}
{Zhang}, C., {Zhang}, H., {Wu}, Q., {Ho}, L.~C., \& {Wang}, J.-M. 2026{\natexlab{a}}, arXiv e-prints, arXiv:2606.02773, \dodoi{10.48550/arXiv.2606.02773}

\bibitem[{{Zhang} {et~al.}(2026{\natexlab{b}}){Zhang}, {Egami}, {Sun}, {Lin}, {Lyu}, {Zhu}, {Rinaldi}, {Sun}, {Bunker}, {Bhatawdekar}, {Helton}, {Maiolino}, {Ma}, {Robertson}, {Tacchella}, {Venturi}, {Williams}, \& {Willott}}]{Zhang2026b}
{Zhang}, J., {Egami}, E., {Sun}, F., {et~al.} 2026{\natexlab{b}}, \apj, 997, 250, \dodoi{10.3847/1538-4357/ae2681}

\bibitem[{{Zhang} {et~al.}(2026{\natexlab{c}}){Zhang}, {Ding}, {Yang}, {Lambrides}, {Akins}, {Battisti}, {Casey}, {Chen}, {Cox}, {Faisst}, {Franco}, {Haghjoo}, {Ho}, {Inayoshi}, {Jin}, {Karmen}, {Koekemoer}, {Kartaltepe}, {Liao}, {Gozaliasl}, {Onoue}, {Kokorev}, {Roy}, {Rich}, {Silverman}, {Tanaka}, {You}, {Yesuf}, \& {Zavala}}]{Zhang2025}
{Zhang}, Y., {Ding}, X., {Yang}, L., {et~al.} 2026{\natexlab{c}}, Nature Astronomy, \dodoi{10.1038/s41550-026-02945-z}

\bibitem[{{Zhang} {et~al.}(2026{\natexlab{d}}){Zhang}, {Jiang}, {Liu}, {Ho}, \& {Inayoshi}}]{NL_LRDS}
{Zhang}, Z., {Jiang}, L., {Liu}, W., {Ho}, L.~C., \& {Inayoshi}, K. 2026{\natexlab{d}}, \apj, 998, 170, \dodoi{10.3847/1538-4357/ae33c4}

\end{thebibliography}

\null\clearpage
\begin{appendix}

\section{Dependence of the Spectral Fitting Ranges}
\label{apdx:fittingregion}

\begin{figure}
    \centering
    \includegraphics[width=0.95\columnwidth]{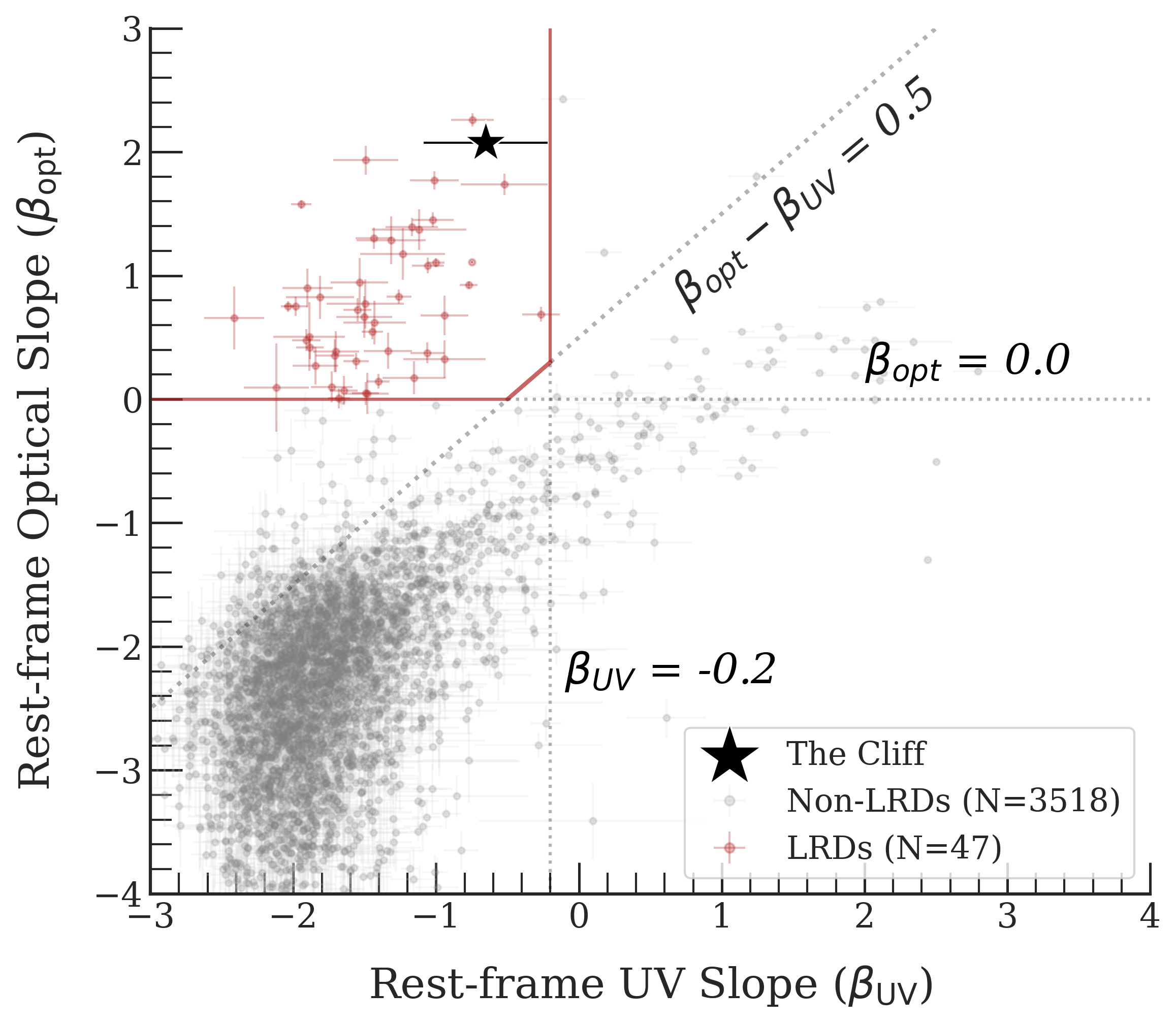} 
    \caption{The $\bopt$ versus $\buv$ distribution for all sources in our sample, adopting rest-frame UV and optical fitting regions similar to the methodology of \citet{deGraaff2025b}. The dotted gray lines indicate the LRD selection boundaries defined by \citet{HvidingRUBIES}, while the solid red lines forming the border of the LRD selection criteria adopted by \citet{deGraaff2025b}. The UV–optical regime is parameterized with a broken power-law anchored at the Balmer limit, spanning the wavelength range $0.15$–$0.60\,\mu\mathrm{m}$. The LRDs are shown as red points, while the gray points correspond to non-LRD sources. The Cliff is plotted explicitly using a black five-pointed-star.} 
    \label{fig:plot2_new}
\end{figure}

As detailed in \S\ref{subsec:curvefit}, we parametrize the power-law fits to the UV-to-optical spectral region by adopting a broken fitting interval of $0.15$–$0.36\,\mu\mathrm{m}$ for the UV regime and $0.40$–$0.60\,\mu\mathrm{m}$ for the optical regime. These fitting intervals were selected to (i) minimize the impact of noise, (ii) enable broader redshift coverage of the sample, and (iii) ensure a robust characterization of the optical continuum despite the presence of the Balmer break in evolved galaxies.

Figure~\ref{fig:plot2_new} presents our previously derived $\buv$--$\bopt$ distribution, now re-computed following a methodology similar to that of \citet{deGraaff2025b}. In particular, the rest-frame UV and optical wavelength intervals are fitted with a broken power law anchored at the Balmer limit over the range $0.15$–$0.60\,\mu\mathrm{m}$. This comparison highlights the sensitivity of the inferred slopes to the precise choice of fitting windows. In this configuration, The Cliff exhibits among the highest $\bopt$ values, consistent with its exceptionally strong Balmer break relative to other LRDs.

However, when we apply the exact same wavelength range used by \citet{deGraaff2025b}, a broken power-law fit at the Balmer limit that spans $0.12$–$0.70\,\mu\mathrm{m}$, the resulting distribution more closely resembles that shown in Figure~\ref{fig:plot2}, in which The Cliff no longer exhibits the maximum $\bopt$ value. We emphasize this point to warn that differences in the fitting regions adopted by various studies can introduce systematic discrepancies in the inferred slope values. As our parametrization strategy is specifically designed to robustly capture the continua of evolved galaxies with pronounced Balmer breaks, a clear vertical trend emerges in the $\buv$--$\bopt$ plane, which formed the basis of our morphological analysis.

\section{Extendedness difference between individual and stacked measurements}
\label{apdx:ERdifference}
Although both the median-stacked image and the ensemble of individual measurements exhibit the same increasing trend in the extendedness ratio (ER), the ER derived from the median-stacked image and the median of the individual ER values differ slightly in their absolute values. This appendix is devoted to revealing the origin of this discrepancy, which arises from differences in the underlying methodology.

Table~\ref{tab:psf_gap} presents, for each bin (1 through 4) and for both UV and optical regimes, the median ER derived from Figures~\ref{fig:ER+Median_Stack_UV} and \ref{fig:ER+Median_Stack_Optical} in the first column, the ER measured from the corresponding median stacked image in the second column, and, in the third column, the difference between these two median ER values. The discrepancy between the median ER obtained from the extendedness plots and that derived from the median-stacked images arises from the choice of reference PSF used to compute the ER. For the median ER in the extendedness plots, the ER is first measured for each individual source using its own native PSF associated with the selected filter. After that, these measurements are grouped by bin and the median ER is computed for each bin. For the extendedness plot, the ER can be expressed as follows:

\begin{equation}
    \mathrm{ER}_{\mathrm{own}}^{2}
    = \frac{\sigma_{\mathrm{obs}}^{2}}{\sigma_{\mathrm{srcPSF}}^{2}}
    = 1 + \frac{\sigma_{\mathrm{int}}^{2}}{\sigma_{\mathrm{srcPSF}}^{2}}.
    \label{eq:er_own}
\end{equation}

For the case of the stacked image, all sources must be convolved to match the PSF of the longest-wavelength band, which is F200W for the UV regime and F444W for the optical regime. This convolution introduces an additional $\sigma_{\mathrm{kernel}}$, which drives the overall median ER closer to unity, as indicated in equation~(\ref{eq:er_tgt}): 

\begin{equation}
    \mathrm{ER}_{\mathrm{tgt}}^{2}
    = \frac{\sigma_{\mathrm{matched}}^{2}}{\sigma_{\mathrm{tgtPSF}}^{2}}
    = 1 + \frac{\sigma_{\mathrm{int}}^{2}}{\sigma_{\mathrm{tgtPSF}}^{2}}
    = 1 + \frac{\sigma_{\mathrm{int}}^{2}}{\sigma_{\mathrm{srcPSF}}^{2} + \sigma_{\mathrm{kernel}}^{2}}
    \label{eq:er_tgt}
\end{equation}

Consequently, Table~\ref{tab:psf_gap} shows that the discrepancy generally increases with bin number, because an increasing fraction of sources in each bin are not observed in the longest-wavelength filter in their respective regime (F200W or F444W) and therefore require convolution. This is summarized in Table~\ref{tab:sources_per_filter}. In particular, there is a smaller fraction of UV sources in the F200W band (9/116, i.e. 7.8\%) compared to optical sources in the F444W band (25/130, i.e. 19.2\%). This imbalance causes the difference between the median ER derived from the extendedness plot and the ER derived from the median stacked image to be substantially larger in the UV regime than in the optical regime when examining each bin individually.

Based on the argument involving $\sigma_{\mathrm{kernel}}$, it is reasonable that the gap between the measured values in bins 1 and 2 in the UV regime, as reported in Table~\ref{tab:psf_gap}, remains small and shows a similar value. Both bins contain the same number of sources in the F200W filter (3) and nearly identical numbers of sources in the shorter-wavelength bands (13 for bin 1 and 14 for bin 2), as well as comparable total source counts (16 for bin 1 and 17 for bin 2). 

Since bins 3 and 4 exhibit systematically higher fractions of sources in filters other than F200W, relative to sources using the F200W filter, than those in bins 1 and 2, this supports the interpretation that the larger discrepancies observed in these bins are primarily driven by the relation described in equation~(\ref{eq:er_tgt}). A similar behavior is observed for bins 3 and 4 in the optical regime.

For bins 1 and 2 in the optical regime, the differences are again consistent with the expectations from equation~(\ref{eq:er_tgt}). As shown in Table~\ref{tab:sources_per_filter}, bin 1 contains a larger fraction of sources for which the native filter is not the longest-wavelength filter (F444W): 14/18 (77.8\%) in bin 1, compared to 11/18 (61.1\%) in bin 2. Consequently, a greater number of sources in bin 1 require convolution, which leads to a reduction in the measured ER for the median stacked image. Moreover, because the sources in bins 1 and 2 are already expected to be extremely compact \citep{Labbe2025, Barro2024}, the impact of convolution is small, given that the intrinsic ER values for objects in these bins are already close to unity.

\begin{table}
\caption{Median extendedness ratio (ER) measured at each source's native PSF, compared with the ER of the median-stacked image (PSF-matched to F200W for UV and F444W for optical; see Figs~\ref{fig:ER+Median_Stack_UV} and \ref{fig:ER+Median_Stack_Optical}), and their difference.}
\label{tab:psf_gap}
\centering
\begin{tabular}{lccc}
\toprule
Bin & \makecell{Median ER \\ (own PSF)} & \makecell{Stack ER \\ (target PSF)} & Gap \\
\midrule
Optical 1 & 1.38 & 1.29 & 0.09 \\
Optical 2 & 1.33 & 1.30 & 0.03 \\
Optical 3 & 1.76 & 1.43 & 0.33 \\
Optical 4 & 2.69 & 1.95 & 0.74 \\
\midrule
UV 1 & 2.18 & 1.62 & 0.56 \\
UV 2 & 2.42 & 1.81 & 0.61 \\
UV 3 & 3.03 & 1.90 & 1.13 \\
UV 4 & 4.19 & 2.86 & 1.33 \\
\bottomrule
\end{tabular}
\end{table}

\begin{table}
\caption{Number of sources per filter and per $\bopt$ bin, for the UV and optical samples.}
\label{tab:sources_per_filter}
\centering
\begin{tabular}{lccccc}
\toprule
Filter & Bin 1 & Bin 2 & Bin 3 & Bin 4 & Total \\
\midrule
\multicolumn{6}{l}{\textit{UV}} \\
F115W & 7  & 6  & 8  & 46 & 67  \\
F150W & 6  & 8  & 8  & 18 & 40  \\
F200W & 3  & 3  & 1  & 2  & 9   \\
Total & 16 & 17 & 17 & 66 & 116 \\
\midrule
\multicolumn{6}{l}{\textit{Optical}} \\
F277W & 7  & 5  & 9  & 46 & 67  \\
F356W & 7  & 6  & 6  & 19 & 38  \\
F444W & 4  & 7  & 8  & 6  & 25  \\
Total & 18 & 18 & 23 & 71 & 130 \\
\bottomrule
\end{tabular}
\end{table}

\section{Sources with Very Steep Optical Slopes}
\label{apdx:veryblue}

\begin{figure}[t]
    \centering
    \includegraphics[width=0.95\columnwidth]{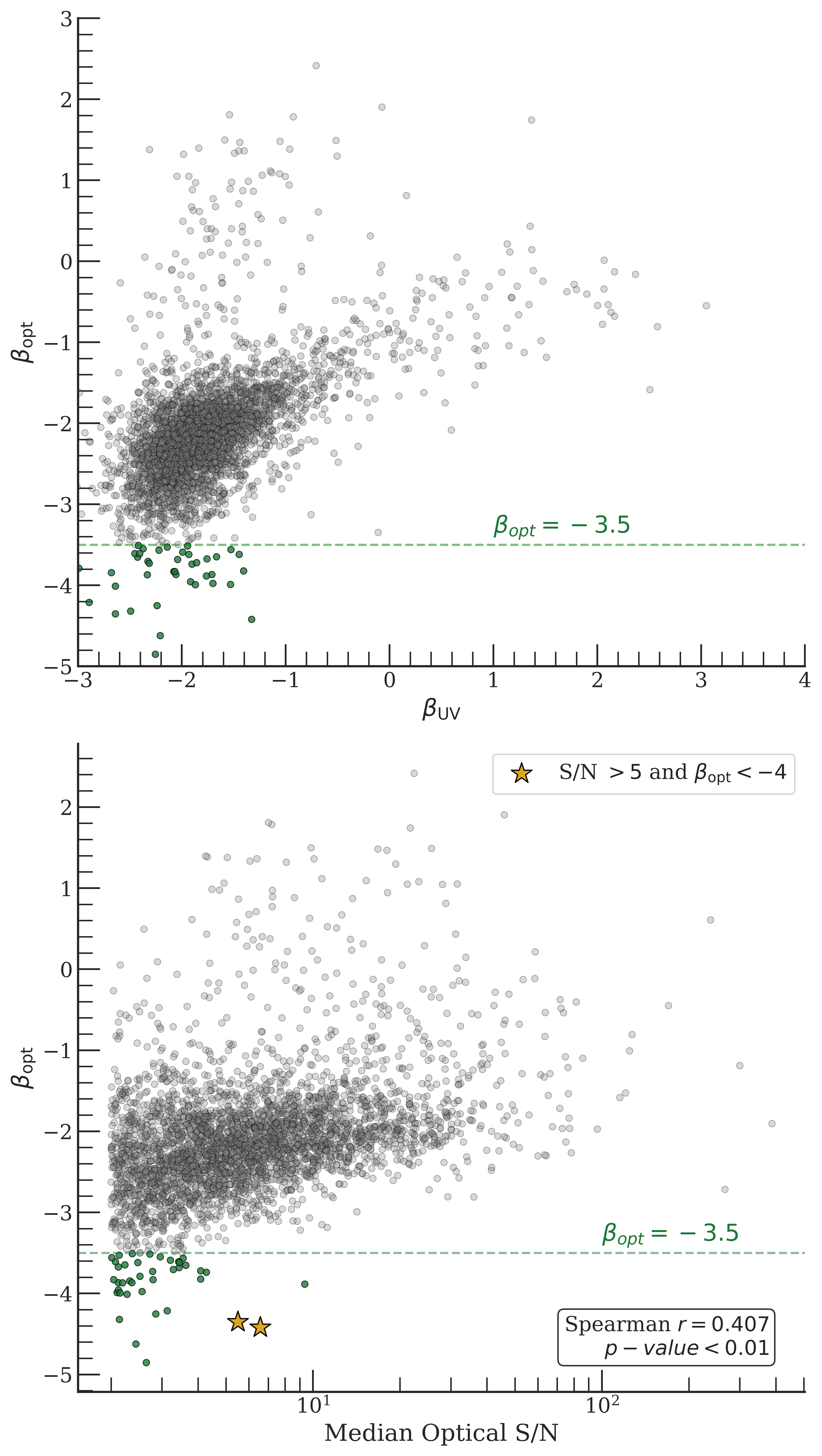} 
    \caption{\textbf{Top:} $\bopt$ versus $\buv$ diagram of our sample. The green points indicate sources that have $\bopt<-3.5$. The rest of the gray points are sources with $\bopt > -3.5$. The dotted green lines indicate the boundary for $\bopt = -3.5$. \textbf{Bottom:} The $\bopt$ values versus the median S/N in the optical region of our sample (logarithmic scale). The dotted green lines indicate the boundary for $\bopt = -3.5$. The golden five-pointed-stars indicate sources that have median optical $\mathrm{S/N} > 5$ and exhibit $\bopt<-4$. The Spearman coefficient of this graph is $0.407$ and the associated p-value less than 0.01.}
    \label{fig:betaopt_sn}
\end{figure}

In Figure~\ref{fig:circ_beta_dist}, there is a pronounced excess of sources with extremely blue $\bopt$ values (i.e. $\bopt < -3.5$). The threshold $\bopt < -3.5$ was defined arbitrarily, based on the radius adopted ($r=1.3$ in the $\bopt$ versus $\buv$ plane) in Figure~\ref{fig:circ_beta_dist}: the central coordinate $\bopt$ is approximately $-2.19$ derived from \S\ref{subsec:kde}, and subtracting $1.3$ yields a limit value of approximately $-3.5$. 

\begin{figure}[t]
    \centering
    \includegraphics[width=\columnwidth]{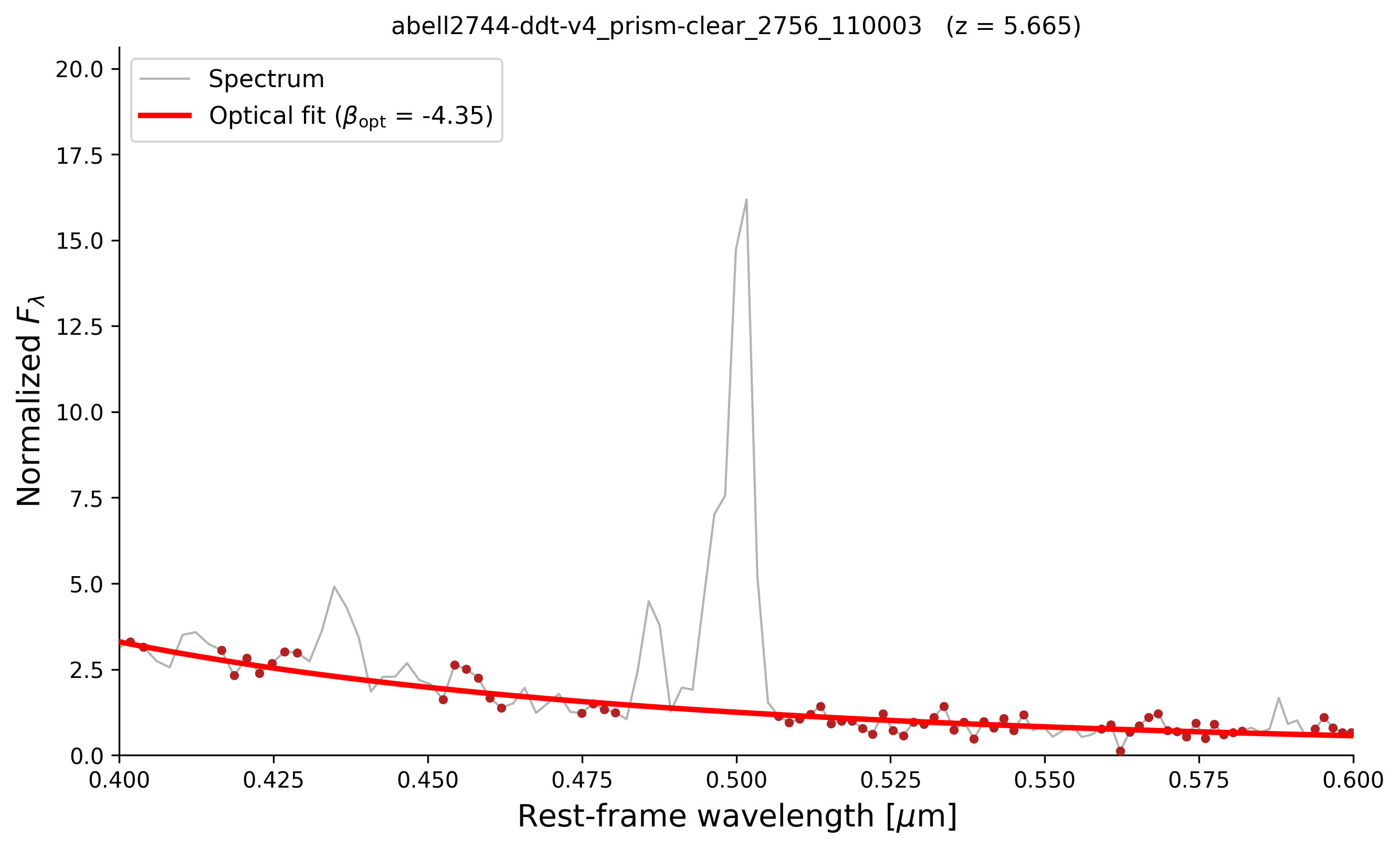}\\[2pt]
    \includegraphics[width=\columnwidth]{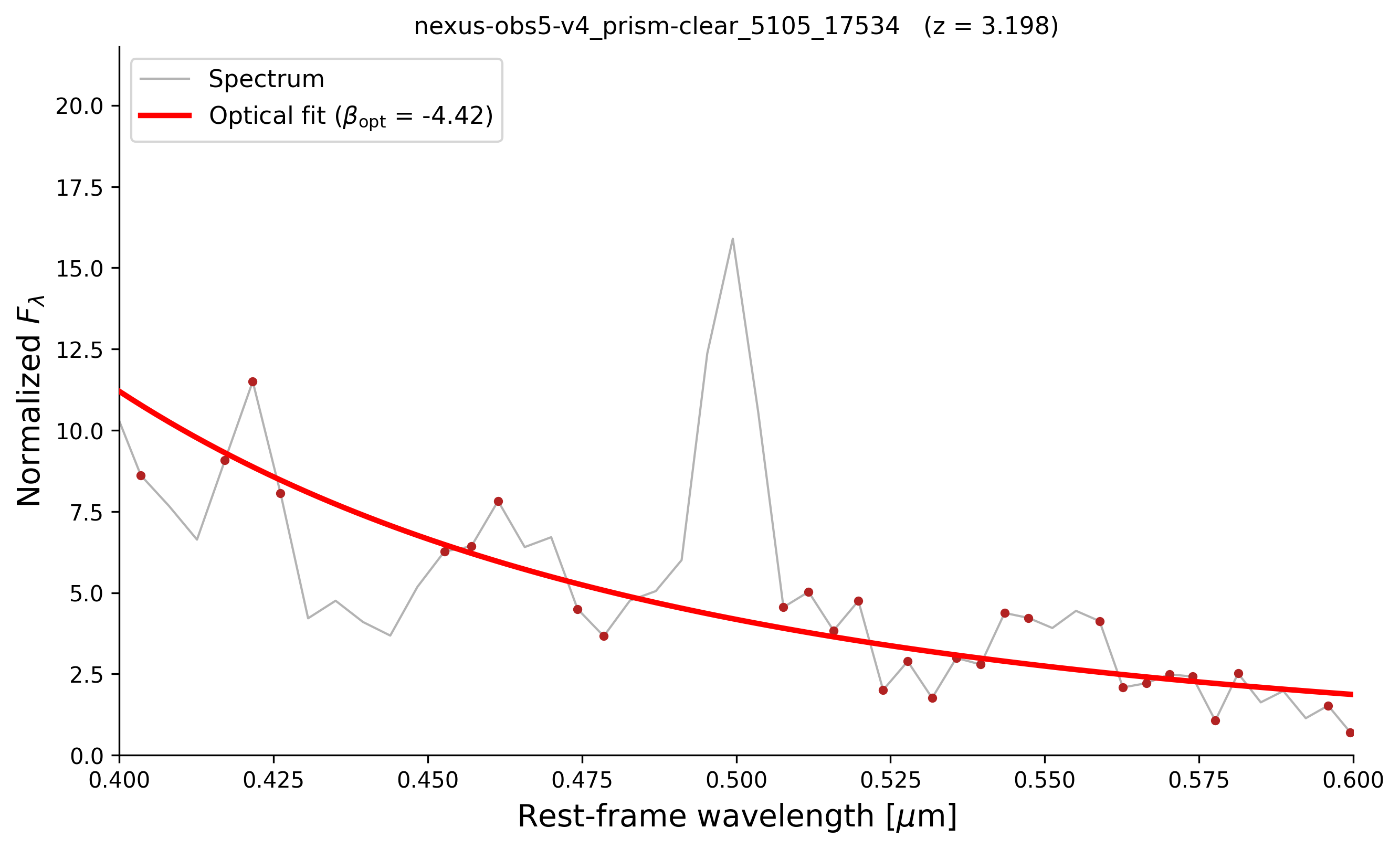}
    \caption{Normalized flux density against rest-frame wavelength over the optical fitting window for the two sources with median optical $\mathrm{S/N}>5$ and $\bopt<-4$. \textbf{Top:} \texttt{abell2744-ddt-v4\_2756\_110003} ($\bopt=-4.35$). \textbf{Bottom:} \texttt{nexus-obs5-v4\_5105\_17534} ($\bopt=-4.42$). The gray line shows the spectrum, the red points the unmasked pixels used in the fit, and the red curve the best-fitting power law.}
    \label{fig:steep_sources}
\end{figure}

These extreme values are most likely caused by the faintness of the sources in the optical continuum, which limits our ability to accurately determine the continuum slope using our curve-fitting method. As shown in Figure~\ref{fig:betaopt_sn}, there is a clear positive correlation between $\bopt$ and the median S/N across the defined optical region, as quantified by a Spearman correlation coefficient of 0.407. This indicates that the extremely negative $\bopt$ values are more plausibly attributed to instrumental sensitivity limitations of JWST/PRISM and the intrinsic faintness of the sources, rather than the sources genuinely exhibiting steep intrinsic optical continuum slopes.

According to the Rayleigh–Jeans limit of Planck’s law \citep{RayleighJeans}, the specific intensity can be approximated as
\begin{equation}
    I(\lambda, T) \approx \frac{2 c k_B T}{\lambda^4},
\label{Rayleigh_Jeans_Limit}
\end{equation}
where $c$ denotes the speed of light, $k_B$ is the Boltzmann constant, and $T$ is the blackbody temperature. In this classical regime, the expected asymptotic slope of the optical continuum, $\bopt$, is therefore $-4$. Consequently, values of $\bopt$ down to $-4$ may be attributed to intrinsically steep, but physically plausible, optical spectra.

However, we identify two sources with median optical S/N ratios (measured before masking the emission lines) of $\mathrm{S/N} > 5$ that exhibit $\bopt < -4$, steeper than any thermal (blackbody) continuum can produce in the Rayleigh-Jeans regime: \texttt{abell2744-ddt-v4\_2756\_110003} ($\bopt=-4.35$; top panel of Figure~\ref{fig:steep_sources}) and \texttt{nexus-obs5-v4\_5105\_17534} ($\bopt=-4.42$; bottom panel of Figure~\ref{fig:steep_sources}).

Since our median S/N estimate was calculated prior to masking the strong emission lines, the S/N of the continuum alone is likely lower than the calculated value. Moreover, visual inspection indicates that the low continuum S/N is the most plausible explanation for the unusually steep $\bopt$ value measured for source \texttt{nexus-obs5-v4\_5105\_17534}. However, source \texttt{abell2744-ddt-v4\_2756\_110003} appears to exhibit an intrinsically steep optical continuum that cannot be reproduced by any single-temperature thermal continuum and may instead indicate the presence of a non-thermal contribution.

Figure~\ref{fig:steep_sources} shows the optical spectra and power-law fits of these two sources, which retain exceptionally steep $\bopt$ values despite our signal-to-noise cut.

\end{appendix}

\end{document}